\documentclass[11pt]{article}  

\input{setup.sty}
\def\Hlarge{\cH_{\textit{large}}}
\def\ln{\log}
\def\OD{\textit{OD}}
\def\rest{\textit{sub}}
\def\DomD{\textit{DomD}}
\def\remain{\textit{remain}}
\def\etail{\e_{\textit{tail}}}
\def\eFLM{\e_{\textit{FLM}}}
\def\eisosmall{\e_{\textit{iso,small}}}
\def\epJLMS{\e_\textit{pJLMS}}
\def\eeJLMS{\e_\textit{eJLMS}^{(s)}}
\def\eiso{\e_{\textit{iso}}}
\def\eOD{\ve_\OD}
\def\eresttr{\ve_{\textit{sub-tr}}}
\def\erestlog{\ve_{\textit{sub-log}}}
\def\erestexp{\ve_{\textit{sub-exp}}^{(s)}}
\def\eenhanced{\ve_{\textit{enhanced}}}
\def\esmoothlog{\ve_{\textit{l-smooth}}}
\def\esmoothexp{\ve_{\textit{e-smooth}}^{(s)}}
\def\esupp{\textit{$\e$-supp}}
\def\ealign{\e_{\textit{align}}}
\def\elalign{\e_{\textit{l-align}}}

\def\cD{\mathcal D}

\title{The JLMS formula in a large code with approximate error correction}

\author[a]{Xi Dong,}
\author[a]{Donald Marolf,}
\author[b,c]{and Pratik Rath}

\affiliation[a]{Department of Physics, University of California, Santa Barbara, CA 93106, USA}
\affiliation[b]{Leinweber Institute for Theoretical Physics and Department of Physics,\\
University of California, Berkeley, California 94720, U.S.A.}
\affiliation[c]{Department of Theoretical Physics, Tata Institute of Fundamental Research, Homi Bhabha Rd,
Mumbai 400005, India}

\emailAdd{xidong@ucsb.edu}
\emailAdd{marolf@ucsb.edu}
\emailAdd{pratik\_rath@berkeley.edu}

\abstract{
Gauge/gravity duality is often described as a quantum error correcting code.  However, as seen in the Jafferis-Lewkowycz-Maldacena-Suh (JLMS) formula, exact quantum error correction with complementary recovery (and thus entanglement wedge reconstruction) emerges only in the limit $G \to 0$.  As a result, precise arguments controlling error terms have focused on what we call `small' codes which, as $G \to 0$,  describe only perturbative excitations near a given classical solution.  Such settings are quite restrictive and, in particular, they prohibit discussion of any modular flow that would change the classical background.  As a result, they forbid consideration of modular flows generated by semiclassical bulk states at order-one modular parameters.
 
In contrast, we present a single `large' code for the bulk theory that can accommodate such flows and, in particular, in the $G \to 0$ limit includes superpositions of states associated with distinct classical backgrounds. This large code is assembled from small codes that each satisfy an approximate Faulkner-Lewkowycz-Maldacena formula. In this extended setting we clarify the meaning of the (approximate) JLMS relation between bulk and boundary modular Hamiltonians and quantify its validity in an appropriate class of states.
}

\begin{document}

\maketitle

\section{Introduction}

\label{sec:intro}

The Jafferis-Lewkowycz-Maldacena-Suh (JLMS) relation \cite{Jafferis:2015del} forms a linchpin connecting many aspects of bulk reconstruction and quantum error correction in AdS/CFT \cite{Almheiri:2014lwa}.  It is generally understood to be valid at a level that includes the leading bulk quantum corrections, and for states in a `code subspace' of the CFT associated with appropriately semiclassical physics in the bulk.   Under such conditions, and for Einstein-Hilbert gravity with minimally-coupled matter, it reads
\begin{equation}\label{eq:JLMS}
P_{code}        \hat{K}_{R} P_{code}=\frac{\hat A(\gamma_R)}{4G}+ \hat{K}_{r},
\end{equation}
where $R$ is a region in the CFT, $\hat A(\gamma_R)$ is the bulk operator defined by the area of the associated Ryu-Takayanagi (RT) or Hubeny-Rangamani-Takayanagi (HRT)  surface $\gamma_R$ \cite{Ryu:2006bv,Ryu:2006ef,Hubeny:2007xt}, $P_{code}$ is the projection onto the chosen code subspace, and $\hat{K}_R$,
$\hat{K}_{r}$ are respectively the CFT modular Hamiltonian in $R$ and the bulk modular Hamiltonian in the entanglement wedge $r$ of $R$.  Throughout this work, we restrict to the case where $\gamma_R$ is compact and does not extend to the boundary so that $\hat A(\gamma_R)$ is finite and does not require renormalization. This may be arranged either by taking $R$ to be an entire connected component of the boundary or by working at finite radial cutoff in the bulk. We also remind the reader that a modular Hamiltonian $\hat K$ is defined in terms of the corresponding density operator (or density matrix) $\rho$ by
\begin{equation}
\label{eq:modularK}
\hat{K} = -\log \rho.
\end{equation}
 For higher-derivative theories and/or non-minimal couplings, the term $\hat A/4G$ should be replaced by the appropriate geometric entropy $\hat \sigma$  that appears in the corresponding generalization of the RT/HRT formula. See \cite{Dong:2013qoa} and references thereto for studies of such corrections.

The JLMS relation \eqref{eq:JLMS} implies that operators in the bulk entanglement wedge can be reconstructed in $R$ via an operator-algebra quantum error-correcting code (OAQEC) \cite{Dong:2016eik,Harlow:2016vwg}.  But holographic codes have additional structure that allows the reconstruction maps to be written in terms of modular flow and related constructions.
This is seen by first performing additional bulk computations to show that the OAQEC is associated with a flat entanglement spectrum, from which it follows that $\hat{K}_R$ commutes with $P_{code}$ \cite{Dong:2018seb,Akers:2018fow,Dong:2019piw}.  As a result, when acting on a state in the code subspace, one may drop the $P_{code}$ projectors from \Eqref{eq:JLMS} to write
\begin{equation}\label{eq:UPJLMS}
    \hat{K}_{R} =\frac{\hat A(\gamma_R)}{4G}+ \hat{K}_{r}.
\end{equation}
This unprojected form can then be exponentiated to relate bulk and boundary modular flows.  Such modular flows are then useful for providing explicit formulae for bulk reconstruction either directly \cite{Faulkner:2017vdd} or via the Petz map \cite{Cotler:2017erl,Chen:2019gbt}.  We will refer to the resulting relation between bulk and boundary modular flows as the `exponentiated JLMS relation,' even though direct exponentiation of \eqref{eq:JLMS} would produce a much less useful result involving extra insertions of $P_{code}.$

The goal of this work is to return to the JLMS relation \eqref{eq:JLMS},  to better understand its status for rather general classes of semiclassical bulk states, and to bound the associated error terms.    We are in particular motivated by a desire to set the stage for using the JLMS formula in a separate analysis of modular flows \cite{ModFlow}, and of the extent to which such flows are approximated by the classical Hamiltonian flow generated by $A(\gamma_R)/4G$.  Recall, however, that order-one modular flows defined by boundary duals of semiclassical bulk states generally do not preserve any given classical background; see \cite{Bousso:2020yxi,Kaplan:2022orm,Dong:2025orj}.  This can be seen, for example, in the familiar case of modular flows defined by thermo-field double states, where an order-one modular flow generates an asymptotic one-sided time-translation by an amount of order the inverse temperature $\beta$. Furthermore, since the classical flow generated by the HRT-area creates a kink (see again \cite{Bousso:2020yxi,Kaplan:2022orm,Dong:2025orj}), 
if an exponentiated JLMS relation does indeed hold in this context, and if it can in some cases be approximated by just the $A/4G$ term, then it is clear that the general case will be far more drastic.

In contrast, standard derivations \cite{Jafferis:2015del,Dong:2016eik} of the JLMS formula \eqref{eq:JLMS} are performed in the context of a code subspace consisting only of perturbative fluctuations about a given classical background. We refer to such code subspaces as `small.'  As in many discussions of JLMS, we assume that we work either with an entire connected component of the CFT spacetime or in a context with a UV cutoff on the CFT so that, at finite bulk Newton's constant $G$, we may treat the relevant von Neumann algebras as being of type I and thus admitting well-defined density operators.  Now, it is true that in the limit $G \rightarrow 0$ the errors in \eqref{eq:JLMS} vanish and perturbative excitations about two distinct backgrounds define  orthogonal quantum states. This then suggests that in this limit one may use the OAQEC formalism \cite{Harlow:2016vwg} to take a direct sum of such code subspaces over all classical geometries and to obtain a version of the JLMS relation valid on the resulting much larger (non-separable!) space.  However,   setting $G$ to zero before taking the direct sum removes any control over the form of the error terms that arise for $G\neq 0$.  Even worse is the fact that density operators for semiclassical states tend to become highly degenerate as $G\rightarrow 0$, with the majority of their eigenvalues vanishing exponentially in $1/G$.  As a result, the modular Hamiltonian $K=-\ln \rho$ diverges and the modular flow defined by such states becomes ill-defined. Taking this full limit also makes it difficult to discuss the operator $\hat A/4G$ in \eqref{eq:JLMS} which manifestly diverges in that limit; see e.g.\ the related discussion in~\cite{Chandrasekaran:2022cip,Chandrasekaran:2022eqq,Jensen:2023yxy,Chen:2024rpx} which considered a class of states with $\cO(G)$ area fluctuations, so that the operator $\frac{\hat A-A_0}{4G}$ remains well-defined as $G\rightarrow 0$.

We resolve this tension below by constructing a `large' code which contains states that can approach arbitrary classical solutions if one chooses to take the limit $G\rightarrow 0$.  However, we do not take this limit in full.  Instead, our construction is performed in the sense of an asymptotic expansion at small $G$. The large code is assembled from small codes to which one can apply the standard derivation of \eqref{eq:JLMS}, though these small codes are somewhat different from those discussed above.  By imposing appropriate error bounds motivated by AdS/CFT, and by keeping careful track of the cumulative errors, for an appropriate class of states we are then able to derive a version of \eqref{eq:JLMS} that holds on our large code with controlled errors that vanish in the limit $G\rightarrow 0$. 

We stress that there are several different sources of potential errors to be addressed.  The most obvious of these arises from the derivation of the JLMS relation \eqref{eq:JLMS} from the Faulkner-Lewkowycz-Maldacena (FLM) form of the quantum-corrected HRT formula \cite{Faulkner:2013ana}.  The FLM formula controls only terms of order $G^0$ so that, assuming a regular expansion in $G$, one will generally find errors at ${\cal O}(G^1)$.  Indeed, such terms are known to be associated with the lack of sharp edges for entanglement wedges at this order and with a corresponding expected failure for exact quantum error correction with complementary recovery (which we will also sometimes call `two-sided recovery'); see e.g.\ \cite{Harlow:2016vwg,Hayden:2018khn}. There are also additional small errors in the exponentiated JLMS formula associated with small departures from flatness in the code entanglement spectrum. Furthermore, to be conservative we should also admit the possibility of further (perhaps non-perturbative) errors that might arise if, for example, perhaps due to subtle wormhole effects, AdS/CFT turns out not to be an exactly isometric code with the bulk inner product defined using the usual gravitational path integral. This approach also allows us to use an approximate low-energy version of the bulk path integral in which, for example, perturbative higher derivative corrections might be included only to some finite order.

In addition, as discussed in \cite{Kudler-Flam:2022jwd}, certain types of states can provide much more sensitivity to such errors than one might naively expect.  We stress in Section~\ref{sec:AdS/CFT} that, when combined with any  bound on the accuracy of the FLM formula, this places a fundamental limit on our ability to derive a useful JLMS formula in completely general settings.   Our goal is thus to describe a particular setting where sensitivity to the above errors is minimized.  We will do so in two different ways. In the first, we require the state $\r$ defining the modular Hamiltonians to satisfy a property which we call log-stability and which in particular puts a positive lower bound on the eigenvalues of the reduced bulk state on $r$ in a small code -- this will be the approach taken in Theorems~\re{thmsmjlms}, \re{thmsmexp}, \re{thmlgjlms}, \re{thmlgjlmsop}, and~\re{thmlgexpop}. In the second, rather than requiring JLMS to hold as an operator equation, we ask JLMS to hold only at the level of expectation values in certain test states $\s$ that behave well with respect to $\r$ -- this will be the approach taken in Theorems~\re{thmsmjlmsrel}, \re{thmsmexprel}, and~\re{thmlgexp}.

In deriving our error bounds, it will be convenient to separate the quantum information arguments from the AdS/CFT arguments.  We will use this approach to sharpen the discussion of error terms even at the level of our small codes. In particular, our main results will be phrased as showing that, when the errors in certain approximate relations are subject to given bounds (generically denoted $\epsilon$), additional approximate relations then hold with other error bounds (set by functions of $\epsilon$).  For example, when the FLM relation holds up to given errors, and for appropriately stable states, we can then bound the errors in the JLMS relation \eqref{eq:JLMS}.  Arguments from AdS/CFT and, in particular, the bulk path integral, are then used to motivate the expected parametric dependence on $G$ of the $\epsilon$-parameters.   In such an AdS/CFT context, a few of the resulting $\epsilon$-parameters ($\eFLM$, $\etail$, $\eisosmall$, $\eiso$, $\elalign$, $\ealign$) are then naturally taken to be perturbatively small in $G$ while others ($\eOD$, $\eresttr$, $\erestlog$, $\erestexp$, $\eenhanced$, $\esmoothlog$, $\esmoothexp$) are expected to be non-perturbatively small.  We therefore distinguish the two sets of parameters by using the symbols $\epsilon$ vs $\varepsilon$ as above.      

We begin in Section~\ref{sec:AdS/CFT} by constructing the relevant small codes in the context of AdS/CFT.  Here we follow the approach of  \cite{Dong:2019piw}, which results in code subspaces somewhat different from those in \cite{Jafferis:2015del,Dong:2016eik}.  This process motivates the later formulation of certain fundamental error bounds used in our main arguments and, in particular, the dependence on $G$ of the parameters $\eFLM$, $\etail$ that will be introduced later.  We also take the opportunity to illustrate the fundamental issue mentioned above whereby seemingly-small error terms can be greatly enhanced by divergences (often associated with factors of $\rho^{-1}$ for a semiclassical density operator $\rho$) that arise as $G\rightarrow 0$. This section then largely concludes the AdS/CFT part of the paper. 

We then proceed to the part of our work addressing more technical quantum information and error manipulation arguments.  The first step is the 
derivation 
in section~%
\ref{sec:smallcodesubspace} of a careful bound on the error terms in the small code JLMS relation \eqref{eq:JLMS} in terms of i) the error $\eFLM$ in the FLM formula and ii) a parameter $\etail$ describing the degree to which a property  we call log-stability holds in the class of states one wishes to consider.  We also derive bounds on errors in the exponentiated version of \eqref{eq:JLMS}.  Section~\ref{sec:largecodesubspace} then assembles our large code and obtains error bounds in that context for both \eqref{eq:JLMS} and its exponentiated version.
Finally, Section~\ref{sec:discussion} provides a brief summary of our work, discussing future directions and connections to the recently studied type II algebras in \cite{Witten:2021unn,Chandrasekaran:2022cip}.


\section{AdS/CFT framework for small codes}
\label{sec:AdS/CFT}

This section provides a brief discussion of the small codes we envision within the framework of AdS/CFT.  In addition, we take this opportunity to quickly review the derivation given in \cite{Dong:2016eik} of the JLMS relation \eqref{eq:JLMS} from the FLM formula.  This review allows us to provide a non-technical discussion of where complications naturally arise when attempting to derive \eqref{eq:JLMS} in large subspaces of states and, in particular, in subspaces containing states that describe well-separated classical backgrounds in the limit $G\rightarrow 0$.

We also briefly describe our strategy for ameliorating such complications.  This will set the stage for a much more technical treatment of error bounds in later sections.  In particular, the discussion here will motivate the manner in which various error parameters introduced in later sections are expected to depend parametrically on $G$ when our results are used in AdS/CFT.  
As noted above, we assume that we work either with an entire connected component of the CFT spacetime or in a context with a UV cutoff on the CFT so that we may treat the associated von Neumann algebras as being of type I and thus admitting well-defined density operators.

As usual, we take the bulk side of the AdS/CFT correspondence to be described by a low-energy effective gravitational theory, which we assume to be valid below some cutoff scale $\L$. We will allow the bulk to be described by a low energy effective action (which may have been truncated to contain only a finite number of higher derivative terms) so that the bulk-to-boundary dictionary need not be exactly isometric.
While the traditional definitions of small codes given in \cite{Almheiri:2014lwa,Dong:2016eik} proceed by applying quantum perturbations to a given semiclassical reference state, we will instead find it convenient to begin with the construction of small codes outlined in \cite{Dong:2019piw}.  The differences between these two procedures will be explained below.

\subsection{Small codes with small area fluctuations}

We now describe the desired small codes by choosing a preferred set of bulk states. We refer to the span of these states as the `small' code Hilbert space. The corresponding subspace of the CFT Hilbert space will be called the small code subspace.\footnote{Throughout this work, a code Hilbert space means what is often called the `logical' Hilbert space for a code, and a code subspace means the corresponding subspace of the `physical' Hilbert space (which is the CFT Hilbert space here).}

The approach of \cite{Dong:2019piw} requires that we first choose the boundary region $R$ and then decompose the desired set of bulk states into subspaces defined by a set of non-overlapping small windows of eigenvalues of the form $[A-\e_{window}/2,A+\e_{window}/2)$ of the operator $\hat A(\g_R)$  describing the area of the HRT surface $\g_R.$  For the moment, let us focus on one such small window.   
We will choose the window width $\e_{window}$ to be of order $G^\b$ with some $\b>1$. The subspace of bulk states in a given area window will be called an area-window Hilbert space $\cH_{\text{area window}}$ (labeled by that area window).

The benefits of this approach for our current goals are closely related to the original motivations for its introduction in \cite{Dong:2019piw}.  In particular, taking $\epsilon_{window}$ to be of order $G^\b$ allowed \cite{Dong:2019piw} to show that the so-called entanglement-spectrum of the associated quantum code is approximately-flat which, as noted in the introduction, is required for a useful exponentiated JLMS formula to hold.\footnote{It would be interesting to better understand the relation to the importance of similar small windows in the construction of type II von Neumann algebras in \cite{Chandrasekaran:2022cip,Chandrasekaran:2022eqq,Chen:2024rpx}.}  We will similarly find it useful in analyzing the exponentiated form of the JLMS formula below.  In addition, for small $\epsilon_{window}$ we may now approximate the HRT-area operator by the c-number $A$,
which leads to further simplifications in our arguments.  

However,
in contrast to the perturbative code subspaces of \cite{Almheiri:2014lwa,Dong:2016eik}, the reader should be aware that our area-window Hilbert spaces $\cH_{\text{area window}}$ allow states with large geometric fluctuations.   In fact, since we required fluctuations of $\hat A(\g_R)/4G$ to vanish at small $G$, our states must have correspondingly {\it large} fluctuations of the conjugate time-shift; see e.g.\ related discussions in \cite{Marolf:2018ldl,Dong:2022ilf,Dong:2023xxe}.  For the moment, we also (at least in principle) allow large fluctuations in quantities localized away from the HRT surface, though such a
choice can make it difficult to satisfy some of the further restrictions that we will require; see the brief discussion at the end of Section~\ref{sec:subtleties}.

As noted in the introduction, an important property of the bulk-to-boundary map (at leading order in $G$) is that it defines a quantum error correcting code with 2-sided recovery.  In order to describe this in our setup,  we require our code Hilbert space to be a sum of tensor products of Hilbert spaces associated respectively with the entanglement wedge $r$ of $R$ and the entanglement wedge $\rb$ of the complementary region $\Rb$; i.e., the area-window Hilbert space $\cH_{\text{area window}}$ must satisfy
\begin{equation}
\label{eq:awsum}
{\cal H}_{\text{area window}} = \bigoplus_{\alpha \in J_{\text{area window}}} \left({\cal H}^\alpha_r \otimes {\cal H}^\alpha_{\rb} \right),
\end{equation}
where $J_{\text{area window}}$ is the set of all indices $\a$ for a given area window.
Each term in this sum will be called a `small' code Hilbert space:
\be
{\cal H}^\a = {\cal H}^\alpha_r \otimes {\cal H}^\alpha_{\rb}.
\ee
We will also later construct a `large' code which includes an additional sum over area-windows:
\be
\Hlarge = \bigoplus_{\text{area window}} {\cal H}_{\text{area window}} = \bigoplus_{\alpha} \left({\cal H}^\alpha_r \otimes {\cal H}^\alpha_{\rb} \right),
\ee
where $\a$ now takes values from the (disjoint) union of all index sets $J_{\text{area window}}$. In particular, the index $\a$ can be chosen to include the specification of the area window along with other labels that distinguish terms in the sum \er{eq:awsum}.

When we take the approach of using log-stability to control the accuracy of JLMS (as mentioned in the introduction), we will assume each small code Hilbert space to contain at most an order-$G^{0}$ number of states\footnote{Though this could be slightly relaxed to be e.g.\ of order $|\ln G|$.} in the limit $G \rightarrow 0$.  At first glance one might expect this to require the UV cutoff $\Lambda$ to be taken to be of order the AdS energy-scale (perhaps with a large order-one coefficient).   We therefore mention that, in appropriate contexts, one can take our small code Hilbert space to be a coarse-grained version of another code that contains many more states.   We will discuss a concrete way of doing so at the end of this subsection by applying the recent construction of \cite{RG}. Such a coarse-graining can allow one to push the strict cutoff parametrically further into the UV.

A useful code subspace will contain states in which the boundary entanglement is both large and highly structured. This would be in tension with taking the width $\epsilon_{window}$ of our area-window to be exponentially small, but there is no tension with our choice to take $\epsilon_{window}$ to vanish as a power of $G$.  For our purposes it will be useful to characterize the form of this entanglement by the existence of an approximate FLM relation.

In particular, focusing now on one such small code Hilbert space $\cH^\a$, we will often consider the case in which, for every density operator $\r$ on $\cH^\a$, the FLM formula holds in the form\footnote{Some of our later results (Theorems~\re{thmsmjlmsrel}, \re{thmsmexprel}, and~\re{thmlgexp}) require the FLM formula (or its R\'enyi version) only for a specific family of states, rather than for all states.}
\be\la{smflm}
S(\r_R) = \fr{A^\a}{4G} + S(\r_r) +o(G^0),
\ee
where the c-number $A^\a$ is the mid-point $A$ of the area window (up to a logarithmic correction arising from the window widths used to constrain the area and other labels specifying $\a$ in the gravitational path integral)\footnote{This correction can be understood as arising from the normalization of the fixed-area path integral: in deriving \er{smflm} using the replica method \cite{Lewkowycz:2013nqa}, the $n$-replica path integral $Z_n$ gets a factor of $\epsilon_{window}$ (made dimensionless using an implicit reference scale) when integrating the HRT-area over the window, and the boundary entropy $S_R = -\lim_{n\to 1} \pa_n \log (Z_n/Z_1^n)$ receives an additive correction of $\log\epsilon_{window}$.}, $\r_R$ is the reduced density operator on $R$ of the full boundary state corresponding to $\r$, and the reduced density operator $\r_r$ on the bulk entanglement wedge $r$ is obtained by tracing $\r$ over $\cH^\a_{\rb}$.  The von Neumann entropy $S(\r_r) = - \tr_r \left(\r_r \ln \r_r \right)$ is then computed using the trace on the Hilbert space $\cH_r^\a$.
When considering the case where the FLM formula \eqref{smflm} holds for all states in the small code, we will assume the $o(G^0)$ error term to be uniform over all such states; i.e., at each $G> 0$ the error term is bounded by some quantity that does not depend on the state $\rho$ but which (using the usual `little $o$' notation) vanishes in the limit $G\rightarrow 0$.

The reader should note two (related) differences between \eqref{smflm} and other common presentations of the FLM formula.  The first is that the right-hand side of \eqref{smflm} does not treat the HRT-area as an operator.  Instead, it refers to the c-number $A^\a$ that depends on the mid-point of the area-window.  The second difference is that, for this reason, we require only that the error term vanishes as $G\rightarrow 0$ and do not specify any particular rate at which it does so.  When the parameter $\beta$ governing the width of our area-window is sufficiently close to $1$, the rate at which our FLM-error vanishes can be an arbitrarily slow power law (though for $\beta >2$ the error is of order $G$ or smaller).

A convenient way to show the existence of small codes satisfying the above requirements, and to also provide an explicit construction of the associated states, follows from combining the low-energy effective bulk gravitational path integral (with a constraint fixing the relevant HRT area) with the notion of RG flow on error correcting codes discussed in \cite{RG}. 
To do so, we first relax the desired bound on the number of states and consider the bulk effective field theory at some desired scale $\Lambda$ which, for the moment, we take to be well above the AdS energy-scale but leave otherwise unconstrained.

The associated small code Hilbert space displays the desired tensor product structure at leading order in small $G$.  However,  it will contain a strictly infinite number of states in each factor due to the infinite volume of the AdS bulk as well as the potential presence of bosonic excitations.  For our results based on log-stability, we truncate each factor to a finite-dimensional space.\footnote{In contrast, the results of Section~\ref{sec:relativelogstable} are based on relative log-stability and do not require this truncation.}  This may be done, for example, by restricting each factor to an appropriate energy window,\footnote{Recall that we consider contexts where $\gamma_R$ is compact and does not reach the boundary.  This implies that $R$ and $\Rb$ are disjoint sets of boundaries, and thus that there are independent notions of bulk energy associated with the asymptotic time translations on $R$ and $\Rb$.} which of course also imposes an effective IR cutoff in the presence of asymptotically AdS boundary conditions.  The results in this work can then apply to states in this truncated Hilbert space, though in appropriate contexts one can then take limits that describe states in the full infinite-dimensional Hilbert space describing physics in the given effective field theory at small $G$.  An example of this sort will be described in the forthcoming work \cite{ModFlow}.

In this context, we can now derive an FLM formula of the form \eqref{smflm}. To do so, 
we note that the replica argument of \cite{Faulkner:2013ana} implies the approximate FLM formula \eqref{smflm} for any dual CFT state for which the corresponding bulk state can be computed by a bulk path integral.  In particular, this is the case even for bulk path integrals that include the insertion of bulk operators (or energy-window projections) in each entanglement wedge, and even in the case where the corresponding CFT operators are quite complicated (e.g., because the bulk operators lie outside the associated causal wedge, or perhaps even in a ``python's lunch'' behind a non-minimal extremal surface \cite{Brown:2019rox}).  The important point here is that the derivation relies only on knowing how to translate replica computations of $\Tr \rho_R^n$ into the language of bulk path-integrals; an explicit translation to the language of boundary path integrals is not required.  

We then note that such a bulk path integral prescription can be given for each state in our truncated small code Hilbert space.
In particular, we recall that values of the HRT-area in states defined by such path integrals can be restricted to the specified window by including an appropriate constraint on the geometries over which one integrates \cite{Akers:2018fow,Dong:2018seb}.  
While this constraint significantly changes the states when the window width $\e_{window}$ becomes small and, in particular, renders the associated time-shift-like operators far from being semiclassical, an FLM formula relating bulk and boundary entropies nevertheless follows from the standard replica argument; see e.g.\ \cite{Dong:2016hjy,Dong:2018seb}.  In particular, as noted in \cite{Dong:2019piw}, while our states are not semiclassical, their failure to be so can be seen only by observables which are not localized to either entanglement wedge (e.g., it {\it can} be seen by two-point functions with one operator in each entanglement wedge and by other operators sensitive to fluctuations in time-shift-like quantities).  As a result, such fluctuations have no effect on the density operator $\rho_r$ associated with the entanglement wedge of $R$, and thus do not affect the usual derivation of \eqref{smflm}.

We thus see that \eqref{smflm} holds separately for each state in the truncated small code Hilbert space.  To connect to the uniform bound over this small code assumed in the paragraph containing \eqref{smflm}, we note that since the explicit terms on both the left- and right-hand sides of \eqref{smflm} are continuous functions of the states,\footnote{Since the dimension of the boundary Hilbert space is infinite, the von Neumann entropy is not continuous on all CFT states.  However, all states in the truncated bulk Hilbert space have bounded energies.  As a result, their images in the CFT will have bounded expectation values of the CFT energy.  In this context, the desired continuity of CFT von Neumann entropy follows from e.g.\ Lemma~15 of \cite{Winter_2016}.}  the error term must be as well. And since the truncated small code Hilbert space has finite dimension, the space of normalized states is compact.  Thus the magnitude of the error term is maximized at some particular state in the small code Hilbert space. We further assume that this maximum is $o(G^0)$ as $G\to0$, as required above.

For our results based on log-stability, it remains only to ensure that the number of states in each small code Hilbert space is $\cO(G^0)$.  This may be done either by explicitly taking the UV cutoff $\Lambda$ to be sufficiently low (so that the entire area-window Hilbert space contains $\cO(G^0)$ states), or by using a higher-scale UV cut-off and choosing an appropriate set of seed states in that Hilbert space for which the coarse-graining algorithm of \cite{RG} constructs an IR code with an $\cO(G^0)$ number of states in each small code Hilbert space (i.e., in each tensor-product term in \eqref{eq:awsum}).\footnote{In \cite{RG} we used the convention $G=1/4$ while here we keep factors of $4G$ explicit.}  Here it is useful to recall from \cite{RG} that this is in particular the result whenever one starts with just a single seed state as, in that case, the resulting spaces ${\cal H}^\alpha_r, {\cal H}^\alpha_{\rb}$ in the IR code are all one-dimensional.

\subsection{Subtleties and error terms in deriving JLMS}
\label{sec:subtleties}

We now wish to follow the logic of \cite{Jafferis:2015del,Dong:2016eik} to derive the JLMS formula on a given small code Hilbert space $\cH^\a$ from small variations of \eqref{smflm}.  In particular, we wish to choose a state (density operator) $\r$ on $\cH^\a$ and then to vary it within the space of such states by some small $\d\r$, as in \cite{Jafferis:2015del,Dong:2016eik}.  The change of the left-hand side of \er{smflm} is then governed by an expansion in $\d\r_R$:
\be\la{sbdyv}
S(\r_R+\d\r_R)-S(\r_R)
=\<K_{\r_R}\>_{\d\r_R}+\cO\((\d\r_R)^2\),
\ee
where $K_{\r_R}=-\log\r_R$ is the boundary modular Hamiltonian for the subregion $R$, angle brackets denote expectation values, and we have noted the existence of second order terms in $\d\r_R$.  The change of the bulk entropy $S(\r_r)$ in \er{smflm} under a corresponding variation $\d\r_r$ is similar\footnote{This variation is consistent with our previous discussion of $S(\r_r)$ in terms of replica path integrals, because expectation values of $K_{\r_r}$ can be defined by analytic continuation of similar replica path integrals.}:
\be\la{sbulkv}
S(\r_r+\d\r_r)-S(\r_r)
=\<K_{\r_r}\>_{\d\r_r}+\cO\((\d\r_r)^2\),
\ee
where $K_{\r_r}=-\log\r_r$ is the bulk modular Hamiltonian for the entanglement wedge $r$.

For simplicity, let us now assume that the Hilbert space of the bulk low-energy effective theory of gravity is isometrically embedded in the boundary Hilbert space.  We can then identify our small code Hilbert space $\cH^\a$ with its image in the boundary Hilbert space. As a result, for the rest of the section the term `code Hilbert space' will be used interchangeably with the term `code subspace'.

If we can safely neglect the second (and higher) order terms in \er{sbdyv} and \er{sbulkv}, those relations become linear in $\d\r$ (with each relation taking the form of the `first law of entanglement').  In this case, and if we can ignore the $o(G^0)$ error term in \eqref{smflm}, the variation of our FLM formula leads to
\be
\<K_{\r_R}\>_{\d\r_R}=\<K_{\r_r}\>_{\d\r_r}
\ee
for every traceless variation $\d\r$ within $\cH^\a$.  Since this holds for every such $\d\r$, in the finite-dimensional, isometric setting considered in this motivational argument we find
\be
P_{code} K_{\r_R} P_{code} = K_{\r_r} + \text{const},
\ee
where $P_{code}$ denotes the projection from the full boundary Hilbert space onto the specified small code Hilbert space. Using the FLM formula to fix the constant, one then obtains the JLMS formula
\be\la{jlmssm}
P_{code} K_{\r_R} P_{code} = \fr{A^\a}{4G} + K_{\r_r}.
\ee
As described in \cite{Dong:2016eik}, the projections $P_{code}$ arise from the fact that it is useful to vary \eqref{smflm} only within the subspace on which that result is known to hold.

We will eventually wish to extend this result to a large code formed by taking the direct sum of many small code Hilbert spaces.   As a result, the large code will include superpositions involving different area-windows that describe well-defined classical backgrounds in the limit $G\rightarrow 0$ and,  in fact, it will include states that approach many distinct classical backgrounds.
However, before constructing this large code, it remains to address the conditions under which one can in fact neglect terms in \er{sbdyv} and \er{sbulkv} that are higher than linear order in $\delta \rho$.
The danger is that such higher order terms contain inverse powers of $\r_r$ or $\r_R$ (collectively called $\r$ for the rest of the subsection) due to variations of $\log \rho$ in $S=-\Tr(\rho \log \rho)$. As a result, such terms can be strictly infinite when $\r$ is not full rank and when $\d \r$ has a nonzero component on the kernel of $\r$. This is the effect that led to the violations of the JLMS relation described in \cite{Kudler-Flam:2022jwd}.  In Appendix~\re{sec:eg}, we present a corresponding example where JLMS fails in this way in a simple random tensor network setting. Since $\rho$ and $\delta \rho$ need not commute, the precise form of the above higher order terms is rather complicated.
 Such terms were discussed in e.g.\ \cite{Faulkner:2014jva,Lashkari:2018tjh}  and will be discussed further in Section~\ref{sec:smallcodesubspace}.

Furthermore, even when the higher order terms in $\d\r$ are not strictly infinite, since they scale with inverse powers of $\rho$ they can easily be much larger than $K=-\ln \rho$.  As a result, the $\cO(\delta \rho^2)$ corrections to the entropy can be large in comparison with the desired $\cO(\delta \rho)$ terms unless the perturbation $\delta \rho$ is taken to be {\it extremely} small.  The danger is then that, for such small $\delta \rho$, variations of the right-hand side of the FLM formula \eqref{smflm} may be dominated by the $o(G^0)$ error term rather than by the desired contribution from the bulk modular Hamiltonian.  In this case, we may again find large violations of \eqref{jlmssm}.   Since the issue involves possible large changes in the {\it derivative} of the entropy, results like the Fannes-Audenaert inequality that bound changes in the entropy itself do not appear to provide useful control.

The potential for large corrections becomes particularly acute if we allow our small code Hilbert space to contain two semiclassical states that are peaked around classical backgrounds that are distinct in $r$ and if we choose $\r$ to be one of the two states.  In such cases, the overlap of the two states is typically exponentially small in $G$.  As a result, such a construction could easily lead to exponentially small eigenvalues of $\rho_r$, so that the error terms in \eqref{sbdyv} and \eqref{sbulkv} contain factors that are exponentially large.  Controlling such terms using the above argument would then require taking $\delta \rho$  to be  exponentially small.  But when $\delta \rho$ is this small, variations in the FLM formula \eqref{smflm} can easily be dominated by variations of the $o(G^0)$ error term.  In this case, the above argument would allow large violations of the JLMS relation \eqref{jlmssm}.  

However, the above discussion also shows that the problem here can be ameliorated by simply restricting the derivation of \eqref{jlmssm} to reduced density operators whose eigenvalues are bounded below (so that terms involving $\rho^{-1}$ have bounded effect).  This is the essence of the log-stability criterion that we will introduce in Section~\ref{sec:smallcodesubspace} below, and which will be applied on small code Hilbert spaces.  Since, in a Hilbert space of dimension $D$, the smallest eigenvalue of a normalized density operator can be no larger than $1/D$, our restriction that (in a small code Hilbert space) $D$ be of order $G^0$ is important in satisfying this criterion.

What remains is then to keep track of the errors in detail, taking into account that we wish to work at small but finite $G$, and to then assemble the above JLMS relations on our small code Hilbert spaces into a corresponding relation on the appropriate large code.  As noted in the introduction, if there were no errors, and if the small codes satisfy appropriate orthogonality conditions (to be explained in the next paragraph), then one could simply take a direct sum of the small-code JLMS relations to derive the desired result on the large code defined by the corresponding direct sum.  Treating corrections perturbatively when there are small eigenvalues will then require additional `alignment' or `smoothness' assumptions as will be explained in Section~\re{sec:largecodesubspace}.

We now discuss the above-mentioned orthogonality conditions required for this direct sum construction. There are in fact two classes of such conditions:
First, to eliminate cross-terms in the density operators $\rho_R, \rho_r$,  states in the tensor factors associated with $\Rb, \rb$ for one small code must be orthogonal to states in the corresponding factor for other small codes.  Second, to write the result in a block diagonal form, 
states in the tensor factors associated with $R, r$ for one small code must be orthogonal to states in the corresponding factor for other small codes.  In Section~\ref{sec:largecodesubspace} below we will thus make a corresponding `approximate subsystem orthogonality' assumption that encompasses both conditions. As a result, if two small codes labeled by $\a, \b$ consist of states with distinct classical backgrounds, our formalism will generally be restricted to cases where the backgrounds on each of the two entanglement wedges $r,\rb$ are {\it separately} distinct.

\section{JLMS with error bounds in a small code }
\label{sec:smallcodesubspace}

We now wish to codify the senses in which an approximate JLMS formula holds on each small code.  As described in the introduction, we are interested both in the original `projected' JLMS relation \eqref{eq:JLMS} (involving non-trivial projections $P_{code}$), and also in the exponentiated JLMS relation.  We will find it convenient to separate our discussions by establishing an approximate version of the projected JLMS relation in Section~\ref{sec:originalJLMS} and then deriving approximate forms of the exponentiated relation in Section~\ref{sec:expJLMS}.

As also discussed in the introduction, exponentiating JLMS requires the additional input that the entanglement spectrum of the code is close to flat.  In Section~\ref{sec:expJLMS} we will encode this input by assuming the R\'enyi version of the FLM relation for appropriate complex replica numbers.

In Sections~\ref{sec:originalJLMS}--\ref{sec:expJLMS}, we will show that \eqref{eq:JLMS} (or exponentiated JLMS) holds with small errors in the sense that the difference between the desired terms on the right and left sides has small operator norm on our small code Hilbert space.  
In Section~\ref{sec:relativelogstable}, we will derive expectation-value versions of these JLMS results under weaker assumptions.

We remind the reader that, for any bounded Hermitian operator $O_{\text{Hermitian}}$,  the operator norm $\|\cdot\|_\infty$ may be defined as
\be\la{onalt}
\|O_{\text{Hermitian}}\|_\infty = \sup \big\{\lt|\Tr(\r\, O_{\text{Hermitian}})\rt|: \r \text{ is a normalized density operator}\big\}.
\ee
Moreover, for any bounded (but not necessarily Hermitian) operator $O$, we have
\be\la{onbound}
\|O\|_\infty \leq 2 \sup \big\{\lt|\Tr(\r\, O)\rt|: \r \text{ is a normalized density operator}\big\},
\ee
which can be verified by decomposing $O= O_1+i O_2$ with $O_1$, $O_2$ both Hermitian, using $\|O\|_\infty \leq \|O_1\|_\infty + \|O_2\|_\infty$, and noting $\lt|\Tr(\r\, O_i)\rt| \leq \lt|\Tr(\r\, O)\rt|$ for $i=1,2$.
The operator norm is also known as the Schatten $p=\infty$ norm; further review of the Schatten $p$-norms for general $p$ can be found in Appendix~\ref{app:Schatten}.

Below we occasionally use expectation values of self-adjoint operators $O$ that need not be bounded.  We continue to denote such an expectation value by $\<O\>_\t$, defining it in such cases by first imposing a spectral cutoff on $O$ and then removing the cutoff; see \er{EVdef} and \er{EVdefgen} in Appendix~\re{app:Schatten}.  If $O$ is bounded on the relevant Hilbert space, this reduces to the ordinary trace $\Tr(\t O)$.

\subsection{Projected JLMS}
\label{sec:originalJLMS}

We will soon derive our approximate version of the projected JLMS relation \eqref{eq:JLMS}.
Before doing so, we need to give four definitions. For clarity, we now change notation and use $\td\r$ to denote the boundary state corresponding to a bulk state $\r$.

The first definition will provide a precise version of the notion of small code described in Section~\ref{sec:AdS/CFT}.  Here the role of the bulk-to-boundary dictionary will be played by a linear map $V_\alpha$ which we allow to be non-isometric, and which we define separately on each small code Hilbert space ${\cal H}^\alpha$.  This potential non-isometricity allows for the possibility that AdS/CFT might fail to be an exact duality, or that one might choose to use a simplified version of the bulk theory that fails to capture the full details of the CFT.  For example, if the bulk gravitational theory leads to the superselection sectors of \cite{Coleman:1988cy,Giddings:1988cx,Marolf:2020xie} associated with particular baby universe states, one might choose to work with the simpler no-boundary state rather than the particular baby universe state dual to a given CFT, or one might similarly neglect contributions from higher topologies in other contexts.

\begin{ndefi}[Small code]\la{defsmallcode}
Let $\cH^\a$ (where $\a$ is a fixed label) and $\tH$ be possibly infinite-dimensional `bulk' and `boundary' Hilbert spaces, respectively, with tensor factorizations $\cH^\a = \cH_r^\a \otimes \cH_\rb^\a$, $\tH = \tH_R \otimes \tH_\Rb$. We refer to the labels $R, \Rb$ as boundary regions and $r,\rb$ as entanglement wedges.
We use a bounded\footnote{The results of this paper can also be formulated without assuming $V_\a$ to be bounded, but we impose boundedness here to avoid additional technicalities.} linear map $V_\a: \cH^\a \to \tH$ to take a state $\r^\a$ on $\cH^\a$ to a corresponding `boundary' state $\td\r^\a$ defined by\footnote{Throughout this work, we call $\r$ a state (or a density operator) if $\r$ is non-negative and of trace $1$. In particular, states are properly normalized unless stated otherwise.}
\be
\label{eq:tilderho}
\td\r^\a= W_\a(\r^\a):=\fr{\hat\r^\a}{\Tr \hat \r^\a},\qu
\hat\r^\a := V_\a \r^\a V_\a^{\dag},
\ee
whenever $\Tr\hat\r^\a>0$.
Here $\hat\r^\a$ depends linearly on $\r^\a$ but is not properly normalized as a state, whereas $\td\r^\a$ is properly normalized but is obtained via the nonlinear map $W_\a$.
We use $\td\r^\a_R := \Tr_\Rb \td\r^\a$, $\r^\a_r := \tr_\rb \r^\a$ to denote the reduced density operators.\footnote{We use $\tr$ to denote the bulk trace and $\Tr$ to denote the boundary trace (in contexts where both traces appear).  Partial traces are denoted $\tr_r$, $\Tr_R$, and so on.}
For a different state $\s^\a$ on $\cH^\a$, its corresponding boundary state is naturally called $\td\s^\a$.
Under the above conditions, we define $\tH^\a_{\text{code}}$ to be the image of $V_\a$ and call it a small code subspace. The map \er{eq:tilderho} has a natural analogue that acts on pure states as
\be
\label{eq:tildepsi}
|\td\psi^\a\rangle= \fr{V_\alpha |\psi^\a\rangle }{ \sqrt{\langle \psi^\a| V_\alpha^\dagger V_\alpha |\psi^\a\rangle}}.
\ee
This normalized vector is defined whenever $V_\a|\psi^\a\rangle\neq0$.
\end{ndefi}

It is worth noting that rescaling $V_\a$ by a nonzero constant does not affect the `bulk-to-boundary' map $W_\a$ nor any statements about boundary states $\td\r^\a$. Thus we are free to rescale $V_\a$ by any convenient positive constant. 
 We will make use of this freedom in Appendix~\re{app:FLMiso}.  

Our second definition simply codifies the sense in which we take an approximate version of the FLM formula to hold for each small code:

\begin{ndefi}[Approximate FLM]\la{defflm}
We say that the approximate FLM formula holds for the $\alpha$-sector on $R$ if every state $\r^\a$ on $\cH^\a$ satisfies
\be\la{flme}
\lt| S(\td\r^\a_R) - \( \fr{A^\a}{4G} + S(\r^\a_r) \) \rt| \le \eFLM
\ee
with a small $\eFLM$ and a real c-number $A^\a$, both independent of $\r^\a$.
\end{ndefi}

We will refer to $A^\alpha$ as the effective area eigenvalue for ${\mathcal H}^\alpha$, though it need not represent a strict geometric area (as we noted in Section~\ref{sec:AdS/CFT}).
Moreover, our argument below applies equally-well to contexts where the bulk theory receives higher derivative corrections so long as one uses the appropriate geometric entropy.
The important point is that every state in ${\mathcal H}^\alpha$ is a (degenerate) eigenstate of the effective-area operator that we will eventually build by combining various $\alpha$-sectors. 

Note that Definition~\re{defflm} implicitly requires  every $\r^\a$ on $\cH^\a$ to correspond to a well-defined boundary state $\td\r^\a$. This further implies that $V_\a$ is injective: if it is not, $V_\a$ would have at least one null state $|\y^\a\> \in \cH^\a$ satisfying $V_\a |\y^\a\> = 0$, and thus the bulk state $\r^\a = |\y^\a\> \<\y^\a|$ would not define any boundary state (because there is no way to normalize $\hat \rho^\a=0$).

Since the goal of this work is to derive JLMS-like formulae that hold to good approximation in the limit $G\rightarrow 0$, it is natural to assume that $V_\alpha$ deviates only slightly from being an isometry.  We therefore make the following definition.

\begin{ndefi}[Approximate isometry of the small code]\la{defappiso}
We say that $V_\a$ is an approximate isometry if
\be\la{appiso}
\lt\|V_\a^{\dag} V_\a -\mathbb 1^\a \rt\|_\infty \leq \eisosmall
\ee
with some small $\eisosmall$. Here $\mathbb 1^\a$ is the identity on $\cH^\a$. Using \Eqref{onalt}, this is equivalent to the condition that for every state $\r^\a$ on $\cH^\a$,
\be\la{appiso2}
\lt| \Tr \hat\r^\a - 1 \rt| = \lt| \tr \(\r^\a \[V_\a^\dag V_\a -\mathbb 1^\a \] \) \rt| \leq \eisosmall.
\ee
\end{ndefi}

When the bulk path integral is defined by a truncated low energy effective action (including, e.g., only a finite number of higher derivative terms), the parameter $\eisosmall$ should be expected to be perturbative (power law) in $G$. 
We have thus followed the convention described in the introduction where we use the symbol $\epsilon$ for errors of such sizes.  
As an aside, it turns out that when $\Rb$ and $\rb$ are empty (i.e., when $R$ and $r$ are the entire system), one can in fact use \eqref{flme} to derive \eqref{appiso} with the rather weak bound $\eisosmall = 2 \sqrt{\eFLM}$.   This interesting result is derived in Appendix~\ref{app:FLMiso} but will play no special role in our analysis below.

Our fourth definition defines the log-stability criterion that we will use to select states about which variations of $S(\tilde \rho_R^\alpha)$ and $S(\rho_r^\alpha)$ are well-behaved.  As suggested in Section~\ref{sec:AdS/CFT}, in the bulk it is just a lower bound on the eigenvalues of $\rho^\a_r$.  However, we will also need a condition on the boundary density operator $\td\rho^\a_R$ that will play a corresponding role for $S(\tilde \rho_R^\alpha)$. Since $\td\rho_R^\alpha$ generally has eigenvalues that are exponentially small in $G$, we should not attempt to restrict its smallest eigenvalue.  Instead, we use a condition on $\tilde \rho_R^\alpha$ that plays an analogous role when $\rho^\alpha$ is varied within the code Hilbert space ${\cal H}^\alpha$, and which we expect to follow from the above bulk condition in the context of AdS/CFT (in the case where the parameter $\etail$ introduced below is not exponentially small).  We nevertheless list it here as a separate requirement so that our results can be applied to more general discussions of quantum codes and to take into account potentially subtle cases in the AdS/CFT argument.  

As usual, the AdS/CFT argument is based on analytic continuation of replica computations involving gravitational path integrals. Since it is of a rather different character than our main discussion, we relegate it to Appendix~\ref{sec:extra}.  However, we provide some further comments below our definition showing that the bound on eigenvalues required by Condition~1 would in fact imply Condition~2 in the context of an exact quantum code with exact two-sided recovery.  

\begin{ndefi}\la{defsmallstable}
A state $\r^\a$ on $\cH^\a$ is called log-stable if it satisfies the following two conditions with some small $\etail>0$:
\begin{enumerate}[label=Condition \arabic*:,leftmargin=*]
\item Every eigenvalue of $\r_r^\a$ is at least $\etail$.
\item $\< (\td\r^\a_R)^{-1}\>_{(\td\s^\a_R)^2} \leq 1/\etail$ for every state $\s^\a$ on $\cH^\a$.\footnote{\la{foot:support}Throughout this paper, the inverse and logarithm of a positive operator $O$ are defined on the support of $O$. Thus Condition~2 requires the support of $\td\s_R^\a$ to be contained in the support of $\td\r_R^\a$ for every $\s^\a$ under consideration. Note that we call an operator positive if it is positive semi-definite.}
\end{enumerate}
\end{ndefi}

Note that Condition~1 can hold only if $\cH_r^\a$ is finite-dimensional (indeed, its dimension is at most $1/\etail$).

As advertised above, we now offer the following comments on Condition~2 from the code perspective. First, we note that Condition~2 is natural in the sense that it is a boundary analogue of Condition~1. In particular, Condition~1 is equivalent to Condition~2 with $R$ replaced by $r$ (and with tildes removed), as shown by the following lemma.

\begin{nlemma}\la{lemmacond1}
Condition~1 of Definition~\re{defsmallstable} is equivalent to the following:
\begin{enumerate}[label=Condition \arabic*':,leftmargin=*]
\item $\<(\r^\a_r)^{-1}\>_{(\s^\a_r)^2} \leq 1/\etail$ for every state $\s^\a$ on $\cH^\a$.
\end{enumerate}
\end{nlemma}

\begin{proof}
Condition~1 implies that $(\r^\a_r)^{-1}$ is defined on the entire $\cH^\a_r$ with
\be
\lt\|(\r^\a_r)^{-1}\rt\|_\infty\leq\fr{1}{\etail},
\ee
and hence
\be
\<(\r^\a_r)^{-1}\>_{(\s^\a_r)^2}
\leq\fr{1}{\etail}\tr_r\!\left[(\s_r^\a)^2\right]
\leq\fr{1}{\etail}.
\ee
Conversely, Condition~1' implies Condition~1 by choosing $\s_r^\a$ to be the density operator for an arbitrary eigenstate of $\r_r^\a$.
\end{proof}

Our second comment on Condition~2 is that, as a result of Lemma~\ref{lemmacond1}, if we have an exact FLM formula, and therefore an exact quantum error-correcting code with complementary recovery, then Condition~2 would follow directly from Condition~1.   To see this, simply recall from~\cite{Harlow:2016vwg} that in such cases we may always write $\td\r^\a_R = U_R (\r^\a_r \otimes \c^\a) U_R^{-1}$ for any state $\r^\a$ on $\cH^\a$, where $\c^\a$ is a properly normalized density operator and $U_R$ is an appropriate unitary.  As a result,
we would have
\be\la{treq}
\<(\td\r^\a_R)^{-1}\>_{(\td\s^\a_R)^2}
=
\<(\r^\a_r)^{-1}\>_{(\s^\a_r)^2}.
\ee
for any two states $\r^\a$ and $\s^\a$ on $\cH^\a$. Combining \eqref{treq} with the result of Lemma~\re{lemmacond1} then shows Condition~1 to imply Condition~2 in this context.   Of course, we are interested in the broader context in which we have only an approximate FLM formula and thus an approximate code, but it is perhaps reasonable to expect that \Eqref{treq} holds approximately so that Condition~2 is still implied by Condition~1 (up to an $\cO(1)$ constant in front of $\etail$). Indeed we provide a direct AdS/CFT argument for \Eqref{treq} in Appendix~\ref{sec:extra}. It would be interesting to derive such a result from an approximate-code perspective as well.

We are now nearly ready to derive an approximate version of the JLMS formula.  However, before doing so, we first pause to establish the following useful lemma.

\begin{nlemma}\la{lemmacondee}
Let $\r$, $\s$ be states in some Hilbert space, not necessarily of finite dimension.\footnote{We will make use of this lemma in both the bulk and the boundary Hilbert spaces.} Suppose
\be\la{trrssbd}
\<\r^{-1}\>_{\s^2}\leq 1/\etail.
\ee
For any $\e\in(0,1]$, define
\begin{equation}
\label{eq:redef}
\r_\e:=\r+\e(\s-\r).
\end{equation}
Then as long as $S(\r)$ and $S(\r_\e)$ are both finite, we have
\be\la{Svar2}
\bigl|S(\r_\e)-S(\r)-\e\<K_\r\>_{\s-\r}\bigr|
\leq \fr{\e^2}{\etail},
\ee
where $K_\r := -\log \r$.
\end{nlemma}

\begin{proof}
Let $D_1(\r_\e\|\r)$ denote the relative entropy.  By Lemma~\re{lemmainversehs} in Appendix~\re{app:Schatten}, the support of $\s$ is contained in the support of $\r$, and $\r^{-1/2}\s$ is a Hilbert--Schmidt operator with
\be
\left\|\r^{-1/2}\s\right\|_2^2
=\<\r^{-1}\>_{\s^2}.
\ee
It follows that
\be\la{eq:reexpression}
\r^{-1/2}\r_\e=(1-\e)\r^{1/2}+\e\r^{-1/2}\s
\ee
is a Hilbert--Schmidt operator.  Applying Lemma~\re{lemmainversehs} again gives
\be
\left\|\r^{-1/2}\r_\e\right\|_2^2
=\<\r^{-1}\>_{\r_\e^2}.
\ee

Monotonicity of the Petz R\'enyi divergence $D^{\mathrm{Petz}}_\alpha(\r_\e\|\r)$ in $\alpha$ gives\footnote{For the infinite-dimensional statement, see Theorems~2.6, 5.1, and~5.5 of \cite{Androulakis:2022tfz}.}
\be
0\leq D_1(\r_\e\|\r)
\leq D^{\mathrm{Petz}}_2(\r_\e\|\r)
=\log\<\r^{-1}\>_{\r_\e^2}
=\log\left\|\r^{-1/2}\r_\e\right\|_2^2.
\ee
Using \er{eq:reexpression} we find
\be
\left\|\r^{-1/2}\r_\e\right\|_2^2
=1-\e^2+\e^2\left\|\r^{-1/2}\s\right\|_2^2
=1-\e^2+\e^2\<\r^{-1}\>_{\s^2}.
\ee
Using \er{trrssbd} we thus find
\be\la{d1bound}
0\leq D_1(\r_\e\|\r)
\leq\log\left(1-\e^2+\fr{\e^2}{\etail}\right).
\ee
Since
\begin{equation}
\<K_\r\>_{\r_\e}=S(\r_\e)+D_1(\r_\e\|\r),
\qqu
\<K_\r\>_\r=S(\r),
\end{equation}
we find
\be
S(\r_\e)-S(\r)-\e\<K_\r\>_{\s-\r}
=-D_1(\r_\e\|\r),
\ee
which gives the promised bound
\be
\bigl|S(\r_\e)-S(\r)-\e\<K_\r\>_{\s-\r}\bigr| \leq \log\left(1-\e^2+\fr{\e^2}{\etail}\right) \leq \fr{\e^2}{\etail}.
\ee
\end{proof}

We now proceed to the main result of this section, which is Theorem~\re{thmsmjlms} below. 
The proof of the theorem is rather technical, relying on a sequence of algebraic manipulations and bounds.  Some readers may thus wish to skip the details on a first pass and to skip to Section~\ref{sec:expJLMS} after reading the statement of the theorem.
In parallel with \eqref{eq:JLMS}, the result \eqref{smjlms} restricts the projection of $K_{\td\r^\a_R}$ to the code subspace defined by the image of $V_\a$.  We thus refer to this theorem as describing an approximate projected JLMS relation.

\begin{theorem}[Approximate projected JLMS in a small code]\la{thmsmjlms}
Suppose that the approximate FLM formula $\eqref{flme}$ holds in the $\alpha$-sector on a subsystem $R$, and that $V_\a$ is an approximate isometry satisfying \er{appiso}.
Then for every log-stable state $\r^\a$ (see Definition~\re{defsmallstable}), the JLMS formula holds approximately in the sense that we have
\be\la{smjlms}
\lt\| V_\a^{\dag} K_{\td\r^\a_R} V_\a - V_\a^{\dag} V_\a \(\fr{A^\a}{4G} + K_{\r^{\a}_r} \) \rt\|_\infty \lesssim \epJLMS,
\ee
with $\epJLMS$ defined by
\be\la{epJLMSdef}
\epJLMS := \sqrt{\fr{\eFLM}{\etail}} + \eisosmall |\ln \etail|,
\ee
where we have used the convention that $K_{\td\r^\a_R}, K_{\r^{\a}_r}$ implicitly include factors of the identity operators ${\mathds 1}_{\Rb},{\mathds 1}_{\rb}$ on $\tH_{\Rb}, \cH_{\rb}^\a$ so that they define operators on $\tH, \cH^\a$;
moreover, the operator $V_\a^\dag K_{\td\r_R^\a}V_\a$ is taken to mean $T_\a^\dag T_\a$ with $T_\a:=K_{\td\r_R^\a}^{1/2}V_\a$ if $K_{\td\r_R^\a}$ is unbounded in such a way that $K_{\td\r_R^\a}V_\a$ is not everywhere defined.
Here and below, $\lesssim$ means that the inequality holds up to state-independent $\cO(1)$ coefficients, which are also independent of $\alpha$ whenever multiple sectors are considered. We drop such coefficients to simplify both the bookkeeping and the final expression.
\end{theorem}

\begin{proof}
We first prove that for every state $\s^\a$ on $\cH^\a$,
\be\la{smjlms2}
\left|
\<K_{\td\r_R^\a}\>_{\hat\s_R^\a}
-\fr{A^\a}{4G}\Tr\hat\s^\a
-\<K_{\r_r^\a}\>_{\s_r^\a}
\right|
\lesssim\epJLMS.
\ee
To prove this, consider the variation of the approximate FLM formula \er{flme} from $\r^\a$ to $\r_\e^\a:=\r^\a+\d\r^\a$, where $\d\r^\a=\e(\s^\a-\r^\a)$ for some small positive $\e$ to be determined shortly.
On the bulk side, we use Lemma~\re{lemmacond1}, Lemma~\re{lemmacondee}, and Condition~1 of Definition~\re{defsmallstable} to find\footnote{Here and below, equations of the form $X=Y+\cO(\ve)$ are taken to mean $|X-Y|\lesssim\ve$.}
\be\la{vndr}
S(\r_r^\a+\d\r_r^\a)-S(\r_r^\a)
=\<K_{\r_r^\a}\>_{\d\r_r^\a}
+\cO\left(\fr{\e^2}{\etail}\right).
\ee

On the boundary side, first define $\d\td\r^\a$ as the exact corresponding variation of the normalized boundary state:
\be\la{dtdrdef}
\d\td\r^\a := \td\r_\e^\a-\td\r^\a,\qquad
\hat\r_\e^\a:=V_\a\r_\e^\a V_\a^\dag.
\ee
Using the finite variation of the nonlinear map $\td\r^\a=W_\a(\r^\a)$ in \er{eq:tilderho} and partial tracing over $\Rb$, we find
\be\la{traRe}
\d\td\r_R^\a
=\e\left(\td\s_R^\a-\td\r_R^\a\right)
\fr{\Tr\hat\s^\a}{\Tr\hat\r_\e^\a}.
\ee
Applying \er{appiso2} to both $\r^\a$ and $\s^\a$ gives
\be\la{Trhat}
\Tr\hat\r^\a,\,\,
\Tr\hat\s^\a,\,\,
\Tr\hat\r_\e^\a
=1+\cO(\eisosmall),
\ee
and hence
\be\la{Trhatratio}
\fr{\Tr\hat\s^\a}{\Tr\hat\r_\e^\a}
=1+\cO(\eisosmall).
\ee
Thus \er{traRe} gives $\d\td\r_R^\a=\cO(\e)\(\td\s_R^\a-\td\r_R^\a\)$.  Using Lemma~\re{lemmacondee} together with Condition~2 of Definition~\re{defsmallstable}, we therefore find
\be\la{vndR}
S(\td\r_R^\a+\d\td\r_R^\a)-S(\td\r_R^\a)
=\<K_{\td\r_R^\a}\>_{\d\td\r_R^\a}
+\cO\left(\fr{\e^2}{\etail}\right).
\ee

Now, since the approximate FLM formula \er{flme} holds for both $\r^\a$ and $\r^\a +\d\r^\a$, its variation must therefore satisfy
\be\la{flmevar}
\Bigl|
\left[S(\td\r_R^\a+\d\td\r_R^\a)-S(\td\r_R^\a)\right]
-
\left[S(\r_r^\a+\d\r_r^\a)-S(\r_r^\a)\right]
\Bigr|
\leq2\eFLM.
\ee
Inserting \er{vndr} and \er{vndR} into \er{flmevar} thus yields
\be\la{trdiff}
\left|
\<K_{\td\r_R^\a}\>_{\d\td\r_R^\a}
-\<K_{\r_r^\a}\>_{\d\r_r^\a}
\right|
\lesssim\eFLM+\fr{\e^2}{\etail}.
\ee

We may now use $\d\r^\a=\e(\s^\a-\r^\a)$ and \er{traRe} to write the left-hand side of \er{trdiff} as
\ba
&\left|
\<K_{\td\r_R^\a}\>_{\d\td\r_R^\a}
-\<K_{\r_r^\a}\>_{\d\r_r^\a}
\right|\\
={}&\e\left|
\left(\<K_{\td\r_R^\a}\>_{\td\s_R^\a-\td\r_R^\a}\right)
\fr{\Tr\hat\s^\a}{\Tr\hat\r_\e^\a}
-
\<K_{\r_r^\a}\>_{\s_r^\a-\r_r^\a}
\right|\\
={}&\e\fr{\Tr\hat\s^\a}{\Tr\hat\r_\e^\a}
\left|
\<K_{\td\r_R^\a}\>_{\td\s_R^\a-\td\r_R^\a}
-
\left(\<K_{\r_r^\a}\>_{\s_r^\a-\r_r^\a}\right)
\fr{\Tr\hat\r_\e^\a}{\Tr\hat\s^\a}
\right|\\
={}&\e\fr{\Tr\hat\s^\a}{\Tr\hat\r_\e^\a}
\bigl|
\<K_{\td\r_R^\a}\>_{\td\s_R^\a-\td\r_R^\a}
-
\<K_{\r_r^\a}\>_{\s_r^\a-\r_r^\a}
\bigr|
+\cO(\e\,\eisosmall|\ln\etail|),\la{trdiffexp}
\ea
where on the last line we used \er{Trhatratio} and also used Condition~1 in Definition~\re{defsmallstable} to bound
\be\la{trsKbd}
0\leq\<K_{\r_r^\a}\>_{\t_r^\a}\leq \lt\|K_{\r^\a_r}\rt\|_\infty \leq |\ln\etail|
\ee
for every state $\t_r^\a$ on $\cH_r^\a$.
Using \er{trdiffexp} and \er{Trhatratio}, we find that \er{trdiff} becomes
\be\la{trdiffsm2}
\left|
\<K_{\td\r_R^\a}\>_{\td\s_R^\a-\td\r_R^\a}
-
\<K_{\r_r^\a}\>_{\s_r^\a-\r_r^\a}
\right|
\lesssim\fr{\eFLM}{\e}+\fr{\e}{\etail}
+\eisosmall|\ln\etail|.
\ee

Now writing the FLM formula \er{flme} for $\r^\a$ itself as
\be
\left|
\<K_{\td\r_R^\a}\>_{\td\r_R^\a}
-\fr{A^\a}{4G}
-\<K_{\r_r^\a}\>_{\r_r^\a}
\right|
\leq\eFLM,
\ee
and combining it with \er{trdiffsm2}, we find
\be\la{trdiffsm3}
\left|
\<K_{\td\r_R^\a}\>_{\td\s_R^\a}
-\fr{A^\a}{4G}
-\<K_{\r_r^\a}\>_{\s_r^\a}
\right|
\lesssim\fr{\eFLM}{\e}+\fr{\e}{\etail}
+\eisosmall|\ln\etail|.
\ee
We now write the left-hand side of \er{trdiffsm3} as
\ba
&\left|
\<K_{\td\r_R^\a}\>_{\td\s_R^\a}
-\fr{A^\a}{4G}
-\<K_{\r_r^\a}\>_{\s_r^\a}
\right|\\
={}&\fr{1}{\Tr\hat\s^\a}
\left|
\<K_{\td\r_R^\a}\>_{\hat\s_R^\a}
-\fr{A^\a}{4G}\Tr\hat\s^\a
-(\Tr\hat\s^\a)\<K_{\r_r^\a}\>_{\s_r^\a}
\right|\\
={}&\fr{1}{\Tr\hat\s^\a}
\left|
\<K_{\td\r_R^\a}\>_{\hat\s_R^\a}
-\fr{A^\a}{4G}\Tr\hat\s^\a
-\<K_{\r_r^\a}\>_{\s_r^\a}
\right|
+\cO(\eisosmall|\ln\etail|),\la{trdiff3l}
\ea
where on the last line we used \er{Trhat} and \er{trsKbd}.
Using \er{trdiff3l} and \er{Trhat}, we find that \er{trdiffsm3} becomes
\be\la{TrVsV}
\left|
\<K_{\td\r_R^\a}\>_{\hat\s_R^\a}
-\fr{A^\a}{4G}\Tr\hat\s^\a
-\<K_{\r_r^\a}\>_{\s_r^\a}
\right|
\lesssim\fr{\eFLM}{\e}+\fr{\e}{\etail}
+\eisosmall|\ln\etail|.
\ee
We are free to choose $\e$, so to optimize the bound we take $\e\sim\sqrt{\eFLM\etail}$, which yields
\be\la{smjlms3}
\left|
\<K_{\td\r_R^\a}\>_{\hat\s_R^\a}
-\fr{A^\a}{4G}\Tr\hat\s^\a
-\<K_{\r_r^\a}\>_{\s_r^\a}
\right| \lesssim \sqrt{\fr{\eFLM}{\etail}} + \eisosmall |\ln \etail| = \epJLMS.
\ee
This is exactly our desired inequality \er{smjlms2}.

It remains to obtain the operator-norm bound \er{smjlms}.  For every normalized $|\psi^\a\>\in\cH^\a$, \er{smjlms3} holds for $\s^\a=|\psi^\a\>\<\psi^\a|$.  Its first term can be written as
\be
\<K_{\td\r_R^\a}\>_{\hat\s_R^\a}
=\<K_{\td\r_R^\a}\>_{\hat\s^\a}
=\<K_{\td\r_R^\a}\>_{V_\a|\psi^\a\>\<\psi^\a|V_\a^\dag}
=\left\|K_{\td\r_R^\a}^{1/2}V_\a|\psi^\a\>\right\|^2,
\ee
where in the first step we used the partial-trace identity \er{EVpartialtrace} and in the last step we used \er{EVvector}.  This result and the definition $T_\a=K_{\td\r_R^\a}^{1/2}V_\a$ allow us to rewrite \er{smjlms3} as
\be
\left|
\bigl\|T_\a|\psi^\a\>\bigr\|^2
-\lt\<\psi^\a\middle|\left(\fr{A^\a}{4G}V_\a^\dag V_\a+K_{\r_r^\a}\right)\middle|\psi^\a\rt\>
\right|
\lesssim\epJLMS.
\ee
Since this holds for all normalized $|\psi^\a\>\in\cH^\a$, using Lemma~\re{lemmaopnorm} in Appendix~\ref{app:Schatten} gives
\be\la{smjlmsopbounded}
\left\|
V_\a^\dag K_{\td\r_R^\a}V_\a
-
\fr{A^\a}{4G}V_\a^\dag V_\a
-
K_{\r_r^\a}
\right\|_\infty
\lesssim\epJLMS.
\ee

Finally, we insert the desired $V^\dagger_\alpha V_\alpha$ in the $K_{\r^\a_r}$ term by
using \er{appiso} and \er{trsKbd} to write
\be\la{VVKbound}
\lt\|(V_\a^\dag V_\a-\mathbb 1^\a)K_{\r_r^\a}\rt\|_\infty \leq \lt\|V_\a^\dag V_\a - \mathbb 1^\a \rt\|_\infty \lt\|K_{\r^\a_r}\rt\|_\infty \leq \eisosmall |\ln \etail|.
\ee
The triangle inequality \er{triangle} then converts \er{smjlmsopbounded} into the approximate JLMS formula \er{smjlms}.
\end{proof}

In the approximate JLMS formula \er{smjlms}, it is important that the $A^\a/4G$ term appear with a factor of $V_\a^\dag V_\a$.  However, it is not in fact  necessary for the $K_{\r^{\a}_r}$ term to come with this factor.
Indeed, the proof above also established \er{smjlmsopbounded}, or equivalently
\be
\lt\| V_\a^{\dag} \(K_{\td\r^\a_R} - \fr{A^\a}{4G}\) V_\a - K_{\r^{\a}_r} \rt\|_\infty \lesssim \epJLMS,
\ee
which may be viewed as an alternative -- but equally correct -- version of the approximate JLMS formula.

The standard holographic derivation \cite{Faulkner:2013ana} of the approximate FLM relation \eqref{flme} gives an error that vanishes as $G\rightarrow 0$, though it says little about the rate at which this occurs.  Thus we should take $\eFLM = o(G^0)$, using the so-called little $o$ notation.  We may then usefully apply Theorem~\re{thmsmjlms},
finding that the JLMS formula holds up to parametrically small errors at most of order $\sqrt{\eFLM/\etail} + \eisosmall |\ln \etail| \ll 1$,
as long as we restrict to log-stable states with $\etail \gg \eFLM$ -- e.g.\ for $\etail$ an $o(G^0)$ value much greater than $\eFLM$ -- and as long as $\eisosmall$ is much smaller than $1/|\ln \etail|$ (a fairly weak condition).

One might hope that the above analysis can be improved in a manner that replaces use of the approximate FLM formula with an appropriate formula in terms of a quantum extremal surface (QES).  By doing so, one might expect to reduce the error $\eFLM$ from $o(G^0)$ to $e^{-\cO(1/G)}$.  If this can be done, the lower bound $\etail$ on the eigenvalues of $\r^\a_r$ could then be taken to be any number much larger than $e^{-\cO(1/G)}$.  However, we will not pursue this idea here.

\subsection{Exponentiated JLMS}
\label{sec:expJLMS}

We now proceed to our approximate version of the exponentiated JLMS relations.  As noted earlier, for the exponentiated relation to hold with general exponents $is$ (for real $s$), the entanglement spectrum of the code must be flat.  When this holds exactly, it is equivalent to flatness of the R\'enyi entropies for certain code parts of the state (see e.g.\ discussion in \cite{Dong:2018seb}), and thus to R\'enyi versions of the FLM formula for general complex replica numbers $n$.  We will thus find it convenient to encode approximate flatness of the code entanglement spectrum by simply assuming our code to satisfy approximate versions of R\'enyi FLM.  In particular, this will allow us to derive our approximate exponentiated JLMS relation with exponent $is$ (with $s$ real) using approximate R\'enyi FLM for only the particular replica number $n=1+is$. 

To this end, we make the following definition.

\begin{ndefi}[Approximate R\'enyi FLM]\la{defflmre}
For a fixed, generally complex $n\neq 1$, we say that the approximate R\'enyi FLM formula holds in the $\alpha$-sector on $R$ if every state $\r^\a$ on $\cH^\a$ satisfies
\be\la{flmereexp}
\lt| \fr{1}{n-1}\[
\Tr_R \!\(\td\r^\a_R\)^n
- e^{(1-n)\fr{A^\a}{4G}} \tr_r \!\(\r^\a_r\)^n
\] \rt| \le \xi(n) \eFLM,
\ee
with state-independent values of $A^\a$, $\xi(n)$, and $\eFLM$. Here $A^\a$ is a real c-number\footnote{Note that if R\'enyi FLM holds exactly (with zero error) for $n=1+is$ at even two real and incommensurate values of $s$, the code must have an exactly flat entanglement spectrum.
To see this note that, when $\r_r^\a$ is pure, R\'enyi FLM
gives $\Tr_R \(\td\r^\a_R\)^{1+is} = e^{-is \fr{A^\a}{4G}}$. This is a pure phase and so has magnitude $1$. But since $\td\r^\a_R$ is normalized, it then follows that all eigenvalues of $s \log \td\r^\a_R$ are equal modulo $2\pi$. Two incommensurate values of $s$ therefore force a flat entanglement spectrum.}.
At $n=1$, we define the condition to be \er{flme}, which agrees with the formal $n\to1$ limit of \er{flmereexp} if $\xi(1)=1$.
For later convenience, we define $\xi_s$ to be $\xi(n)$ evaluated at $n=1+is$.
\end{ndefi}

For $n=1+is$ with real $s\neq 0$, one may replace \er{flmereexp} by the stronger condition
\be\la{flmere}
\lt| S_n(\td\r^\a_R) - \( \fr{A^\a}{4G} + S_n(\r^\a_r) \) \rt| \le \xi(n) \eFLM,
\ee
where $S_n(\r) := \fr{1}{1-n}\ln \Tr \r^n$ is the R\'enyi entropy of order $n$.
Note that, for complex $n$, we generally need to choose a branch cut when defining the $\ln$ function in $S_n$, and \er{flmere} is taken to hold if it can be made true with some choice of branch cuts (which may be different for $S_n(\td\r^\a_R)$ and $S_n(\r^\a_r)$).
    
To see that \er{flmere} implies \er{flmereexp}, note that any complex numbers $z_1,z_2$ with $|e^{z_1}|,|e^{z_2}|\leq1$ obey
\be\la{z12ineq}
\lt|e^{z_1} - e^{z_2} \rt| \leq \lt|z_1 - z_2 \rt|,
\ee
which follows from
\be
e^{z_1}-e^{z_2}=(z_1-z_2)\int_0^1 e^{(1-t)z_2+t z_1}\,dt,
\ee
and the fact that the integrand has magnitude at most one.
For real $s\neq 0$, take $z_1=-isS_{1+is}(\td\r^\a_R)$ and $z_2=-is\bigl(\frac{A^\a}{4G}+S_{1+is}(\r^\a_r)\bigr)$.
Then both $e^{z_1}=\Tr_R(\td\r^\a_R)^{1+is}$ and $e^{z_2}=e^{-is\fr{A^\a}{4G}}\tr_r(\r^\a_r)^{1+is}$ have magnitude at most one because the states are normalized and $A^\a$ is real.
Dividing \er{z12ineq} by $|s|$ gives \er{flmereexp} from \er{flmere}.

Before deriving our exponentiated JLMS, we first establish the following useful lemma.

\begin{nlemma}\la{lemmacondre}
Let $\r$, $\s$ be states in some Hilbert space, not necessarily of finite dimension. Suppose
\be\la{trrssbd2}
\<\r^{-1}\>_{\s^2}\leq1/\etail.
\ee
Then for every $\e \in [0,1)$ and every real $s$, we have
\be
\lt|\fr{1}{1+is} \[\Tr \(\r_\e^{1+is}\) - \Tr \(\r^{1+is}\)\] - \e \Tr \[\r^{is} (\s - \r)\] \rt| \leq \fr{|s| \e^2}{2(1-\e)\etail},
\ee
where $\r_{\e}:=\r+\e(\s-\r)$.
\end{nlemma}

This lemma is the R\'enyi entropy analogue of Lemma~\re{lemmacondee}. Since its proof is somewhat technical, we delegate the proof to Appendix~\re{app:lemmaproof}.

Our main result for exponentiated JLMS on a small code subspace is then given by the theorem below.  The proof is structurally similar to the argument for projected JLMS given in Section~\ref{sec:originalJLMS} and is again rather technical.  Many readers may thus wish to skip ahead to Section~\ref{sec:largecodesubspace} (or even to Section~\ref{sec:discussion}) on a first reading.

\begin{theorem}[Approximate exponentiated JLMS in a small code]\la{thmsmexp}
Let $s$ be a real number. Suppose that the approximate R\'enyi FLM formula~\eqref{flmereexp} for $n=1+is$ holds in the $\alpha$-sector on $R$, and that $V_\a$ is an approximate isometry satisfying \er{appiso}.
Then for every log-stable state $\r^\a$ (as defined in Definition~\re{defsmallstable}), the exponentiated JLMS formula holds approximately in the sense that we have
\be\la{smexp}
\lt\| V_\a^\dag \, e^{-is K_{\td\r^{\a}_R}} \, V_\a - e^{-is \(\fr{A^\a}{4G} + K_{\r^{\a}_r} \)} \rt\|_\infty \lesssim \eeJLMS,
\ee
with $\eeJLMS$ defined by
\be\la{eeJLMSdef}
\eeJLMS := \sqrt{\xi_s\, |s| \fr{\eFLM}{\etail}} + \xi_s\, |s| \eFLM + \eisosmall.
\ee
Here $\lesssim$ again means that we drop $\cO(1)$ coefficients.  Note, however, that we do not require the parameter $s$ to be $\cO(1)$.  Instead, we include any terms that might become relevant if $s$ is taken to be large in a limit where other parameters become small.  In particular, we have kept the $|s| \eFLM$ term explicit even though $\eFLM \ll \sqrt{\frac{\eFLM}{\etail}}$.
\end{theorem}

\begin{proof}
Using \Eqref{onbound}, we only need to show that for every state $\s^\a$ on $\cH^\a$, we have
\be
\label{eq:thm2goal}
\lt| \tr \(\s^\a \[V_\a^\dag \, e^{-is K_{\td\r^{\a}_R}} \, V_\a - e^{-is \(\fr{A^\a}{4G} + K_{\r^{\a}_r} \)} \] \) \rt| \lesssim \eeJLMS.
\ee
This relation can be rewritten in the form
\be\la{smexp2}
\lt| \tr \(\s^\a \[V_\a^\dag \, \(\td\r^{\a}_R\)^{is} \, V_\a - e^{-is \fr{A^\a}{4G}} \(\r^{\a}_r\)^{is} \] \) \rt| \lesssim \eeJLMS.
\ee
To prove that this is the case, consider the bulk variation from $\r^\a$ to $\r_\e^\a:=\r^\a+\d\r^\a$, where $\d \r^\a=\e(\s^\a-\r^\a)$ for some small $\e$ to be determined shortly. The corresponding boundary variation $\d\td\r^\a$ is given by \er{dtdrdef}.
We first apply Lemma~\re{lemmacondre} (and the log-stability of $\r^\a$) to both the bulk and the boundary, finding
\be\la{trdisr}
\lt| \fr{1}{1+is}\[\tr_r \!\(\r^\a_r+\d\r^\a_r\)^{1+is} - \tr_r \!\(\r^\a_r\)^{1+is}\] - \tr_r \!\[(\r^\a_r)^{is} \d\r^\a_r\] \rt| \lesssim \fr{|s| \e^2}{\etail}
\ee
and
\be\la{trdisR}
\lt| \fr{1}{1+is}\[\Tr_R \!\(\td\r^\a_R+\d\td\r^\a_R\)^{1+is} - \Tr_R \!\(\td\r^\a_R\)^{1+is}\] - \Tr_R \!\[(\td\r^\a_R)^{is} \d\td\r^\a_R\] \rt| \lesssim \fr{|s| \e^2}{\etail}.
\ee

The approximate R\'enyi FLM formula \er{flmereexp} for $n=1+is$ gives
\be\la{flmereexp2}
\lt| \Tr_R \!\(\td\r^\a_R\)^{1+is} - e^{-is\fr{A^\a}{4G}} \tr_r \!\(\r^\a_r\)^{1+is} \rt| \leq |s|\, \x_s \eFLM,
\ee
where the $s=0$ case follows directly from $\Tr_R\td\r_R^\a=\tr_r\r_r^\a=1$.

The inequality \er{flmereexp2} holds for every $\r^\a$ on $\cH^\a$, so it also holds for $\r^\a +\d\r^\a$. Using it in both ways in \er{trdisR}, we thus find
\be
\lt| \fr{e^{-is\fr{A^\a}{4G}}}{1+is}\[\tr_r \!\(\r^\a_r+\d\r^\a_r\)^{1+is} - \tr_r \!\(\r^\a_r\)^{1+is}\] - \Tr_R \!\[(\td\r^\a_R)^{is} \d\td\r^\a_R\] \rt|
\lesssim \fr{|s|}{\sqrt{s^2+1}}\, \x_s \eFLM + \fr{|s|\e^2}{\etail}.
\ee
Comparing this inequality with \er{trdisr} then gives
\be
\lt| \Tr_R \!\[(\td\r^\a_R)^{is} \d\td\r^\a_R\] - \tr_r \!\[ e^{-is \fr{A^\a}{4G}} (\r^\a_r)^{is} \d\r^\a_r\] \rt| \lesssim \fr{|s|}{\sqrt{s^2+1}}\, \x_s \eFLM + \fr{|s|\e^2}{\etail}.
\ee
Using $\d\r_r^\a=\e(\s_r^\a-\r_r^\a)$ and $\d\td\r^\a_R=\e(\td\s^\a_R-\td\r^\a_R) \fr{\Tr\hat\s^\a}{\Tr\hat\r_\e^\a}$ from \er{traRe}, together with the fact that both  $\Tr\hat\r_\e^\a$ and $\Tr\hat\s^\a$ are of the form $1+\cO(\eisosmall)$, we may now rewrite the above bound in the form
\bm
\label{eq:AA}
\e \lt| \Tr_R \!\[(\td\r^\a_R)^{is} \(\td\s^\a_R \Tr\hat\s^\a - \td\r^\a_R\)\] - \tr_r \!\[ e^{-is \fr{A^\a}{4G}} (\r^\a_r)^{is} (\s_r^\a -\r_r^\a)\] \rt| \\
\lesssim \fr{|s|}{\sqrt{s^2+1}}\, \x_s \eFLM + \fr{|s|\e^2}{\etail} + \e\, \eisosmall.
\em
Combining \eqref{eq:AA} with \eqref{flmereexp2} we obtain
\be\la{eq:TrRdiff}
\lt| \Tr_R \!\[(\td\r^\a_R)^{is} \td\s^\a_R (\Tr\hat\s^\a)\] - \tr_r \!\[ e^{-is \fr{A^\a}{4G}} (\r^\a_r)^{is} \s_r^\a \] \rt|
\lesssim \fr{|s|\x_s \eFLM}{\e\sqrt{s^2+1}} +\fr{|s|\e}{\etail} + |s| \x_s \eFLM + \eisosmall.
\ee
Using $\td\s^\a (\Tr\hat\s^\a) = \hat\s^\a = V_\a \s^\a V_\a^{\dag}$ then yields a bound of the desired form
\be\la{eq:trdiff}
\lt| \tr \(\s^\a \[V_\a^\dag \, \(\td\r^{\a}_R\)^{is} \, V_\a - e^{-is \fr{A^\a}{4G}} \(\r^{\a}_r\)^{is} \] \) \rt|
\lesssim \fr{|s|\x_s \eFLM}{\e\sqrt{s^2+1}} +\fr{|s|\e}{\etail} + |s| \x_s \eFLM + \eisosmall.
\ee
What remains is just to choose $\e$. To optimize the bound, we take $\e \sim \sqrt{\xi_s \eFLM \etail/ \sqrt{s^2+1}}$.
After discarding $\cO(1)$ coefficients, we find
\be
\lt| \tr \(\s^\a \[V_\a^\dag \, \(\td\r^{\a}_R\)^{is} \, V_\a - e^{-is \fr{A^\a}{4G}} \(\r^{\a}_r\)^{is} \] \) \rt|
\lesssim \sqrt{\xi_s\, |s| \fr{\eFLM}{\etail}} + \xi_s\, |s| \eFLM + \eisosmall = \eeJLMS.
\ee
This is the promised inequality  \eqref{eq:thm2goal}, thus proving the approximate exponentiated JLMS formula \er{smexp}.
\end{proof}

It is worth noting that, unlike the projected formula \er{smjlms},  the exponentiated formula \er{smexp} does not require a factor of $V_\a^\dag V_\a$ to multiply the term containing $A^\a$ (namely, $e^{-is \big(\fr{A^\a}{4G} + K_{\r^{\a}_r} \big)}$).

\subsection{Generalization to relative log-stability}
\label{sec:relativelogstable}

Here we generalize the projected and exponentiated JLMS results (Theorems~\re{thmsmjlms} and~\re{thmsmexp}) of the previous two subsections. If we only require these JLMS results to hold at the level of expectation values in a particular test state $\s^\a$, we may relax the log-stability condition to a notion of relative log-stability with respect to that state $\s^\a$, as defined in a precise way below.

{
\renewcommand{\thendefi}{\ref*{defsmallstable}b}
\begin{ndefi}\la{defsmallstablerel}
A state $\r^\a$ on $\cH^\a$ is called relatively log-stable with respect to a state $\s^\a$ on $\cH^\a$ if they satisfy the following two conditions with some small $\etail>0$:
\begin{enumerate}[label=Condition \arabic*:,leftmargin=*]
\item $\<(\r_r^\a)^{-1}\>_{(\s_r^\a)^2}\leq1/\etail$.
\item $\<(\td\r_R^\a)^{-1}\>_{(\td\s_R^\a)^2}\leq1/\etail$.
\end{enumerate}
\end{ndefi}
\addtocounter{ndefi}{-1}
}

Relative log-stability with respect to a given state $\s^\a$ can hold even when $\cH_r^\a$ is infinite-dimensional.  It is clear that requiring relative log-stability with respect to every state $\s^\a$ on $\cH^\a$ with the same $\etail$ is equivalent to log-stability.

{
\renewcommand{\thetheorem}{\ref*{thmsmjlms}b}

\begin{theorem}[Approximate projected JLMS with relative log-stability in a small code]\la{thmsmjlmsrel}
Suppose that we have a small code defined as in Definition~\re{defsmallcode}, and two states $\r^\a$, $\s^\a$ on $\cH^\a$ that satisfy the following three conditions:
\begin{enumerate}[label=(\roman*)]
\item $V_\a$ acts on $\r^\a$, $\s^\a$ as an approximate isometry in the sense that $|\Tr \hat\r^\a - 1|, |\Tr \hat\s^\a - 1| \leq \eisosmall$ (but this is not required for other states).
\item The approximate FLM formula $\eqref{flme}$ holds in the relevant $\alpha$-sector on $R$ for all mixtures of $\r^\a$ and $\s^\a$, i.e., for all states $\r^\a +\e(\s^\a - \r^\a)$ with $\e \in [0,1]$ (but not necessarily for other states), and does so with the same $A^\a$ and $\eFLM$.
\item $\r^\a$ is relatively log-stable with respect to $\s^\a$ (as defined in Definition~\re{defsmallstablerel}).
\end{enumerate}
Then the JLMS formula holds approximately in the sense that we have
\be\la{smjlmsrel}
\left|
\<K_{\td\r_R^\a}\>_{\hat\s_R^\a}
-\fr{A^\a}{4G}\Tr\hat\s^\a
-\<K_{\r_r^\a}\>_{\s_r^\a}
\right|\lesssim\epJLMS^{b\,(\a)},
\ee
with $\epJLMS^{b\,(\a)}$ defined by
\be\la{epJLMSbdef}
\epJLMS^{b\,(\a)} := \sqrt{\fr{\eFLM}{\etail}} + \eisosmall \Bigl(\bigl|\ln \etail\bigr| + S(\r^{\a}_r) + S(\s^{\a}_r)\Bigr).
\ee
\end{theorem}
\addtocounter{theorem}{-1}
}

\begin{proof}
The proof follows the proof of Theorem~\re{thmsmjlms} through \er{smjlms3}, with the following two modifications. First, we derive \er{vndr} and \er{vndR} using Definition~\re{defsmallstablerel} instead of Definition~\re{defsmallstable}. Second, we replace \er{trsKbd} by
\be\la{trsKbd2}
\<K_{\r_r^\a}\>_{\r_r^\a}=S(\r_r^\a),\qquad
\<K_{\r_r^\a}\>_{\s_r^\a}
=S(\s_r^\a)+D_1(\s_r^\a\|\r_r^\a)
\leq S(\s_r^\a)+|\ln\etail|,
\ee
where the final inequality follows from \er{d1bound} at $\e=1$.
\end{proof}

In the result above, the $b$ superscript in the bound $\epJLMS^{b\,(\a)}$ emphasizes that the bound is for Theorem~\re{thmsmjlmsrel}, to be distinguished from the bound $\epJLMS$ for Theorem~\re{thmsmjlms}.

The generalization of Theorem~\re{thmsmexp} (exponentiated JLMS) to the case of relative log-stability is as follows.

{
\renewcommand{\thetheorem}{\ref*{thmsmexp}b}

\begin{theorem}[Approximate exponentiated JLMS with relative log-stability in a small code]\la{thmsmexprel}
Let $s$ be a real number. Suppose that we have a small code defined as in Definition~\re{defsmallcode}, and two states $\r^\a$, $\s^\a$ on $\cH^\a$ that satisfy the following three conditions:
\begin{enumerate}[label=(\roman*)]
\item $V_\a$ acts on $\r^\a$, $\s^\a$ as an approximate isometry in the sense that $|\Tr \hat\r^\a - 1|, |\Tr \hat\s^\a - 1| \leq \eisosmall$ (but this is not required for other states).
\item The approximate R\'enyi FLM formula~\eqref{flmereexp} for $n=1+is$ holds in the $\alpha$-sector on $R$ for all mixtures of $\r^\a$ and $\s^\a$, i.e., states $\r^\a +\e(\s^\a - \r^\a)$ with $\e \in [0,1]$ (but not necessarily for other states), with the same $A^\a$, $\x_s$, and $\eFLM$.
\item $\r^\a$ is relatively log-stable with respect to $\s^\a$ (as defined in Definition~\re{defsmallstablerel}).
\end{enumerate}
Then the exponentiated JLMS formula holds approximately in the sense that we have
\be\la{smexprel}
\lt| \tr \(\s^\a \[V_\a^\dag \, e^{-is K_{\td\r^{\a}_R}} \, V_\a - e^{-is \(\fr{A^\a}{4G} + K_{\r^{\a}_r} \)} \] \) \rt| \lesssim \eeJLMS,
\ee
with $\eeJLMS$ defined by \er{eeJLMSdef}.
\end{theorem}
\addtocounter{theorem}{-1}
}

\begin{proof}
The proof is the same as the proof of Theorem~\re{thmsmexp}, except that we derive \er{trdisr} and \er{trdisR} using Definition~\re{defsmallstablerel} instead of Definition~\re{defsmallstable}.
\end{proof}

Note that the proof of this theorem relies on approximate R\'enyi FLM only for mixtures of $\r^\a$ and $\s^\a$, and that it relies on $V_\a$ being an approximate isometry only for $\r^\a$, $\s^\a$.

\section{A large code and its JLMS relations with error bounds}

\label{sec:largecodesubspace}

We are now ready to assemble our `large' code from a set of small codes.  Indeed, for the moment we simply take an arbitrary collection of orthogonal small code Hilbert spaces $\cH^\a$ and define
\be\la{lcode}
\Hlarge = \bigoplus_\a {\mathcal H}^\a = \bigoplus_\a \( \cH^\a_r \otimes \cH^\a_{\rb}\).
\ee
In AdS/CFT, in order for the modular Hamiltonian and modular flows defined in the above $\Hlarge$ to match those defined in the bulk effective theory, the Hilbert space factors $\cH^\a_r$ with different $\a$ would need to be {\it precisely} orthogonal with respect to the inner product of the bulk effective theory (and similarly for $\cH^\a_\rb$).\footnote{If such orthogonality is not exact but only approximate, we can use the results of this section to bound the differences in the modular Hamiltonian and modular flows between $\Hlarge$ and the bulk effective theory. To do so, we need only interpret the relation between $\Hlarge$ and the bulk effective field theory as defining a code and take $\tH$ to denote the Hilbert space of the latter.}  In many cases, this will be straightforward to arrange by e.g.\ associating each $\a$ with a non-overlapping window of eigenvalues of self-adjoint central operators, including for example the area windows discussed in Section~\ref{sec:AdS/CFT} (perhaps along with analogous windows for other central observables as well).
However, even in such cases, since we have allowed failures of exact complementary recovery, the corresponding \textit{subsystem} orthogonality condition in the boundary theory (defined using the boundary inner product) may not hold exactly.\footnote{A tensor network demonstrating this is in Appendix~\ref{sec:eg}.  A more realistic example was studied in JT gravity in \cite{Kudler-Flam:2022jwd}.}

As noted in both the introduction and Section~\ref{sec:AdS/CFT}, if such boundary subsystem orthogonality were to hold exactly, then the JLMS relations in the large code would be precisely a direct sum of the small-code JLMS relations. They would thus hold to the same extent as the small-code JLMS relations.
The purpose of the current section is thus to show that, so long as appropriate notions of approximate subsystem orthogonality (and further technical assumptions) are satisfied, the JLMS relations in the large code hold to a correspondingly good approximation.  As in our small codes, we will in fact derive two notions of approximate JLMS relations.  The first is again a generalization of the projected JLMS formula \eqref{eq:JLMS}, while the second is again an analogue of the exponentiated JLMS.  These are addressed in Sections~\ref{sec:lpJLMS}-\ref{sec:expJLMSop} below.  However, we first provide the above-mentioned definitions (and a few related results) in Section~\ref{sec:lcdefs}.

\subsection{Definitions and properties for large codes}
\label{sec:lcdefs}

We begin by defining our concepts of `large' bulk Hilbert space and the associated `large' code.  This also provides an opportunity to make simple observations that follow from such definitions and to introduce relevant notation.

\begin{ndefi}[Large bulk Hilbert space and large code]\la{deflargecode}
Let $\a$ label a possibly infinite set of mutually orthogonal bulk Hilbert spaces $\cH^\a$ satisfying the properties listed in Definition~\re{defsmallcode} for an $\a$-independent pair of boundary regions $R,\Rb$ and bulk labels\footnote{The fact that the labels $r,\rb$ are independent of $\a$ in no way forbids the associated bulk physics from depending on $\a$. Physical properties of bulk states in $r$ can depend strongly on $\a$.} $r,\rb$. 
We then define the `large' bulk Hilbert space $\Hlarge$ as the (orthogonal) direct sum of all $\cH^\a$:
\be\la{hdecom}
\Hlarge := \bigoplus_\a \cH^\a.
\ee
We will refer to the $\cH^\a$ as $\a$-sectors below.

For each $\alpha$,
Definition~\re{defsmallcode} requires that there are Hilbert space factors $\cH^\a_r$ and $\cH^\a_{\rb}$ of $\cH^\a$ associated with the entanglement wedges $r$ and $\rb$.  We require the factors $\cH^\a_r$ and $\cH^{\a'}_r$ to be orthogonal for $\a\neq \a'$, and similarly for $\cH^\a_{\rb}$ and $\cH^{\a'}_{\rb}$, in the sense that we again use orthogonal direct sums to define $\cH_r := \bigoplus_\a \cH^\a_r, \cH_{\rb} := \bigoplus_\a \cH_{\rb}^\a$ and treat $\Hlarge$ as living in the diagonal subspace of $\cH_r\otimes \cH_{\rb}$.

Using the decomposition \eqref{hdecom}, any state $\r$ on $\Hlarge$ can be written in the associated block form:
\be\la{rdecom}
\r =
\begin{pmatrix}
\r^{\a\a} & \r^{\a\b} & \cd\\
\r^{\b\a} & \r^{\b\b} & \cd\\
\cd & \cd & \cd
\end{pmatrix}
=\sum_\a p_\a\r^\a+\r^\OD,
\qqu
p_\a:=\tr\r^{\a\a},
\ee
where $\sum_\a p_\a=1$ and the operator $\r^\OD:=\r-\bigoplus_\a\r^{\a\a}$ is the `off-diagonal' part of $\r$. When $p_\a>0$, $\r^\a:=\r^{\a\a}/p_\a$ is a (properly normalized) state on $\cH^\a$; when $p_\a=0$, $\r^\a$ may be chosen arbitrarily. It will be useful to define the $\e$-support of $\r$ as
\be\la{esuppdef}
\esupp(\r) := \{\a: p_\a>\e\}.
\ee

Recalling that each of our $V_\alpha$ maps to the same ($\alpha$-independent) boundary Hilbert space $\tH$, we now assume that the maps $V_\alpha$ combine to define a new bounded\footnote{Our results can also be formulated without assuming $V$ to be bounded, but we impose boundedness to avoid additional technicalities.} linear map $V:\Hlarge\to\tH$ whose restriction to each $\cH^\alpha$ is $V_\alpha$. Thus to any state $\r$ on $\Hlarge$ with $\Tr\hat\r>0$, we associate a corresponding boundary state $\td\r$ on $\tH$ defined by
\be\la{rtdhat}
\td\r= W(\r):=\fr{\hat\r}{\Tr \hat \r},\qu {\rm with} \qu
\hat\r := V \r V^{\dag}.
\ee
Here $\hat\r$ depends linearly on $\r$ but is not properly normalized as a state, whereas $\td\r$ is properly normalized but is obtained via the nonlinear map $W$.
We use $\td\r_R$, $\r_r$ to denote the associated reduced density operators on $R$, $r$ defined by tracing over $\tH_{\Rb}$, $\cH_{\rb}$. In particular, since bulk states in distinct $\cH_{\rb}^\alpha$ are orthogonal, the reduced density operator $\r_r$ is block diagonal on $\cH_r$ and can be written in the form
\be\la{rrdecom}
\r_r=\bigoplus_\a p_\a \r^\a_r.
\ee

Finally, we define the area operator $\hat A$ as
\be\la{aopdef}
\hat A \eq \bigoplus_\a A^\a \mathbb 1^\a.
\ee
This clearly commutes with block diagonal operators such as $\r_r$ and $K_{\r_r}$. Below, $\fr{\hat A}{4G}+K_{\r_r}$ denotes the self-adjoint operator obtained by adding the two operators sector by sector.

For a different state $\s$ on $\Hlarge$, we use the same conventions, writing
\be\la{sdecom}
\s=\sum_\a q_\a\s^\a+\s^\OD,\qqu
q_\a:=\tr\s^{\a\a}.
\ee
Its corresponding boundary state is naturally called $\td\s$.
We define $\tH_{\text{code}}$ to be the image of $V$ and call it the large code subspace.
\end{ndefi}

We now introduce the notion of approximate subsystem orthogonality motivated near the end of Section~\ref{sec:AdS/CFT}.

\begin{ndefi}[Approximate subsystem orthogonality]\la{defortho} 
We will in fact define two flavors of this term which will be called `log approximate subsystem orthogonality' and `exponentiated approximate subsystem orthogonality' and which differ in the details of Condition~3 below.
In either case, for given boundary regions $R, \Rb$, we say that a large code satisfies  approximate subsystem orthogonality when the following three conditions hold:
\begin{enumerate}[label=Condition \arabic*:,wide=0pt,font=\bfseries]
\item \mbox{}\\*
Small code subspaces $\tH^\a_{\code}$ with different $\a$ are approximately orthogonal on $\ol R$ in the following sense. For every state $\r$ on $\Hlarge$, write it as in~\er{rdecom} and define $\hat\r^\OD:= V \r^\OD V^\dag$.  We require $\hat\r^\OD_R := \Tr_{\ol R} \hat\r^\OD$ to be small in the sense that
\be\la{condrb}
\lt\|\hat\r^\OD_R\rt\|_1 \le \eOD,
\ee
with some small, state-independent $\eOD$.  (Recall from Appendix~\ref{app:Schatten} that for a self-adjoint trace-class operator $O$ with eigenvalues $\lambda_i$ we have $\lt\|O\rt\|_1=\sum_i|\lambda_i|$.)

\item \mbox{}\\*
Recalling that $\tH = \tH_R \otimes \tH_{\Rb}$, we require the small code subspaces $\tH_{\code}^\a$ with different $\a$ to be approximately orthogonal on $R$ in the following sense. There exists an (orthogonal) direct sum decomposition\footnote{We could allow $\tH_R = (\bigoplus_\a \tH_{R\a}) \bigoplus \tH_{\text{remain}}$ with some $\tH_{\text{remain}}$, but for simplicity we choose to absorb $\tH_{\text{remain}}$ into one of the $\tH_{R\a}.$}
\be\la{hrdsum}
\tH_R = \bigoplus_\a \tH_{R\a}
\ee
where $R\a$ are just labels for orthogonal subspaces of $\tH_R$ (and do not mean boundary subregions inside $R$), such that for every $\a$ and every state $\r^\a$ on $\cH^\a$, the resulting $\hat\r^\a_R =\Tr_{\Rb} (V_\a\r^\a V_\a^\dag)$ is approximately within $\tH_{R\a}$.  By this we mean that, letting $P_{R\a}$ denote the projection onto $\tH_{R\a}$, we require the part of $\hat\r^\a_R$ supported on this subspace to dominate in the following sense: writing
\be\la{condr0}
\hat\r^\a_R = \hat\r^\a_{R\a} + \hat\r^\a_{R,\rest}, \qu \text{with} \qu 
\hat\r^\a_{R\a} := P_{R\a}\hat\r^\a_RP_{R\a},\qu
\hat\r^\a_{R,\rest} := \hat\r^\a_R-\hat\r^\a_{R\a},
\ee
we require the sub-leading part $\hat\r^\a_{R,\rest}$ to satisfy
\bg\la{condr1}
\lt\|\hat\r^\a_{R,\rest} \rt\|_1 \leq \eresttr,
\eg
for some small, state- and $\a$-independent $\eresttr$.
Thus $\hat\r^\a_{R\a}$ can be viewed as an (unnormalized) density operator on $\tH_{R\a}$; it is the dominant block of $\hat\r^\a_R$.

\item \mbox{}\\*
We additionally require $\hat\r^\OD_R$ and $\hat\r^\a_{R,\rest}$ above to be small in the following sense. For every state $\r$ on $\Hlarge$, decompose it as $\sum_{\a} p_\a \r^\a + \r^\OD$ according to \er{rdecom} and use \er{condr0} to write the corresponding $\hat\r_R = \sum_{\a} p_\a \hat\r^\a_R +\hat\r^\OD_R$ as
\be\la{decomrest}
\hat\r_R = \hat\r^{\DomD}_R + \hat\r_{R}^\remain, \qu
\hat\r^{\DomD}_R := \bigoplus_\a p_\a \hat\r^\a_{R\a},\qu
\hat\r_{R}^\remain := \sum_\a p_\a \hat\r^\a_{R,\rest} +\hat\r^\OD_R.
\ee
We call $\hat\r^{\DomD}_R$ the `dominant diagonal'  part of $\hat\r_R$, as it is block-diagonal according to the decomposition \er{hrdsum} (though $\hat\r_{R}^\remain$ also has block-diagonal parts). Let $P_{R,\esupp(\r)}$ denote the projection onto $\bigoplus_{\a\in\esupp(\r)}\tH_{R\a}$, where the notion of $\esupp$ is determined by some parameter $\e$. For every state $\s$ on $\Hlarge$, define
\be\la{sigmasuppdef}
\hat\s_{R,\esupp(\r)}:=P_{R,\esupp(\r)}\hat\s_RP_{R,\esupp(\r)}.
\ee
For log approximate subsystem orthogonality, we require 
\be
\la{condr2e}
\left| \<K_{\hat\r_R}\>_{\hat\s_{R,\esupp(\r)}}
-\<K_{\hat\r_R^{\DomD}}\>_{\hat\s_{R,\esupp(\r)}}
\right| \leq\erestlog
\ee
for all states $\r$, $\s$ on $\Hlarge$ for which both expectation values above are finite,\footnote{This is in contrast with our other inequalities, which implicitly require all expectation values therein to be finite unless explicitly stated otherwise. For example, \er{laligned} below implicitly requires the finiteness of all four expectation values therein.} with some small, state-independent values of $\e$ and $\erestlog$, while for exponentiated approximate subsystem orthogonality, we require
\be
\la{condr3e}
\Bigl|\Tr_R\!\left[\hat\s_{R,\esupp(\r)}
\left((\hat\r_R)^{is}-(\hat\r_R^{\DomD})^{is}\right)\right]\Bigr|
\leq\erestexp
\ee
for all states $\r$, $\s$ on $\Hlarge$, with some small, state-independent values of $\e$ and $\erestexp$ (the latter of which can depend on the real parameter $s$).\footnote{In parallel with the convention described in footnote~\re{foot:support}, imaginary powers $O^{is}$ of a positive operator $O$ are defined on its support. Whenever this support is smaller than the relevant Hilbert space, statements involving $O^{is}$ or the corresponding modular-flow operator in definitions, lemmas, and theorems are understood to hold for every unitary extension to that Hilbert space.\la{foot:uextension}}
Here and below, we define $K_{O} := -\ln O$ for any positive operator $O$ (even if it does not have trace $1$).

In the special case where the state $\r$ is chosen to be $\r^\a$ (in a single sector $\cH^\a$), for any $\e\in[0,1)$ the $\e$-support consists of only that $\a$. We also choose $\s$ to be an arbitrary state $\s^\a$ on the same $\cH^\a$ and define
\be\la{sigmablockdef}
\hat\s^\a_{R\a}:=P_{R\a}\hat\s_R^\a P_{R\a}.
\ee
The log condition \er{condr2e} then takes the $\e$-independent form
\be
\la{condr2}
\left|
\<K_{\hat\r_R^\a}\>_{\hat\s^\a_{R\a}}
-\<K_{\hat\r_{R\a}^\a}\>_{\hat\s^\a_{R\a}}
\right| \leq\erestlog
\ee
for all states $\r^\a$, $\s^\a$ on $\cH^\a$ for which both expectation values above are finite.
The exponentiated condition \er{condr3e} takes the $\e$-independent form
\be
\la{condr3}
\Bigl|\Tr_R\!\left[\hat\s^\a_{R\a}
\left((\hat\r_R^\a)^{is}-(\hat\r_{R\a}^\a)^{is}\right)\right]\Bigr|
\leq\erestexp
\ee
for all states $\r^\a$, $\s^\a$ on $\cH^\a$.
\end{enumerate}
\end{ndefi}

We will use the log form of approximate subsystem orthogonality \eqref{condr2e} to derive our projected JLMS results (Theorem~\re{thmlgjlms} in Section~\ref{sec:lpJLMS} and Theorem~\re{thmlgjlmsop} in Section~\ref{sec:lpJLMSop}).  We will use the exponentiated form 
\er{condr3e} to derive the exponentiated JLMS results (Theorem~\re{thmlgexp} in Section~\ref{sec:expJLMSaligned} and Theorem~\re{thmlgexpop} in Section~\ref{sec:expJLMSop}).

In AdS/CFT, we expect the error bounds $\eOD$, $\eresttr$, $\erestlog$, and $\erestexp$ to be nonperturbatively small in $G$, as gravitational path integral calculations do not seem to produce these errors to any perturbative order in $G$. This is because different area windows do not overlap, so that the corresponding area-window constraints are strictly incompatible with each other.

Before proceeding, we comment on how to understand \er{condr2e} and \er{condr3e} intuitively.
They state that, in a certain sense, $\hat\r_R$ and $\hat\r_R^{\DomD}$ remain close after taking a logarithm or imaginary power, so that the correction $\hat\r_R^\remain=\hat\r_R-\hat\r_R^{\DomD}$ is not greatly amplified when restricted to $\esupp(\r)$.
At a heuristic level, expanding the left-hand sides of \er{condr2e}, \er{condr3e} to linear order in $\hat\r_R^\remain$ produces a factor of $(\hat\r^{\DomD}_R)^{-1} = \bigoplus_\a (p_\a \hat\r^\a_{R\a})^{-1}$.  Since $\a\in \esupp(\r)$ bounds the potentially large term $p_\a^{-1}$ to be smaller than $1/\e$, we now understand the conditions to require that $\hat\r_{R}^{\remain}$ be sufficiently small even when multiplied by $1/\e$ (and other factors generated by the expansion). As discussed in the previous paragraph, in AdS/CFT the object $\hat\r_{R}^{\remain}$ is expected to be nonperturbatively small in $G$ so, as long as we choose $\e$ to be a power of $G$ (no matter how large the power is), we expect that we may take $\erestlog$, $\erestexp$ to be nonperturbatively small in $G$ as well.

We now return to our main line of discussion.
For some of our results, we will also need the full map $V$ to be an approximate isometry.  While this in fact follows from approximate subsystem orthogonality when the individual $V_\alpha$ are approximate isometries with errors bounded by an $\a$-independent $\eisosmall$ (see Lemma~\ref{lemmaappisolg} below), it is useful to first parameterize the degree of isometry of $V$ with a new parameter $\eiso$ by stating a new definition:

\begin{ndefi}[Approximate isometry of the large code]\la{defappisolg}
We say that $V$ is an approximate isometry if
\be\la{appisolg}
\lt\|V^{\dag} V -\mathbb 1 \rt\|_\infty \leq \eiso
\ee
with some small $\eiso$ and where $\mathbb 1$ is the identity on $\Hlarge$. Using \er{onalt}, this is equivalent to the condition that for every state $\r$ on $\Hlarge$ we have
\be\la{appisolg2}
\lt| \Tr \hat\r - 1 \rt| = \lt| \tr \(\r \[V^\dag V -\mathbb 1 \] \) \rt| \leq \eiso.
\ee
\end{ndefi}

As mentioned earlier, one may bound $\eiso$ in terms of $\eisosmall$ and $\eOD$ as follows.

\begin{nlemma}[Approximate isometry of the large code]\la{lemmaappisolg}
Suppose that Condition~1 of Definition~\re{defortho} holds, and let $\r$ be a state on $\Hlarge$ such that, for every $\a$ with $p_\a>0$,
\be\la{eisosector}
\bigl|\Tr\hat\r^\a-1\bigr|\leq\eisosmall,
\ee
with an $\a$-independent value of $\eisosmall$.
Then
\be\la{eiso}
\bigl|\Tr\hat\r-1\bigr|\leq\eisosmall+\eOD=: \eiso.
\ee
In particular, if each $V_\a$ is an approximate isometry satisfying \er{appiso} with the same $\eisosmall$, then $V$ must be an approximate isometry with $\eiso$ defined above.
Note that this bound introduces no additional dependence on the dimensions of $\Hlarge$ and $\tH$.
\end{nlemma}

\begin{proof}
We first sandwich~\er{rdecom} between $V$ and $V^\dag$ and use \er{rtdhat}, finding
\be\la{rdecomhat}
\hat\r = \sum_{\a} p_\a \hat\r^\a + \hat\r^\OD,
\ee
where we used $V \r^\a V^\dag= V_\a \r^\a V_\a^\dag = \hat\r^\a$. Using \er{eisosector}, we find
\be
\lt| \sum_\a p_\a \Tr \hat\r^\a - 1 \rt| \leq  \sum_\a p_\a \lt| \Tr \hat\r^\a - 1 \rt| \leq \eisosmall.
\ee
We then use \er{condrb} to write
\be
\lt|\Tr \hat\r^\OD\rt| = \lt|\Tr_R \hat\r^\OD_R\rt| \leq \lt\|\hat\r^\OD_R\rt\|_1 \le \eOD,
\ee
and thus obtain the desired inequality \er{eiso}.

If each $V_\a$ satisfies \er{appiso}, then \er{eiso} is satisfied for every state $\r$ on $\Hlarge$, and $V$ is therefore an approximate isometry.
\end{proof}

In the next two subsections, we will derive approximate versions of the projected JLMS and exponentiated JLMS formulas in the large code that apply when the operators act on special states $\s$ that are `aligned' with the state $\r$ used to define the modular Hamiltonians in the senses described by Definitions~\ref{defaligned} and~\ref{deflaligned} below. In the remaining two subsections, we will derive operator-norm versions of the projected and exponentiated JLMS formulas.  These latter results hold for all states $\s$ but require the state $\r$ to satisfy certain additional `smoothness' conditions.

\begin{ndefi}[Aligned states]\la{defaligned}
For two states $\r$, $\s$ on $\Hlarge$, we say that $\s$ is (approximately) aligned with $\r$ if there exist some small $\e$ and $\ealign$ such that
\be\la{aligned}
\sum_{\a \notin \esupp(\r)} q_\a \leq \ealign,
\ee
i.e., $\s$ is approximately within the $\e$-support of $\r$.
\end{ndefi}

The notion of aligned states above will be sufficient for our exponentiated JLMS results in the large code, but for our projected JLMS results we will need the following variant.

\begin{ndefi}[Log-aligned states]\la{deflaligned}
For two states $\r$, $\s$ on $\Hlarge$, we say that $\s$ is (approximately) log-aligned with $\r$ if there exist some small $\e$ and $\elalign$ such that
\be\la{laligned}
\left|\(\<K_{\hat\r_R}\>_{\hat\s_R}
-\<K_{\hat\r_R}\>_{\hat\s_{R,\esupp(\r)}}\)
-\(\<K_{\hat\r_R^{\DomD}}\>_{\hat\s_R}
-\<K_{\hat\r_R^{\DomD}}\>_{\hat\s_{R,\esupp(\r)}}\)\right|
\leq\elalign.
\ee
Here $\hat\s_{R,\esupp(\r)}$ is defined in \er{sigmasuppdef}.
\end{ndefi}

A useful way of understanding these two alignment conditions is as follows. The ordinary alignment condition \er{aligned}, together with \er{condrb}, \er{condr1}, and \er{eisosector} applied to $\s$, implies that the trace norm of the operator $\hat\s_R -\hat\s_{R,\esupp(\r)}$ is bounded by $\cO(\ealign)+\cO(\eOD)+\cO(\eresttr)$. The log-alignment condition \er{laligned}, on the other hand, requires the corresponding differences of expectation values of $K_{\hat\r_R}$ and $K_{\hat\r_R^{\DomD}}$ to be close.

We will make use of the following lemmas in the next two subsections.

\begin{nlemma}\la{lemmalaligned}
For two states $\r$, $\s$ on $\Hlarge$, if \er{condr2e} holds with some $\erestlog$, $\e$, and if $\s$ is log-aligned with $\r$ (as in Definition~\re{deflaligned}) with some $\elalign$ and the above-mentioned value of $\e$, then we have
\be\la{laligned2}
\left|\<K_{\hat\r_R}\>_{\hat\s_R}-\<K_{\hat\r_R^{\DomD}}\>_{\hat\s_R}\right|
\leq\elalign+\erestlog.
\ee
\end{nlemma}

\begin{proof}
This follows immediately by combining \er{laligned} with \er{condr2e}. Here, \er{laligned} implies the finiteness condition required to apply \er{condr2e}.
\end{proof}

\begin{nlemma}\la{lemmaaligned}
For two states $\r$, $\s$ on $\Hlarge$, suppose that \er{condrb} holds for $\s$ with some $\eOD$, and that for every $\a$ with $q_\a>0$, \er{condr1} holds for $\s^\a$ with the same $\eresttr$ and $|\Tr\hat\s^\a-1|\leq\eisosmall$ holds with the same $\eisosmall$. Suppose also that \er{condr3e} holds for $\r$, $\s$ with some $\erestexp$ and $\e$.  If $\s$ is aligned with $\r$ (as in Definition~\re{defaligned}) with some $\ealign$ and the above-mentioned value of $\e$, then we have
\be\la{aligned2}
\Bigl|\Tr_R\!\left[\hat\s_R\left((\hat\r_R)^{is}-(\hat\r_R^{\DomD})^{is}\right)\right]\Bigr|
\lesssim\ealign+\eOD+\eresttr+\erestexp.
\ee
\end{nlemma}

\begin{proof}
Applying \er{decomrest} to $\s$, we find
\be
\hat\s_R
=\sum_\a q_\a\left(\hat\s^\a_{R\a}+\hat\s^\a_{R,\rest}\right)
+\hat\s_R^\OD.
\ee
Using $\|X-PXP\|_1\leq2\|X\|_1$ for any projection $P$ and trace-class $X$, we obtain
\be
\lt\|\hat\s_R-\hat\s_{R,\esupp(\r)}\rt\|_1
\leq\sum_{\a\notin\esupp(\r)}q_\a\lt\|\hat\s^\a_{R\a}\rt\|_1
+2\sum_\a q_\a\lt\|\hat\s^\a_{R,\rest}\rt\|_1
+2\lt\|\hat\s_R^\OD\rt\|_1.
\ee
The first term is bounded using
\be
\lt\|\hat\s^\a_{R\a}\rt\|_1
\leq\lt\|\hat\s^\a\rt\|_1
\leq1+\eisosmall,
\ee
together with the alignment condition \er{aligned}.
We then combine this with \er{condr1} for each $\s^\a$ with $q_\a>0$, \er{condrb} for $\s$, and $\sum_\a q_\a=1$ to obtain
\be
\lt\|\hat\s_R-\hat\s_{R,\esupp(\r)}\rt\|_1
\lesssim\ealign+\eresttr+\eOD.
\ee
Using $\big\|\left(\hat\r_R\right)^{is}\big\|_\infty = \big\|\left(\hat\r_R^{\DomD}\right)^{is}\big\|_\infty =1$ (which follows from positivity of $\hat\r_R$ and $\hat\r_R^{\DomD}$), we find
\be\la{bdytraligned}
\Bigl|\Tr_R\!\left[
\left(\hat\s_R-\hat\s_{R,\esupp(\r)}\right)
\left((\hat\r_R)^{is}-(\hat\r_R^{\DomD})^{is}\right)
\right]\Bigr|
\lesssim\ealign+\eOD+\eresttr.
\ee
Combining this with \er{condr3e} immediately gives \er{aligned2}.
\end{proof}

\subsection{Projected JLMS for log-aligned states}
\label{sec:lpJLMS}

We are now almost ready to derive our large code approximate projected JLMS result for log-aligned states.  However, before doing so, we require one final definition.

\begin{ndefi}[Enhanced log-stability]\la{defenhancedstable}
For a state $\r$ on $\Hlarge$, we say that it has enhanced log-stability with respect to a state $\s$ on $\Hlarge$ if $p_\a>0$ and $\r^\a$ satisfies log-stability (see Definition~\re{defsmallstable}) for all $\a$ with some $\a$-independent $\etail$, and $\r$, $\s$ further satisfy
\ba\la{enhanced1}
\left|
\bigl\<K_{\hat\r_R^{\DomD}}\bigr\>_{\hat\s_R^\OD}
-
\biggl\<\fr{\hat A}{4G}+K_{\r_r}\biggr\>_{\s^\OD V^\dag V}
\right|
&\leq\eenhanced,\\
\la{enhanced2}
\left|\sum_{\a:q_\a>0} q_\a\left\{
\bigl\<K_{\hat\r_R^{\DomD}}\bigr\>_{\hat\s^\a_{R,\rest}}
-\bigl\<K_{p_\a\hat\r_R^\a}\bigr\>_{\hat\s^\a_{R,\rest}}
\right\}\right|
&\leq\eenhanced,
\ea
with some small $\eenhanced$. Here $\hat\s^\a_{R,\rest}$ is defined by \er{condr0} with $\r^\a$ replaced by $\s^\a$.
\end{ndefi}

Before proceeding, we comment on why we believe that the above conditions can hold in the AdS/CFT context.
First define $(V^\dag V)^D:=\bigoplus_\a V_\a^\dag V_\a$, $(V^\dag V)^\OD:=V^\dag V-(V^\dag V)^D$ as the block-diagonal and off-block-diagonal parts of $V^\dag V$, respectively. Since $\fr{\hat A}{4G}+K_{\r_r}$ is block diagonal, $\s^\OD V^\dag V$ in \er{enhanced1} can be replaced by $\s (V^\dag V)^\OD$, whose trace norm is bounded by $\eOD$ due to
\be
\lt\|(V^{\dag} V)^\OD\rt\|_\infty = \sup_\s \lt|\tr\[\s (V^{\dag} V)^\OD\] \rt| = \sup_\s \lt|\Tr \hat\s^\OD \rt| = \sup_\s \lt|\Tr_R  \hat\s_R^\OD  \rt| \leq \sup_\s \lt\|\hat\s^\OD_R\rt\|_1 \le \eOD,
\ee
where we used \er{onalt} and \er{condrb}. Similarly, \er{condrb} and \er{condr1} give $\|\hat\s_R^\OD\|_1\leq\eOD$ and $\|\hat\s^\a_{R,\rest}\|_1\leq\eresttr$.  As mentioned earlier, we believe that $\eOD$ and $\eresttr$ are exponentially small in $G$. The content of \er{enhanced1} and \er{enhanced2} is thus that these exponentially small errors do not become much larger than $\eenhanced$ after taking the differences of the expectation values there.
We remind the reader that we will need this enhanced log-stability criterion only to prove our projected JLMS results below and not for the exponentiated JLMS results in Section~\ref{sec:expJLMSaligned} and Section~\ref{sec:expJLMSop}.

We now present our approximate projected JLMS relation in a large code for log-aligned states and the associated proof.  As usual, the argument consists of carefully combining the above error bounds and working through the details.

\begin{theorem}[Approximate projected JLMS in a large code for log-aligned states]\la{thmlgjlms}
Suppose that we have a large code defined by a collection of $\cH^\a$ (as in Definition~\ref{deflargecode}) that satisfies the following two conditions:
\begin{enumerate}[label=(\alph*)]
    \item For each $\a$, we have an approximate FLM formula and an approximate isometry $V_\alpha$ as defined by some $\a$-independent values of $\eFLM$ and $\eisosmall$.
    \item The log version of approximate subsystem orthogonality holds (see Definition~\re{defortho}) for some $\eOD$, $\eresttr$, $\erestlog$, and $\e$.
\end{enumerate}
Suppose also that we have two states $\r$, $\s$ on $\Hlarge$ that satisfy the following two conditions:
\begin{enumerate}[label=(\roman*)]
    \item $\r$ has enhanced log-stability with respect to $\s$ (as in Definition~\re{defenhancedstable}) with some $\etail$ and $\eenhanced$.
    \item $\s$ is log-aligned with $\r$ as defined by some $\elalign$ and the above-mentioned value of $\e$.
\end{enumerate}
Then these states satisfy the approximate projected JLMS formula
\be\la{lgjlms}
\left|
\<K_{\td\r_R}\>_{\hat\s_R}
-
\biggl\<\fr{\hat A}{4G}+K_{\r_r}\biggr\>_{\s V^\dag V}
\right|
\lesssim\epJLMS+\eiso+\eenhanced+\erestlog+\elalign,
\ee
where $\epJLMS$ is defined by \er{epJLMSdef} and $\eiso$ is given by \er{eiso}.
\end{theorem}

\begin{proof}
We bound the left-hand side of \er{lgjlms} by a sum of terms, each of which is small.

First, since $\s$ is log-aligned with $\r$ and the assumptions in Lemma~\re{lemmalaligned} are satisfied, we have \er{laligned2}, which we reproduce here:
\be\la{laligned3}
\left|\<K_{\hat\r_R}\>_{\hat\s_R}
-\<K_{\hat\r_R^{\DomD}}\>_{\hat\s_R}\right|
\leq\elalign+\erestlog.
\ee
Recalling the relation $\td\r_R=\hat\r_R/\Tr\hat\r$ from \er{rtdhat}, we use Lemma~\re{lemmaappisolg} for both $\r$ and $\s$ to obtain
\be\la{trdiff1}
\bigl|\<K_{\td\r_R}\>_{\hat\s_R}-\<K_{\hat\r_R}\>_{\hat\s_R}\bigr|
=\bigl|\ln(\Tr\hat\r)\bigr|\Tr\hat\s
\lesssim\eiso,
\ee
where $\eiso$ is given by \er{eiso}.
Furthermore, by the assumptions stated in the theorem, we see that for every $\a$ the state $\r^\a$ is log-stable and (according to Theorem~\re{thmsmjlms}) satisfies the approximate projected JLMS formula \er{smjlms}, which upon taking expectation values in $\s^\a$ gives
\be\la{smjlmsev}
\left|
\<K_{\td\r_R^\a}\>_{\hat\s_R^\a}
-\lt\<\fr{A^\a}{4G} + K_{\r_r^\a}\rt\>_{\s^\a V_\a^\dag V_\a}
\right| \lesssim \epJLMS.
\ee
For the rest of the proof, write
\be
\s^D:=\sum_\a q_\a\s^\a,\qquad
\hat\s_R^D:=\sum_\a q_\a\hat\s_R^\a.
\ee
By \er{sdecom}, $\s=\s^D+\s^\OD$ and $\hat\s_R=\hat\s_R^D+\hat\s_R^\OD$.  

We now show that both expectation values in \er{condr2} are finite for every $\a$ with $q_\a>0$.
First, \er{smjlmsev} and \er{enhanced2} imply the finiteness of $\<K_{\hat\r_R^\a}\>_{\hat\s_R^\a}$ and $\<K_{\hat\r_R^\a}\>_{\hat\s_{R,\rest}^\a}$, and thus also of $\<K_{\hat\r_R^\a}\>_{\hat\s_{R\a}^\a}$, the first expectation value in \er{condr2}.
Moreover, \er{laligned3} and \er{enhanced1} imply the finiteness of $\<K_{\hat\r_R^{\DomD}}\>_{\hat\s_R}$ and $\<K_{\hat\r_R^{\DomD}}\>_{\hat\s_R^\OD}$, and thus also of $\<K_{\hat\r_R^{\DomD}}\>_{\hat\s_R^D}$.
Note that we have
\be\la{phatrhobound}
\bigl\|p_\a\hat\r_{R\a}^\a\bigr\|_\infty \leq \bigl\|p_\a\hat\r_R^\a\bigr\|_\infty \leq \bigl\|\hat\r_R^\a\bigr\|_1 = \Tr_R \hat\r_R^\a \leq 1+\eisosmall,\qu \forall\, \a.
\ee
Therefore, $\hat\r_R^{\DomD}$ is also bounded above by $1+\eisosmall$, and
\be\la{kdomdbound}
K_{\hat\r_R^{\DomD}}+\ln(1+\eisosmall) \geq 0.
\ee
Thus the finiteness of $\<K_{\hat\r_R^{\DomD}}\>_{\hat\s_R^D}$ ensures that each $\<K_{\hat\r_R^{\DomD}}\>_{\hat\s_R^\a}$ with $q_\a>0$ is finite as well. Since $P_{R\a}$ commutes with $K_{\hat\r_R^{\DomD}}$, $\<K_{\hat\r_R^{\DomD}}\>_{\hat\s_{R\a}^\a}$ is finite. This shows the finiteness of $\<K_{\hat\r_{R\a}^\a}\>_{\hat\s_{R\a}^\a}$, the second expectation value in \er{condr2}.

Having shown that \er{condr2} applies for each such $\a$, we now sum these inequalities with weights $q_\a$ and add \er{enhanced2} to obtain
\be\la{trdiff20}
\left|
\sum_{\a:q_\a>0}q_\a
\left\{
\<K_{\hat\r_R^{\DomD}}\>_{\hat\s_R^\a}
-\<K_{p_\a\hat\r_R^\a}\>_{\hat\s_R^\a}
\right\}
\right|
\lesssim\eenhanced+\erestlog.
\ee
Using \er{kdomdbound}, $\Tr_R\hat\s_R^\a\leq 1+\eisosmall$, and the finiteness of $\<K_{\hat\r_R^{\DomD}}\>_{\hat\s_R^D}$ established above, we can sum the first terms in \er{trdiff20} to obtain
\be\la{trdiff2}
\left|
\<K_{\hat\r_R^{\DomD}}\>_{\hat\s_R^D} -
\sum_\a q_\a
\<K_{p_\a\hat\r_R^\a}\>_{\hat\s_R^\a}
\right|
\lesssim\eenhanced+\erestlog.
\ee
Moreover, since $K_{p_\a\hat\r_R^\a}$ is also bounded from below by $-\ln(1+\eisosmall)$ due to \er{phatrhobound}, the negative contribution to the sum in \er{trdiff2} converges, and therefore the sum converges absolutely:
\be\la{trdiff2abs}
\sum_\a q_\a
\left|\<K_{p_\a\hat\r_R^\a}\>_{\hat\s_R^\a}\right|<\infty.
\ee

Using \er{trdiff1} with $\r$ replaced by $\r^\a$, we find that \er{smjlmsev} implies
\be\la{smjlms4}
\left|
\<K_{p_\a\hat\r_R^\a}\>_{\hat\s_R^\a}
-\lt\<\fr{A^\a}{4G} + K_{p_\a\r_r^\a}\rt\>_{\s^\a V_\a^\dag V_\a}
\right| \lesssim \epJLMS +\eiso.
\ee
Since this bound is uniform in $\a$, \er{trdiff2abs} implies
\be\la{trdiff3abs}
\sum_\a q_\a
\left|
\lt\<\fr{A^\a}{4G} + K_{p_\a\r_r^\a}\rt\>_{\s^\a V_\a^\dag V_\a}
\right|<\infty.
\ee

We now show
\be\la{bulkdiagev}
\biggl\<\fr{\hat A}{4G}+K_{\r_r}\biggr\>_{\s^D V^\dag V}
=
\sum_\a q_\a
\lt\<\fr{A^\a}{4G} + K_{p_\a\r_r^\a}\rt\>_{\s^\a V_\a^\dag V_\a}.
\ee
Recall from \er{EVdefgen} that each expectation value is defined using a spectral cutoff imposed by a spectral projection $P_\L$ of $\fr{\hat A}{4G}+K_{\r_r}$.
Since $\fr{\hat A}{4G}+K_{\r_r}$ is block diagonal by \er{rrdecom} and \er{aopdef}, so is $P_\L$, with blocks denoted $P_\L^{(\a)}$. At any finite cutoff, the expectation value on the left-hand side of \er{bulkdiagev} becomes 
\be\la{bulkdiagevcut}
\tr \[\s^D V^\dag V \biggl(\fr{\hat A}{4G}+K_{\r_r}\biggr) P_\L\] =\sum_\a q_\a \tr \[\s^\a V_\a^\dag V_\a \(\fr{A^\a}{4G} + K_{p_\a\r_r^\a}\) P_\L^{(\a)}\].
\ee
Each term in the sum is bounded by
\ba
\left|
\tr \[\s^\a V_\a^\dag V_\a \(\fr{A^\a}{4G} + K_{p_\a\r_r^\a}\) P_\L^{(\a)}\]
\right| &\leq
\lt\| V_\a^\dag V_\a \rt\|_\infty \lt\| \fr{A^\a}{4G} -\ln p_\a + K_{\r_r^\a} \rt\|_\infty\\
&\leq
(1+\eisosmall)
\left(
\left|\fr{A^\a}{4G}-\ln p_\a\right|
+\bigl|\ln\etail\bigr|
\right),
\la{bulkdiagcutbd}
\ea
where we used \er{trsKbd}.
Similarly, we note
\be\la{bulkdiaglowerbd}
\left|
\lt\<\fr{A^\a}{4G} + K_{p_\a\r_r^\a}\rt\>_{\s^\a V_\a^\dag V_\a}
\right|
\geq
(1-\eisosmall) \left|\fr{A^\a}{4G}-\ln p_\a\right|
-(1+\eisosmall)\, \bigl|\ln\etail\bigr|.
\ee
Combining \er{bulkdiagcutbd} with \er{bulkdiaglowerbd}, we find that the absolute value of each summand on the right-hand side of \er{bulkdiagevcut} satisfies a cutoff-independent bound that is summable by \er{trdiff3abs}. Dominated convergence therefore allows the cutoff to be removed before summing over $\a$ in \er{bulkdiagevcut}, thus showing \er{bulkdiagev}.

Now summing \er{smjlms4} over $\a$ with weights $q_\a$ and using \er{bulkdiagev}, we find
\be\la{trdiff3}
\left|
\sum_\a q_\a
\<K_{p_\a\hat\r_R^\a}\>_{\hat\s_R^\a}
- \biggl\<\fr{\hat A}{4G}+K_{\r_r}\biggr\>_{\s^D V^\dag V}
\right| \lesssim \epJLMS +\eiso.
\ee
The promised result \er{lgjlms} then follows by combining \er{trdiff1}, \er{laligned3}, \er{trdiff2}, \er{trdiff3}, and \er{enhanced1}.
\end{proof}

\subsection{Exponentiated JLMS for aligned states}
\label{sec:expJLMSaligned}

We now derive our approximate exponentiated JLMS relation for aligned states.  As one might expect, the proof consists of carefully combining the error bounds associated with the exponentiated versions of the properties defined in Section~\ref{sec:lcdefs}.

\begin{theorem}[Approximate exponentiated JLMS in a large code for aligned states]\la{thmlgexp}
Let $s$ be real.  Suppose that we have a large code defined by a collection of $\cH^\a$ (as in Definition~\ref{deflargecode}) that satisfies the exponentiated version of approximate subsystem orthogonality (see Definition~\re{defortho}) for some $\eOD$, $\eresttr$, $\erestexp$, and $\e$.
Suppose also that we have two states $\r$, $\s$ on $\Hlarge$ that satisfy the following three conditions:
\begin{enumerate}[label=(\roman*)]
    \item For every $\a$, we have $|\Tr\hat\r^\a-1|\leq\eisosmall$ whenever $p_\a>0$ and $|\Tr\hat\s^\a-1|\leq\eisosmall$ whenever $q_\a>0$, with some $\a$-independent $\eisosmall$.
    \item For every $\a\in\esupp(\r)$ (as defined by the above-mentioned value of $\e$) with $q_\a>0$, the approximate R\'enyi FLM formula~\eqref{flmereexp} for $n=1+is$ holds on $R$ for all mixtures of $\r^\a$ and $\s^\a$  with the same $A^\a$, $\x_s$, and $\eFLM$ (the latter two being independent of $\a$) and, moreover, $\r^\a$ is relatively log-stable with respect to $\s^\a$ (as defined in Definition~\re{defsmallstablerel}) with an $\a$-independent $\etail$.
    \item $\s$ is aligned with $\r$ as defined by some $\ealign$ and the above-mentioned value of $\e$.
\end{enumerate}
Then these states satisfy the approximate exponentiated JLMS formula\footnote{As specified in footnote~\re{foot:uextension}, this bound is required to hold for every unitary extension of the boundary and bulk flows.}
\be
\la{lgexp}
\lt| \tr \(\s \[ V^\dag e^{-is K_{\tdr_R}} V - e^{-is \(\fr{\hat A}{4G} + K_{\r_r}\)} \] \) \rt| \lesssim \eeJLMS + |s|\, \eiso + \eOD + \eresttr + \erestexp + \ealign,
\ee
with $\eeJLMS$ defined by \er{eeJLMSdef} and $\eiso$ given by \er{eiso}.
\end{theorem}

\begin{proof}
Upon using \er{rtdhat}, we only need to show
\be  \la{lgexp2}
\lt| \Tr \(\hat\s \tdr_R^{is}\) - \tr \(\s e^{-is \(\fr{\hat A}{4G} + K_{\r_r}\)}\) \rt| \lesssim \eeJLMS + |s|\, \eiso + \eOD + \eresttr + \erestexp + \ealign.
\ee
To establish this relation, we first study its left-hand side. Using \er{rrdecom}, \er{aopdef}, and \er{sdecom}, we may write 
\be\la{bulktr}
\tr \(\s e^{-is \(\fr{\hat A}{4G} + K_{\r_r}\)}\)
= \sum_{\a\in\esupp(\r)} q_\a \tr \(\s^{\a} e^{-is \fr{A^\a}{4G}} (p_\a \r^{\a}_r)^{is}\)
+\cO(\ealign),
\ee
where the alignment condition \er{aligned} bounds the total contribution from $\a\notin\esupp(\r)$.

Let us now consider $\Tr \left(\hat\s \tdr_R^{is}\right)$.  Applying Lemma~\re{lemmaappisolg} to $\r$ and $\s$ using condition~(i) gives
\be\la{trrhat}
\Tr \hat \r,\,\,
\Tr \hat \s = 1+ \cO(\eiso)
\ee
and hence
\be\la{bdytrris}
\(\Tr \hat \r\)^{is} = 1+ \cO(|s|\, \eiso).
\ee
Recalling now the relation $\td\r_R = \hat\r_R / \Tr\hat\r$ from \er{rtdhat}, we find
\be\la{bdytr1}
\Tr \(\hat\s \tdr_R^{is}\) = \frac{\Tr \(\hat\s \hat\r_R^{is}\)}{\(\Tr \hat \r\)^{is}} = \Tr \(\hat\s \hat\r_R^{is}\) + \cO(|s|\, \eiso),
\ee
where we used $\lt|\Tr \left(\hat\s \hat\r_R^{is}\right) \rt| \leq \Tr \hat\s = \cO(1)$, with the last step following from \er{trrhat}.
Since $\s$ is aligned with $\r$ and the assumptions in Lemma~\re{lemmaaligned} are satisfied, we have \er{aligned2}, which we reproduce here:
\be\la{bdytr2}
\bigg| \Tr \[\hat\s \(\hat\r_R^{is} - \big(\hat\r_R^{\DomD}\big)^{is}\)\] \bigg| \lesssim \ealign + \eOD + \eresttr + \erestexp.
\ee
Applying \er{decomrest} to $\s$ then yields
\be
\hat\s_R = \hat\s^\DomD_R + \sum_\a q_\a \hat\s^\a_{R,\rest} + \hat\s^\OD_R.
\ee
From \er{condrb}, \er{condr1}, $\lt|\Tr(O O')\rt| \leq \|O\|_1 \|O'\|_\infty$ (see e.g.\ \er{trprodbound}), $\sum_\a q_\a=1$, and $\big\|\left(\hat\r_R^{\DomD}\right)^{is}\big\|_\infty =1$ (which follows from positivity of $\hat\r_R^{\DomD}$), we obtain
\be\la{bdytr3}
\Tr \(\hat\s \(\hat\r_R^{\DomD}\)^{is}\) = \Tr_R \(\hat\s_R \(\hat\r_R^{\DomD}\)^{is}\) = \Tr_R \(\hat\s^{\DomD}_R \(\hat\r_R^{\DomD}\)^{is}\) +\cO(\eOD)+\cO(\eresttr).
\ee
Recalling the definitions of $\hat\r^{\DomD}_R$, $\hat\s^{\DomD}_R$ from \er{decomrest},
\be
\hat\r^{\DomD}_R := \bigoplus_\a p_\a \hat\r^\a_{R\a},\qqu
\hat\s^{\DomD}_R := \bigoplus_\a q_\a \hat\s^\a_{R\a},
\ee
we may write
\ba\la{bdytr4}
\Tr_R \(\hat\s^{\DomD}_R \(\hat\r_R^{\DomD}\)^{is}\) &= \sum_{\a\in\esupp(\r)} q_\a \Tr_{R\a} \(\hat\s^\a_{R\a} \(p_\a \hat\r^\a_{R\a}\)^{is}\)+\cO(\ealign),
\ea
where the alignment condition \er{aligned} and condition~(i) bound the total contribution from $\a\notin\esupp(\r)$.
Applying the exponentiated approximate subsystem orthogonality condition \er{condr3}, we find
\be\la{bdytr5}
\sum_{\a\in\esupp(\r)} q_\a \Tr_{R\a} \(\hat\s^\a_{R\a} \(p_\a \hat\r^\a_{R\a}\)^{is}\) = \sum_{\a\in\esupp(\r)} q_\a \Tr_R \(\hat\s^\a_{R\a} \(p_\a \hat\r^\a_R\)^{is}\) +\cO\(\erestexp\).
\ee
Using \er{condr0}, \er{condr1}, and $\lt|\Tr(O O')\rt| \leq \|O\|_1 \|O'\|_\infty$ (see e.g.\ \er{trprodbound}) then gives
\be
\lt| \Tr_R \(\(\hat\s^\a_R - \hat\s^\a_{R\a}\) \(p_\a \hat\r^\a_R\)^{is} \) \rt| \leq \lt\| \hat\s^\a_R - \hat\s^\a_{R\a} \rt\|_1 \lt\| \(p_\a \hat\r^\a_R\)^{is} \rt\|_\infty \leq \eresttr,
\ee
which yields
\be\la{bdytr6}
\sum_{\a\in\esupp(\r)} q_\a \Tr_R \(\hat\s^\a_{R\a} \(p_\a \hat\r^\a_R\)^{is}\) = \sum_{\a\in\esupp(\r)} q_\a \Tr_R \(\hat\s^\a_R \(p_\a \hat\r^\a_R\)^{is}\) +\cO(\eresttr).
\ee
For every $\a\in\esupp(\r)$, condition~(i) gives \er{trrhat} and \er{bdytrris} with $\r$ replaced by $\r^\a$, and we have
\be
\(\Tr \hat \r^\a\)^{is} = 1+ \cO(|s|\, \eiso).
\ee
Further restricting to $q_\a>0$ and recalling $\td\r^\a_R = \hat\r^\a_R / \Tr\hat\r^\a$ from \er{eq:tilderho} then gives
\be
\Tr_R \(\hat\s^\a_R \(\tdr^\a_R\)^{is}\) = \frac{\Tr_R \(\hat\s^\a_R \(\hat\r^\a_R\)^{is}\)}{\(\Tr \hat \r^\a\)^{is}} = \Tr_R \(\hat\s^\a_R \(\hat\r^\a_R\)^{is}\) + \cO(|s|\, \eiso),
\ee
and thus
\be\la{bdytr7}
\sum_{\a\in\esupp(\r)} q_\a \Tr_R \(\hat\s^\a_R \(p_\a \hat\r^\a_R\)^{is}\) = \sum_{\a\in\esupp(\r)} q_\a \Tr_R \(\hat\s^\a_R \(p_\a \tdr^\a_R\)^{is}\) +\cO(|s|\, \eiso).
\ee
Furthermore, by the assumptions stated in the theorem, we see that for every $\a \in \esupp(\r)$ with $q_\a>0$, the state $\r^\a$ is relatively log-stable with respect to $\s^\a$ and (according to Theorem~\re{thmsmexprel}) satisfies the exponentiated JLMS formula \er{smexprel}, which leads to
\be\la{bdytr8}
\lt|\sum_{\a\in\esupp(\r)} q_\a \Tr_R \(\hat\s^\a_R \(p_\a \tdr^\a_R\)^{is}\) - \sum_{\a\in\esupp(\r)} q_\a \tr \(\s^\a e^{-is \fr{A^\a}{4G}} \(p_\a \r^\a_r\)^{is}\) \rt| \lesssim \eeJLMS.
\ee
The advertised result \er{lgexp2} then follows by
combining \er{bulktr}, \er{bdytr1}, \er{bdytr2}, \er{bdytr3}, \er{bdytr4}, \er{bdytr5}, \er{bdytr6}, \er{bdytr7}, and \er{bdytr8}.
\end{proof}

\subsection{Operator-norm error bounds for projected JLMS}
\label{sec:lpJLMSop}

The previous two subsections used various notions of `alignment' between $\rho$ and $\sigma$ to control errors in the desired JLMS results.  As foreshadowed in the discussion above Definition~\ref{defaligned}, an alternative approach is to impose an additional `smoothness' restriction on $\rho$ and then to allow $\sigma$ to be arbitrary.  In particular, this results in bounds on the norms of operators defined by subtracting the left and right sides of various JLMS-like relations.  We derive such bounds below.

We begin by introducing the smoothness criterion motivated earlier.

\begin{ndefi}[Smoothness]\la{defsmooth}
For a state $\r$ on $\Hlarge$, we say that it is log-smooth if, for every state $\s$ on $\Hlarge$ that is supported in finitely many $\a$-sectors,
\be\la{smcond1}
\left|\<K_{\hat\r_R}\>_{\hat\s_R}-\<K_{\hat\r_R^{\DomD}}\>_{\hat\s_R}\right|
\leq\esmoothlog
\ee
for some $\s$-independent $\esmoothlog$,
and we say that it is exponentiated-smooth if, for every state $\s$ on $\Hlarge$,
\be\la{smcond2}
\Bigl|\Tr_R\!\left[\hat\s_R\left((\hat\r_R)^{is}-(\hat\r_R^{\DomD})^{is}\right)\right]\Bigr|
\leq\esmoothexp
\ee
for some $\esmoothexp$ that may depend on $s$ but not on $\s$.
\end{ndefi}

The relations \er{smcond1}, \er{smcond2} are analogues of \er{condr2e}, \er{condr3e} with $\hat\s_{R,\esupp(\r)}$ replaced by $\hat\s_R$.
We now explain how they are related to smoothness of $p_\a$.
These two relations state that the difference $\hat\r_R^\remain =\hat\r_R-\hat\r_{R}^\DomD$ is small in a certain sense.
At a heuristic level, expanding the left-hand side of \er{smcond1}, \er{smcond2} to linear order in $\hat\r_R^\remain$ produces a factor $(\hat\r_R^{\DomD})^{-1}=\bigoplus_\a(p_\a\hat\r^\a_{R\a})^{-1}$.
Recall the definition $\hat\r_R^\remain:=\sum_\b p_\b\hat\r^\b_{R,\rest}+\hat\r_R^\OD$ which implies, in particular, that the terms involving $p_\b\hat\r^\b_{R,\rest}$ can contain factors $p_\b/p_\a$.
As long as the corresponding matrix elements are nonzero (and even if they are very small), they can contribute to \er{smcond1}, \er{smcond2} with a factor of $p_\b /p_\a$, leading to a danger of violating \er{smcond1}, \er{smcond2} if $p_\b/p_\a$ is very large (which can happen because $p_\a$ can be arbitrarily small).
However, this danger may be avoided if $p_\a$ is sufficiently smooth as a function of $\a$ (in the sense of varying sufficiently slowly) so that $p_\b /p_\a$ can only be very large when $\a$ and $\b$ are far enough apart that those matrix elements (which control the degree of subsystem orthogonality between $\a$ and $\b$) become sufficiently small to overcome $p_\b /p_\a$ (as may be expected from gravitational path integral calculations).
In the rest of this subsection, we use the log-smooth condition above to derive the operator-norm version of our projected JLMS result in a large code, leaving the exponentiated JLMS result to the next subsection.

\begin{theorem}[Operator-norm version of approximate projected JLMS in a large code]\la{thmlgjlmsop}
Suppose that we have a large code defined by a collection of $\cH^\a$ (as in Definition~\ref{deflargecode}) that satisfies the following two conditions:
\begin{enumerate}[label=(\alph*)]
    \item For each $\a$, we have an approximate FLM formula and an approximate isometry $V_\alpha$ as defined by some $\a$-independent values of $\eFLM$ and $\eisosmall$.
    \item The log version of approximate subsystem orthogonality holds (see Definition~\re{defortho}) for some $\eOD$, $\eresttr$, $\erestlog$, and $\e$.
\end{enumerate}
Suppose also that we have a state $\r$ on $\Hlarge$ that satisfies the following two conditions:
\begin{enumerate}[label=(\roman*)]
    \item $\r$ has enhanced log-stability (as in Definition~\re{defenhancedstable}) with respect to every state $\s$ on $\Hlarge$ that is supported in finitely many $\a$-sectors, with some $\s$-independent values of $\etail$ and $\eenhanced$.
    \item $\r$ is log-smooth in the sense of satisfying \er{smcond1} with some $\esmoothlog$.
\end{enumerate}
Then one has the approximate projected JLMS formula
\be\la{lgjlmsop}
\lt\| V^\dag K_{\td\r_R} V - V^{\dag} V \biggl(\fr{\hat A}{4G} + K_{\r_r} \biggr) \rt\|_\infty \lesssim \epJLMS + \eiso + \eenhanced + \erestlog + \esmoothlog,
\ee
with $\epJLMS$ defined by \er{epJLMSdef} and $\eiso$ given by \er{eiso}. Note that this bound does not depend on the choice of $\e$.
Here, to allow for unbounded bulk or boundary operators, the operator inside the norm is taken to mean the unique bounded extension of $T^\dag T-V^\dag V \(\fr{\hat A}{4G}+K_{\r_r}\)$, with $T$ defined as $K_{\td\r_R}^{1/2}V$ restricted to vectors supported in finitely many $\a$-sectors.
\end{theorem}

\begin{proof}
Repeating the proof of Theorem~\re{thmlgjlms}, but using the log-smoothness condition \er{smcond1} in place of \er{laligned3}, gives for every state $\s$ supported in finitely many $\a$-sectors
\be\la{lgjlms3}
\left|
\<K_{\td\r_R}\>_{\hat\s_R}
-
\biggl\<\fr{\hat A}{4G}+K_{\r_r}\biggr\>_{\s V^\dag V}
\right|
\lesssim\epJLMS+\eiso+\eenhanced+\erestlog+\esmoothlog.
\ee
In particular, for every normalized $|\psi\>$ supported in finitely many $\a$-sectors, this holds for $\s=|\psi\>\<\psi|$.
Using \er{EVpartialtrace} and \er{EVvector}, 
for such $\sigma$ we write \er{lgjlms3} as
\be
\left|
\bigl\|T|\psi\>\bigr\|^2
-
\biggl\<\psi\bigg|V^\dag V \biggl(\fr{\hat A}{4G}+K_{\r_r}\biggr)\bigg|\psi\biggr\>
\right|
\lesssim\epJLMS+\eiso+\eenhanced+\erestlog+\esmoothlog.
\ee
Using Lemma~\re{lemmaopnorm} in Appendix~\re{app:Schatten} then gives \er{lgjlmsop}.
\end{proof}

\subsection{Operator-norm error bounds for exponentiated JLMS}
\label{sec:expJLMSop}

Our final result will be to use the exponentiated-smooth condition introduced in the previous subsection to derive the operator-norm version of our exponentiated JLMS result in a large code.

\begin{theorem}[Operator-norm version of approximate exponentiated JLMS in a large code]\la{thmlgexpop}
Let $s$ be real.  Suppose that we have a large code defined by a collection of $\cH^\a$ (as in Definition~\ref{deflargecode}) that satisfies the following two conditions:
\begin{enumerate}[label=(\alph*)]
    \item For each $\a$, we have an approximate R\'enyi FLM formula~\eqref{flmereexp} for $n=1+is$ and an approximate isometry $V_\alpha$ as defined by some $\a$-independent values of $\x_s$, $\eFLM$, and $\eisosmall$.
    \item The exponentiated version of approximate subsystem orthogonality holds (see Definition~\re{defortho}) for some $\eOD$, $\eresttr$, $\erestexp$, and $\e$.
\end{enumerate}
Suppose also that we have a state $\r$ on $\Hlarge$ that satisfies the following two conditions:
\begin{enumerate}[label=(\roman*)]
    \item $p_\a>0$ and $\r^\a$ satisfies log-stability (see Definition~\re{defsmallstable}) for all $\a$ with some $\a$-independent $\etail$.
    \item $\r$ is exponentiated-smooth in the sense of satisfying \er{smcond2} with some $\esmoothexp$.
\end{enumerate}
Then one has the approximate exponentiated JLMS formula
\be
\la{lgexpop}
\lt\| V^\dag e^{-is K_{\tdr_R}} V - e^{-is \(\fr{\hat A}{4G} + K_{\r_r}\)} \rt\|_\infty \lesssim \eeJLMS + |s|\, \eiso + \eOD + \eresttr + \erestexp + \esmoothexp,
\ee
with $\eeJLMS$ defined by \er{eeJLMSdef} and $\eiso$ given by \er{eiso}. Note that this bound does not depend on the choice of $\e$.
\end{theorem}

\begin{proof}
Using \Eqref{onbound}, we only need to prove that for every state $\s$ on $\Hlarge$, we have
\be\la{lgexp3}
\lt| \tr \(\s \[ V^\dag e^{-is K_{\tdr_R}} V - e^{-is \(\fr{\hat A}{4G} + K_{\r_r}\)} \] \) \rt| \lesssim \eeJLMS + |s|\, \eiso + \eOD + \eresttr + \erestexp + \esmoothexp.
\ee
The rest of the proof is the same as the proof of Theorem~\re{thmlgexp}, except for two differences. First, we use the exponentiated-smoothness condition \er{smcond2} in place of \er{bdytr2}.
Second, since here $\r^\a$ is log-stable for all $\a$, we replace every sum over $\a\in\esupp(\r)$ in that proof with a sum over all $\a$ and omit the corresponding $\ealign$ error terms.
\end{proof}

\section{Discussion}
\label{sec:discussion}

The goal of our work was to construct a `large' code in which both projected and
exponentiated JLMS relations hold to good accuracy despite the fact that certain entanglement eigenvalues of the relevant states may be exponentially small.  We found this to be possible even in contexts in which
two-sided recovery is only approximate, and even where the recovery errors are only power-law small, so that our analysis can apply to AdS/CFT at finite values of the bulk Newton's constant $G$.   However, in order to tame the sort of potentially-large effects emphasized in \cite{Kudler-Flam:2022jwd}, for our operator-norm results we found it important to restrict the bulk side of our code to states which respect cutoffs that remain finite in AdS units even in the limit $G\rightarrow 0$.  Our technical arguments were phrased in terms of abstract and somewhat general codes for which AdS/CFT provides examples with additional special properties.

The input to our construction was an appropriate family of  `small' codes that map a so-called `bulk' Hilbert space into a so-called `boundary' Hilbert space. For our operator-norm results, the map was required to be an approximate isometry and satisfy an approximate FLM or R\'enyi FLM relation on each small code. As discussed in Section~\re{sec:AdS/CFT}, in AdS/CFT such codes are naturally constructed by projecting semiclassical states (with perturbative corrections) onto appropriate windows of HRT-area and taking $G$ to be small.

We then showed in Sections~\re{sec:originalJLMS}--\re{sec:expJLMS} that the desired (projected and exponentiated) approximate JLMS relations hold on these small codes for so-called `log-stable' bulk states whose reduced density operators on $r$ have no eigenvalues that are too small, and for which an analogous condition holds in the boundary theory. We also argued in Appendix~\ref{sec:extra} that in AdS/CFT contexts the condition on the boundary density operator in fact follows from the bulk condition.  For these operator-norm results, a positive lower bound on these bulk eigenvalues appears to be necessary to avoid the potential issues emphasized in \cite{Kudler-Flam:2022jwd}.  This condition necessarily limits the dimension of the Hilbert space $\cH_r^\a$ for each small code and,  in AdS/CFT contexts, we took this dimension to remain of order $G^0$ in the limit $G \rightarrow 0$.  We will comment further on this restriction below.
We also generalized these JLMS results in Section~\re{sec:relativelogstable} so that they hold at the level of expectation values in certain test states when relative log-stability is satisfied. In these results, $\cH_r^\a$ may be infinite-dimensional; moreover, approximate isometry is required only for the two states in question, and the appropriate FLM or R\'enyi FLM relation is required only for their mixtures.

However, assuming a notion of subsystem orthogonality between the various small codes, Section~\re{sec:largecodesubspace} was then able to sew together a large collection of such small codes into a `large' code in which both projected and exponentiated JLMS again hold to good accuracy for appropriate states.  We emphasize that, in principle, this collection can be either countably infinite or of arbitrary finite size.     Much smaller entanglement eigenvalues are allowed in the large code, so long as the small parameter is associated with having a small total probability in a given small code.  However, our results in Sections~\ref{sec:lpJLMS} and \ref{sec:expJLMSaligned} required that the density operator $\sigma$ on which the modular Hamiltonian acts be sufficiently `aligned' with the state $\rho$ used to define the modular Hamiltonian.   To derive projected JLMS on this large code, we also needed some additional conditions that enhance our notion of log-stability, though these are not needed for exponentiated JLMS.

Alternatively, as shown in Sections~\ref{sec:lpJLMSop} and \ref{sec:expJLMSop}, one may remove the restriction that $\sigma$ be appropriately aligned with $\rho$ when $\r$ satisfies the smoothness conditions of Definition~\ref{defsmooth}.
Heuristically, these conditions concern smoothness with respect to a notion of distance between small codes that is chosen to reflect the decay of certain subdominant matrix elements.  However, the technical definition was formulated without introducing any explicit such distance.

Now, as emphasized in Section~\re{sec:AdS/CFT},  the R\'enyi FLM formula for our small codes requires that they have approximately-flat entanglement spectra.  In AdS/CFT, such codes are naturally constructed from fixed-area states (and their generalizations in the presence of higher-derivative interactions).   In contrast, states in our large code will generally be far from having a flat entanglement spectrum in the sense that the various small codes it contains can have dramatically-different entanglement eigenvalues on the boundary.  As discussed in \cite{Akers:2018fow,Dong:2018seb,Dong:2019piw}, this must be the case in any code large enough to describe standard semiclassical states in AdS/CFT (for which the R\'enyi entropies of distinct replica numbers $n$ require distinct classical backgrounds).  It also appears to play a key role in recent constructions of type II$_\infty$ von Neumann algebras in perturbative gravity \cite{Witten:2021unn,Chen:2024rpx}.  In particular, any type II$_\infty$ von Neumann algebra is the tensor product of a type II$_1$ algebra with a type I$_\infty$ algebra.  This shows that the Hilbert spaces constructed in \cite{Witten:2021unn,Chen:2024rpx}  can be written as a direct sum of Hilbert spaces $\cH^\a$ on which the stated algebra is truncated to a type II$_1$ algebra.  Such $\cH^\a$ thus contain states whose entropy is maximal within the given $\cH^\a$, with more general states having entropy that is less by some {\it finite} amount.  Since the algebra has infinite dimension, this would be impossible if every degree of freedom were associated with less-than-maximal entanglement.  In this sense, one may say that the entanglement of generic states in $\cH^\a$ is `almost maximal,' and thus that each $\cH^\a$ is analogous to one of our small codes (where the bulk part of the state can have arbitrary entanglement but the entropy is typically dominated by the underlying code which has a nearly flat entanglement spectrum).\footnote{We emphasize that this is only an analogy; these $\cH^\a_{vN}$ differ in detail from our small codes $\cH^\a$, in part because our construction has a cutoff in the bulk.}  The Hilbert space on which any type II$_\infty$ algebra acts is then analogous to our large code.  Indeed, it was argued in \cite{Chen:2024rpx} that the construction of type II$_\infty$ algebras in cosmology should generally be performed in a direct-sum of fixed-area Hilbert spaces much like what we considered here.\footnote{In the de Sitter case \cite{Chandrasekaran:2022cip}, the resulting algebra describing perturbations of empty dS is type II$_1$ in part because there is no such fixed-area Hilbert space with area larger than the area of the empty dS horizon.  (The discrepancy with the more general type II$_1$ claims of \cite{Jensen:2023yxy} raises questions about whether the assumptions of that work are generally satisfied.)}

As noted above, when the FLM error parameter $\eFLM$ is non-zero, our operator-norm results based on log-stability are useful only when each $\cH_r^\a$ is finite-dimensional.  In particular, in the AdS/CFT context we took its dimension to remain of order $G^0$ in the limit $G\rightarrow 0.$  This then required the use of various cutoffs.  It is natural to ask if one can improve these results using von Neumann algebra techniques, and by perhaps emphasizing relative entropies between appropriately-aligned states rather than von Neumann entropies.   

We leave this general question as an interesting direction for future investigation.  Nevertheless, it is important to ask whether the results stated here in terms of codes with bulk cutoffs are in fact of use in studying general semiclassical bulk states (which may not necessarily respect the cutoffs imposed above).  We will return to this question in the forthcoming paper \cite{ModFlow}, where we will show that the answer is affirmative in AdS/CFT.


\acknowledgments


We would like to thank Simon Caron-Huot, Arvin Shahbazi Moghaddam, and Tom Faulkner for useful discussions. This material is based upon work supported by the Air Force Office of Scientific Research under Award Number FA9550-19-1-0360. This material is also based upon work supported by the U.S. Department of Energy, Office of Science, Office of High Energy Physics, under Award Number DE-SC0011702. This work was supported in part by the Leinweber Institute for Theoretical Physics; and
by the Department of Energy, Office of Science, Office of High Energy Physics under
Award DE-SC0025293 and QuantISED
award DE-SC0019380. We also acknowledge support from the University of California. This work is supported by the Department of Atomic Energy, Government of India, under Project Identification Number RTI-4012.

\appendix

\section{Violations of JLMS} 
\label{sec:eg}

In this appendix, we present some simple examples to demonstrate how the JLMS formula can receive large corrections. Small eigenvalues can make the JLMS formula highly sensitive to small corrections. We demonstrate this in a tensor network model, illustrating the basic mechanism of such violations. This illuminates the general mechanism for violations of the JLMS formula in examples like the one in \cite{Kudler-Flam:2022jwd} constructed using the West Coast Model of \cite{Penington:2019kki}, but the tensor network model allows for a more controlled calculation.

\begin{figure}
\centering
        \includegraphics[width=0.4
        \textwidth]{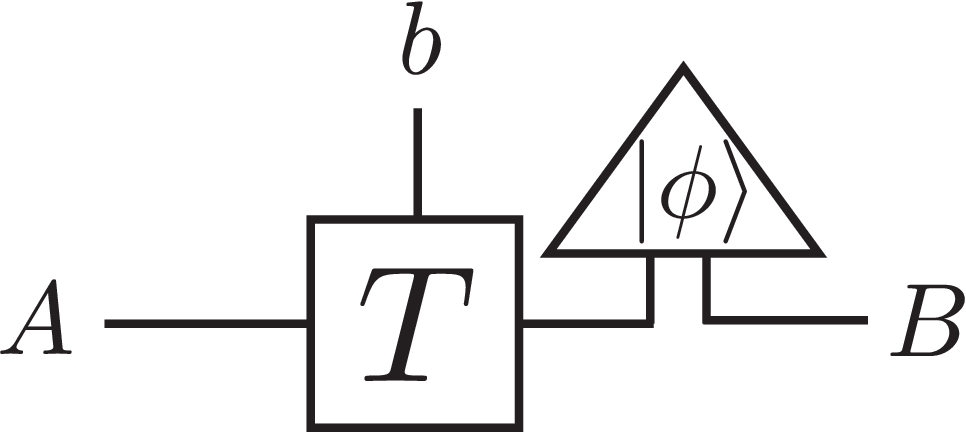}
        \caption{The QEC we consider is a network made with random tensor $T$ with the $B$ leg projected onto a state $\ket{\phi}$ with non-maximal entanglement.}
        \label{fig:RTN}
\end{figure}

We consider a QEC defined by slightly modifying the random tensor network defined in \cite{Hayden:2016cfa}. In particular, we allow the edges of the network to not project the state onto the maximally entangled Bell pairs used there, and we instead project onto edge states with non-trivial entanglement spectra.\footnote{Such constructions were also discussed in \cite{Dong:2021clv,Cheng:2022ori} and are closely related to the equilibrium approximation considered in \cite{Vardhan:2021mdy,Vardhan:2021npf}.} The simplest such model for a QEC is a network with one bulk leg and two boundary legs
as shown in \figref{fig:RTN}.  One of the boundary legs is maximally entangled as usual while the other, denoted $\ket{\phi}$, has a non-trivial entanglement spectrum. For simplicity, we choose the spectrum of the non-maximally entangled leg to consist of two eigenvalues $\frac{p_1}{d_1},\frac{p_2}{d_2}$ with corresponding degeneracies $d_1,d_2$.

In more detail, the boundary state obtained by projecting the bulk leg onto state $\ket{i}$ can be written as
\begin{equation}
        \ket{\psi_i} = \sum_{j=1}^{d_A} \left(\sum_{k=1}^{d_1} \sqrt{\frac{p_1}{d_1}} T_{ijk} \ket{j}_A \ket{k}_B + \sum_{k=d_1+1}^{d_B} \sqrt{\frac{p_2}{d_2}}T_{ijk} \ket{j}_A \ket{k}_B\right),
\end{equation}
where $T_{ijk}$ is a Haar random tensor. The probabilities and weights are normalized so that $p_1+p_2=1$, and we have used $d_1+d_2=d_B$. For such a state, the reduced density operator on subregion $A$ is given by
\begin{equation}
        \rho_A = \sum_{j,j'=1}^{d_A}\left(\sum_{k=1}^{d_1} \frac{p_1}{d_1} T_{ijk}T_{ij'k}^* \ket{j}_A \bra{j'}_A + \sum_{k=d_1+1}^{d_B} \frac{p_2}{d_2} T_{ijk} T_{ij'k}^* \ket{j}_A \bra{j'}_A \right).
\end{equation}
This shows that the density operator takes the form $\rho_A = p_1 \rho_1 + p_2 \rho_2$, where $\rho_1$ and $\rho_2$ are reduced density operators obtained from random tensor networks with maximally entangled bonds.

\begin{figure}
\centering
        \includegraphics[width=0.4
        \textwidth]{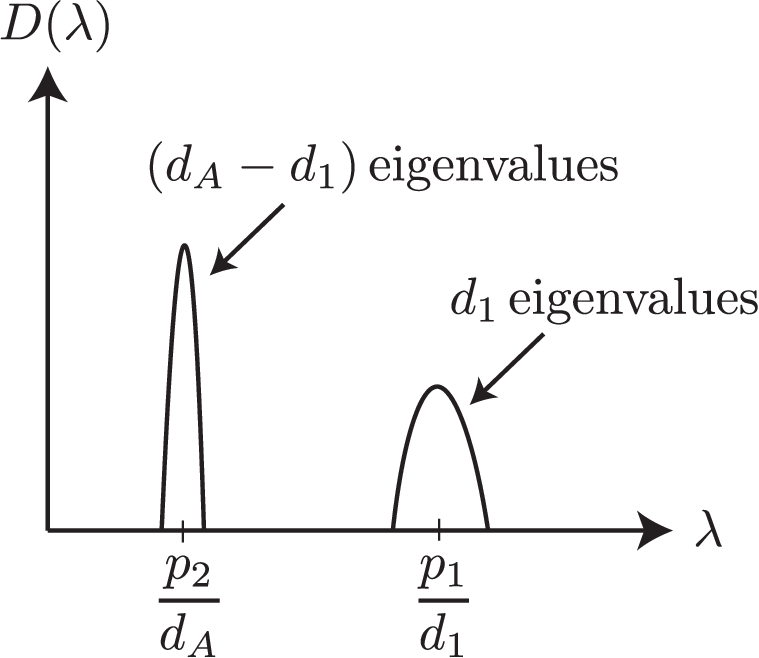}
        \caption{The approximate description of the spectrum of $\rho_A$ in the regime of interest. The important feature is that there are no zero eigenvalues.}
        \label{fig:plot}
\end{figure}

We can now use the techniques of \cite{Hayden:2016cfa}, which allows one to compute the averaged moments of $\rho_{1,2}$ as a sum over elements of the permutation group. The moments can in fact be repackaged into the resolvent function of the density operator defined as
\begin{equation}
        R_A(\lambda) = \frac{d_A}{\lambda} + \sum_{n=1}^{\infty} \frac{\Tr(\rho_A^n)}{\lambda^{n+1}},
\end{equation}
which can be used to obtain the spectrum of $\rho_A$. 

Computing the resolvent by summing over all the non-crossing diagrams is by now a familiar technique that leads to a Schwinger-Dyson equation for the resolvent as demonstrated in various cases in \cite{Penington:2019kki,Akers:2020pmf,Akers:2021pvd,Akers:2022max,Vardhan:2021mdy,Vardhan:2021npf,Shapourian:2020mkc,Dong:2021oad}. Instead of going through the detailed calculation, we refer the reader to \cite{Akers:2020pmf} where the same calculation was performed.\footnote{In that context, the result was interpreted as the spectrum arising from an incompressible bulk state consisting of two sectors in the wavefunction, whereas we are considering the sectors as part of the in-plane legs. Nevertheless, the results are identical.}
By doing the sum over moments, one obtains a Schwinger-Dyson equation for the resolvent given by
\begin{equation}
        \lambda R = d_A + \frac{p_1 R}{d_A - \frac{p_1 R}{d_1}} + \frac{p_2 R}{d_A-\frac{p_2 R}{d_2}}.
\end{equation}
Solving this cubic equation gives us the resolvent and density of states. 

For our purposes, the important fact will be that we will focus on a regime where $d_2\gg d_A \gg d_1$ and $p_1\approx 1$, so that the bulk leg is contained in the entanglement wedge of $A$, though the dimension of Hilbert space $B$ is much larger than $A$. The generic expectation in such a situation is that it leads to an absence of any zero-eigenvalues. This can be explicitly demonstrated in this example. For example, we plot a schematic spectrum in \figref{fig:plot}.

This demonstrates that orthogonal pure states get mapped to non-orthogonal density operators on the boundary subregion $A$. This is crucially true because this QEC only has approximate complementary recovery unlike the exact codes in \cite{Pastawski:2015qua}. This is a generic feature expected of holographic codes and, thus, such a code is sufficient to illustrate the general point we wanted to make. The absence of zero eigenvalues also demonstrates that one has to be careful when considering the JLMS formula relating the bulk and boundary modular Hamiltonians. For pure states in the bulk, the modular Hamiltonian would have infinite eigenvalues when regarded on the full bulk Hilbert space, whereas the boundary modular Hamiltonian has only finite eigenvalues in the regime above. This mismatch leads to the large corrections demonstrated in \cite{Kudler-Flam:2022jwd}, which can be avoided by considering log-stable states as discussed in the main text.

The spectrum in the West Coast model is approximately given by a cutoff thermal spectrum, although there is an uncontrolled region near the edge of the spectrum. We can think of this as arising from the contributions of different degeneracy sectors with different values of energy/black hole area on the JT gravity side. In these sectors, the errors in JLMS are small, and one can in fact see that there is no obvious violation of JLMS in the microcanonical ensemble for the example of pure states in the code Hilbert space as we saw above. However, when we try to combine them together to form a large code to capture the canonical ensemble, we find that small overlaps between the different sectors affect the reduced density operator of the radiation in a way that produces large corrections to the JLMS formula. Again, as we stated before, one can alternatively choose to consider only log-stable states where the effects of such non-perturbative corrections are small and the JLMS formula holds to good accuracy.

\section{Schatten $p$-norms and expectation values}
\label{app:Schatten}

\subsection*{Schatten $p$-norms}
Recall that the Schatten $p$-norm of a bounded operator $O$ is defined to be
\be
\|O\|_p \eq \(\Tr |O|^p\)^{1/p}, \qu
|O| \eq \sqrt{O^\dag O}.
\ee
It satisfies the triangle inequality:
\be\la{triangle}
\|O+O'\|_p \leq \|O\|_p + \|O'\|_p,\qu
\text{for } 1\leq p\leq \infty,
\ee
monotonicity:
\be\la{mono}
\|O\|_1 \geq \|O\|_p \geq \|O\|_{p'} \geq \|O\|_\infty,\qu
\text{for } 1\leq p\leq p'\leq \infty,
\ee
and H\"older's inequality:
\be\la{holder}
\lt\|O O' \rt\|_1 \leq \|O\|_p \|O'\|_q,\qu
\text{for } p, q\in [1,\infty] \text{ such that } \fr{1}{p} + \fr{1}{q} =1.
\ee
In particular, choosing $p=1$ and $q=\infty$ we find 
\begin{equation}
\label{eq:1inf}
\|O O' \|_1 \leq \|O\|_1 \|O'\|_\infty.
\end{equation}
For trace-class $O$, this gives the useful inequality
\be\la{trprodbound}
\lt|\Tr(O O')\rt| \leq \Tr |O O'| = \|O O'\|_1 \leq \|O\|_1 \|O'\|_\infty,
\ee
where the first step follows from a basic inequality $\lt|\Tr X\rt|\leq\Tr|X|$ for trace-class $X$, and the last step follows from H\"older's inequality \er{eq:1inf}.

We work mostly with the operator norm $\|\cdot\|_\infty$ and trace norm $\|\cdot\|_1$ in the main text. For a bounded \textit{Hermitian} $O$, one may alternatively find the operator norm using 
\be\la{onalt2}
\|O\|_\infty = \sup \big\{\lt|\Tr(\r O)\rt|: \r \text{ is a normalized density operator}\big\}.
\ee
To see this, notice that for any bounded (but not necessarily Hermitian) $O$ we have
\be\la{robound}
\lt|\Tr(\r O)\rt|\leq \|\r\|_1 \|O\|_\infty = \|O\|_\infty.
\ee
Moreover, when $O$ is Hermitian, this inequality can be saturated or approached arbitrarily closely by choosing $\r$ to be a suitable pure state.

The following lemma lets us obtain an operator-norm bound without assuming in advance that the operator is bounded.

\begin{nlemma}\la{lemmaopnorm}
Let $T:\cD\to\cH'$ and $O:\cD'\to\cH$ be linear operators that need not be bounded, where $\cH,\cH'$ are Hilbert spaces, $\cD\subseteq\cD'\subseteq\cH$, and $\cD$ is dense in $\cH$.  Suppose
\be\la{diagbound}
\left|\bigl\|T|\psi\>\bigr\|^2-\<\psi|O|\psi\>\right|
\leq C
\ee
for every normalized $|\psi\>\in\cD$.  Then $T^\dag T$ is defined on $\cD$, and $T^\dag T-O$ has a unique bounded extension $E$, with $\|E\|_\infty\leq2C$.
\end{nlemma}

\begin{proof}
For $|\phi\>,|\psi\>\in\cD$, define
\be
e(\phi,\psi):=\<T\phi|T\psi\>-\<\phi|O|\psi\>.
\ee
The bound \er{diagbound} and the complex polarization identity give
\be
\bigl|e(\phi,\psi)\bigr|\leq2C\|\phi\|\,\|\psi\|.
\ee
Thus, by the corollary to the Riesz Lemma in \cite{reed1972functional}, the sesquilinear form $e$ is represented by a unique bounded operator $E$ with $\|E\|_\infty\leq2C$.  For all $|\phi\>,|\psi\>\in\cD$,
\be
\<T\phi|T\psi\>=\<\phi|(O+E)|\psi\>.
\ee
Therefore, $T|\psi\>$ is in the domain of $T^\dag$ and
\be
T^\dag T|\psi\>=(O+E)|\psi\>
\ee
for every $|\psi\>\in\cD$.  This proves the claim.
\end{proof}

\subsection*{Expectation values}
We use the following convention for expectation values of possibly unbounded operators (either in the bulk or on the boundary).  If $O$ is a possibly unbounded positive self-adjoint operator defined on a dense domain in a Hilbert space $\cH$ and $\t$ is a trace-class (not necessarily self-adjoint) operator supported on $\cH$ (meaning that $\t=P_{\cH}\t P_{\cH}$, where $P_{\cH}$ projects onto $\cH$, even if $\tau$ is defined on a larger space), we define
\be\la{EVdef}
\<O\>_{\t}:=\lim_{\L\rightarrow\infty}\Tr\!\left[\t\,O\,\mathbb 1_{[0,\L]}(O)\right]
\ee
whenever the limit exists and is finite.  Here $\mathbb 1_{[0,\L]}(O)$ denotes the spectral projection of $O$ onto $[0,\L]$ and $\Tr$ denotes the trace on $\cH$.  Note in particular that, if $O$ is the inverse or logarithm of some positive operator $\r$ and is therefore densely defined only on the support of $\r$, then $\cH$ is understood to be that support. As a result, $\<O\>_{\t}$ is well-defined only if $\t$ is supported on $\cH$. 

For a self-adjoint $O$ that need not be positive, we extend the definition \er{EVdef} to
\be\la{EVdefgen}
\<O\>_{\t}:=
\lim_{\L_+,\L_-\rightarrow\infty}
\Tr\!\left[\t\,O\,\mathbb 1_{[-\L_-,\L_+]}(O)\right]
\ee
whenever this limit exists and is finite, with the two cutoffs removed independently.

These definitions agree with the usual trace $\Tr(\t O)$ when $O$ is bounded.  In general, they are linear in the trace-class $\t$ whenever the expectation values involved are defined.  For positive $O$ and a vector $|\psi\>$, $\<O\>_{|\psi\>\<\psi|}$ is finite exactly when $O^{1/2}|\psi\>$ is a Hilbert-space vector, and in that case
\be\la{EVvector}
\<O\>_{|\psi\>\<\psi|}=\bigl\|O^{1/2}|\psi\>\bigr\|^2.
\ee

Shifting $O$ by a real constant $b$ gives
\be\la{EVscalarshift}
\<O+b\mathbb 1\>_{\t}=\<O\>_{\t}+b\Tr\t
\ee
whenever either side is defined.

When $\t$ is positive, partial trace gives
\be\la{EVpartialtrace}
\<O_R\>_{\t_R}=\<O_R\otimes\mathds 1_{\Rb}\>_{\t}
\ee
whenever either side is defined because, at every finite cutoff, partial trace gives the same trace on the reduced and full Hilbert spaces.

We also use the following property of the expectation values defined in \er{EVdef}.

\begin{nlemma}\la{lemmainversehs}
Let $\r$ be a positive self-adjoint operator and let $\s$ be a positive trace-class operator on a possibly infinite-dimensional Hilbert space.  Then
\be
\<\r^{-1}\>_{\s^2}<\infty
\ee
if and only if the support of $\s$ is contained in the support of $\r$ and $\r^{-1/2}\s$ is a Hilbert--Schmidt operator.  In this case,
\be\la{inversehsnorm}
\left\|\r^{-1/2}\s\right\|_2^2
=\<\r^{-1}\>_{\s^2}.
\ee
\end{nlemma}

\begin{proof}
Suppose first that the expectation value is finite.  The support convention for \er{EVdef} implies that the support of $\s^2$, and hence that of $\s$, is contained in the support of $\r$.  Let $P_\L:=\mathbb 1_{[0,\L]}(\r^{-1})$ and $X_\L:=\r^{-1/2}P_\L\s$.  For $\L'\geq\L$, cyclicity of the trace gives
\be
\left\|X_{\L'}-X_\L\right\|_2^2
=\Tr\!\left[\s^2\r^{-1}(P_{\L'}-P_\L)\right]\longrightarrow0
\ee
as $\L,\L'\to\infty$.  Thus $X_\L$ converges in Hilbert--Schmidt norm and hence on every vector $|\psi\>$ as $\L\to\infty$.  Since $P_\L\s|\psi\>\to\s|\psi\>$ and $\r^{-1/2}$ is a closed operator, the limit of $X_\L |\psi\> = \r^{-1/2}P_\L\s |\psi\>$ is $\r^{-1/2}\s|\psi\>$ for every such $|\psi\>$.  Therefore $\r^{-1/2}\s$ is a Hilbert--Schmidt operator.

Conversely, suppose that the support of $\s$ is contained in the support of $\r$ and $X:=\r^{-1/2}\s$ is a Hilbert--Schmidt operator.  Since $P_\L$ commutes with $\r^{-1/2}$,
\be
X_\L=P_\L X\longrightarrow X
\ee
in Hilbert--Schmidt norm as $\L\to\infty$. We therefore obtain
\be
\left\|\r^{-1/2}\s\right\|_2^2
=\lim_{\L\to\infty}\left\|X_\L\right\|_2^2
=\lim_{\L\to\infty}\Tr\!\left[\s^2\r^{-1}P_\L\right]
=\<\r^{-1}\>_{\s^2}.
\ee
This proves \er{inversehsnorm}.
\end{proof}

\section{Approximate isometry from full-system FLM}
\label{app:FLMiso}

As noted in the main text,  when $\Rb$ and $\rb$ are empty (i.e., when $R$ and $r$ are the entire system), it is possible to use the approximate FLM formula \eqref{flme} to derive a rather weak form of \eqref{appiso}  (with the  bound $\eisosmall = 2 \sqrt{\eFLM}$).   In particular, we will now show the following:

\begin{nlemma}[Approximate isometry of the small code from FLM]\la{lemmaappiso}
If the approximate FLM formula \er{flme} holds for the $\alpha$-sector on the entire system (i.e., $\Rb$ and $\rb$ are empty), then $V_\a$ must be an approximate isometry in the sense that, after an appropriate rescaling of $V_\a$, it satisfies \er{appiso} with $\eisosmall = 2 \sqrt{\eFLM}$.
In particular, this bound is independent of the dimensions of $\cH^\a$ and $\tH$, and holds even when either dimension becomes large.
\end{nlemma}

\begin{proof}
First, we rescale $V_\a$ by an appropriate constant so as to minimize $\d:= \|V_\a^{\dag} V_\a -\mathbb 1^\a \|_\infty$. If $\d=0$, the result is immediate. For $\delta \neq 0$, let us suppose for now that both spectral endpoints of $V_\a^\dag V_\a$ correspond to normalizable eigenvectors. Since $\delta$ is minimal, the largest and smallest eigenvalues of $V_\a^{\dag} V_\a -\mathbb 1^\a$ must be $\pm \delta$.  The largest and smallest eigenvalues of $V_\a^{\dag} V_\a$ are then $1 \pm  \d$.  

Let the two corresponding eigenstates be $|\y^\a_\pm\> \in \cH^\a$. Their associated boundary states are
\be
|\td\y^\a_\pm\> = \fr{V_\a |\y^\a_\pm\>}{\sqrt{\<\y^\a_\pm | V_\a^\dag V_\a |\y^\a_\pm\>}} = \fr{V_\a |\y^\a_\pm\>}{\sqrt{1\pm \d}}.
\ee
Clearly, $|\td\y^\a_\pm\>$ are orthogonal to each other: $\<\td\y^\a_+|\td\y^\a_-\> \propto \<\y^\a_+| V_\a^\dag V_\a |\y^\a_-\> \propto \<\y^\a_+|\y^\a_-\> =0$.

Using the approximate FLM formula on the entire system, we may write $\td\r^\a_R$ as $\td\r^\a$ and $\r^\a_r$ as $\r^\a$ in \er{flme}:
\be\la{flmentire}
\lt| S(\td\r^\a) - \( \fr{A^\a}{4G} + S(\r^\a) \) \rt| \leq \eFLM.
\ee
This holds for any state $\r^\a$ on $\cH^\a$.
We now apply it on two such states $\r^\a_1$, $\r^\a_2$. The first is the maximally mixed bulk state on the span of the states $|\y^\a_\pm\>$:
\be\la{r1def}
\r^\a_1 = \fr{1}{2} \(|\y^\a_+\> \<\y^\a_+| + |\y^\a_-\> \<\y^\a_-|\),
\ee
for which the corresponding boundary state is
\be\la{tdr1}
\td\r^\a_1 = \fr{1+\d}{2} |\td\y^\a_+\> \<\td\y^\a_+| + \fr{1-\d}{2} |\td\y^\a_-\> \<\td\y^\a_-|.
\ee
The second bulk state $\r^\a_2$ is
\be\la{r2def}
\r^\a_2 = \fr{1-\d}{2} |\y^\a_+\> \<\y^\a_+| + \fr{1+\d}{2} |\y^\a_-\> \<\y^\a_-|,
\ee
which is designed to correspond to the maximally mixed boundary state on the span of $|\td\y^\a_\pm\>$:
\be\la{tdr2}
\td\r^\a_2 = \fr{1}{2} \(|\td\y^\a_+\> \<\td\y^\a_+| + |\td\y^\a_-\> \<\td\y^\a_-|\).
\ee
Now, applying the approximate FLM formula \er{flmentire} to both  $\r^\a_1$ and $\r^\a_2$ implies
\ba
\( \fr{A^\a}{4G} + S(\r^\a_1) \) - S(\td\r^\a_1) &\le \eFLM, \ \ \ {\rm and} \\
S(\td\r^\a_2) - \( \fr{A^\a}{4G} + S(\r^\a_2) \) &\le \eFLM.
\ea
Adding these two inequalities yields
\be\la{sr12bd}
\[S(\r^\a_1) - S(\r^\a_2)\] + \[S(\td\r^\a_2) - S(\td\r^\a_1)\] \le 2\eFLM.
\ee
Using \er{r1def}--\er{tdr2}, we find
\ba
S(\r^\a_1) - S(\r^\a_2) = S(\td\r^\a_2) - S(\td\r^\a_1) &= \ln 2 + \fr{1-\d}{2} \ln \fr{1-\d}{2} + \fr{1+\d}{2} \ln \fr{1+\d}{2}\\
&= \fr{1-\d}{2} \ln (1-\d) +\fr{1+\d}{2} \ln (1+\d) \\
&\geq \fr{1}{2}\[-\d + \fr{(-\d)^2}{4} +\d + \fr{\d^2}{4}\] = \fr{\d^2}{4},
\label{eq:delta2bound}
\ea
where we used $0\leq \d<1$, as well as $(1+x)\ln (1+x) \geq x +x^2/4$ for any $x\in (-1,1)$.  The latter inequality is easily verified by noting that it holds at the endpoints and that the difference between the left and right sides vanishes as $\cO(x^2)$ as $x\to 0$ and is convex for any $x\in (-1,1)$. Using \eqref{eq:delta2bound} in \er{sr12bd}, we find
\be
\lt\|V_\a^{\dag} V_\a -\mathbb 1^\a \rt\|_\infty = \d \le 2 \sqrt{\eFLM},
\ee
thus proving approximate isometry \er{appiso} with $\eisosmall = 2\sqrt{\eFLM}$.

If either spectral endpoint of $V_\a^\dag V_\a$ is not an eigenvalue with a normalizable eigenvector, we choose pure states in disjoint spectral intervals approaching the two endpoints and apply the argument above to the restriction of $V_\a$ to their two-dimensional span, with its own $\d$-minimizing rescaling. The corresponding values of $\d$ converge to the value for the full $V_\a$, so the same bound follows in the limit.
\end{proof}

\section{Log-stability Condition~2 follows from the gravitational path integral}
\label{sec:extra}

This appendix provides an argument that, in AdS/CFT, Condition~2 of our definition of log-stability (Definition~\ref{defsmallstable}) follows from Condition~1 and the use of the bulk gravitational path integral when the parameter $\etail$ in Condition~1 is not exponentially small in the limit $G\rightarrow 0$.  In particular, given any two bulk states $\rho^\alpha$, $\sigma^\alpha$ in the same small code Hilbert space, we wish to use the gravitational path integral to show the following semiclassical version of \er{treq}:
\be\la{treq2}
\<(\td\r_R^\a)^{-1}\>_{(\td\s_R^\a)^2}
\simeq
\<(\r_r^\a)^{-1}\>_{(\s_r^\a)^2},
\ee
and thus that imposing Condition~1 is sufficient in this approximation. We will of course assume that the states $\rho^\alpha$, $\sigma^\alpha$ are defined by imposing boundary conditions on the gravitational path integral which can be interpreted as activating sources in the dual CFT, and that such path integrals may be evaluated semiclassically.
These sources (or boundary conditions from the bulk perspective) can be chosen to be reflection-symmetric across an appropriate cut of the boundary. Here we take this reflection symmetry to include a complex conjugation of all sources (or boundary conditions).

\begin{figure}[h!]
\centering
        \includegraphics[width=0.9
        \textwidth]{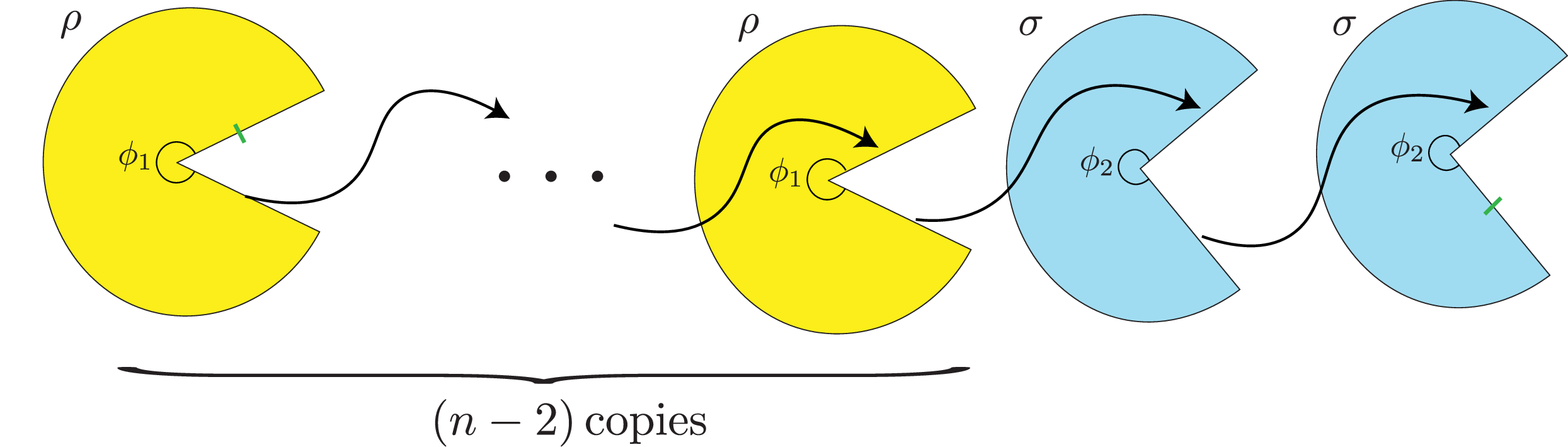}
        \caption{The saddle $g$ that computes $\Tr_R\left( \td\s^\a_R (\td\r^\a_R)^{n-2} \td\s^\a_R\right)$.  It is constructed by gluing together copies of saddles $g_1$ (yellow) and $g_2$ (light blue) around the HRT surface in a cyclic manner.}
        \label{fig:cond2}
\end{figure}

In particular, we assume that the bulk path integrals for $\Tr(\hat\rho^\alpha)$, $\Tr(\hat\sigma^\alpha)$ are both dominated by saddles which we call $g_1$, $g_2$, and that the corresponding bulk traces in the entanglement wedge $r$ (i.e., $\tr_r (\rho_r^\alpha)$, $\tr_r (\sigma_r^\alpha)$) are computed by path integrals for quantum fluctuations around such backgrounds.\footnote{\label{foot:language}Our language here is adapted to the familiar case where such fluctuations are perturbatively small.  But the argument also applies in more general contexts so long as fluctuations of the HRT area $A$ remain restricted to a small window.}  Since these are fixed-area saddles, they need not be smooth and, in particular, the two-dimensional space transverse to the HRT surface will generally have a conical defect.  We parameterize the defect in saddle $g_i$ by its opening angle $\phi_i$ around the HRT surface as shown in \figref{fig:cond2}.  Since the area was the only quantity fixed on the HRT surface, each angle will be constant along the relevant defect; see e.g.\ \cite{Dong:2018seb} for details.

We also assume that the (potentially complex) saddles $g_1$, $g_2$ are invariant under a bulk version of the reflection-symmetry that preserves their boundary conditions, and that the fixed-area HRT surface lies on the reflection-symmetric hypersurface in each saddle. There is thus a preferred way to cut open each saddle along the part of the reflection-symmetric surface extending from the HRT surface to the boundary region  $R$; see \figref{fig:cond2}. Lorentz-signature bulk descriptions of the associated states would then be defined by the data on this cut.   We will call the cut $\Sigma_r$ below.

Associated CFT path integrals then define $\td\rho_R^\alpha$, $ \td\sigma_R^\alpha$, and for positive integer $n\geq 2$, a CFT path integral that computes $\Tr_R\( \td\s^\a_R (\td\r^\a_R)^{n-2} \td\s^\a_R\)$ can be constructed by concatenating appropriate replicas of these boundary conditions.
We now wish to argue that, in the semiclassical limit, the corresponding gravitational path integral gives
\begin{equation}
\label{eq:AppD}
\Tr_R\( \td\s^\a_R (\td\r^\a_R)^{n-2} \td\s^\a_R \)
\simeq
e^{-(n-1)\frac{A^\a}{4G}}
\tr_r\( \s^\a_r (\r^\a_r)^{n-2}\s^\a_r\),
\end{equation}
which may be written as
\be
\<(\td\r_R^\a)^{n-2}\>_{(\td\s_R^\a)^2}
\simeq
e^{-(n-1)\frac{A^\a}{4G}}
\<(\r_r^\a)^{n-2}\>_{(\s_r^\a)^2}.
\ee
Analytically continuing this in $n$ and taking the limit $n\rightarrow 1$ will then yield the desired result \eqref{treq2}.

To do so, we first note that Condition~1 in fact requires the backgrounds $g_1$, $g_2$ to coincide on $\Sigma_r$ (if $\etail$ is not exponentially small).  By this we mean that both the induced fields on $\Sigma_r$ and normal derivatives at $\Sigma_r$ agree in both backgrounds.
This is because both states $\rho^\a$, $\sigma^\a$ are required to be on the same small code Hilbert space, and $\rho_r^\a$ is not allowed to have exponentially small eigenvalues.  Since $\sigma_r^\a$ is normalized, it follows that $\tr_r(\rho_r^\a \sigma_r^\a)$ cannot be exponentially small. In the limit $G \rightarrow 0$, both states must thus correspond to the same classical Lorentz-signature geometry in the domain of dependence of $\Sigma_r$.  However, the backgrounds can differ elsewhere and, in particular, the angles $\phi_1$, $\phi_2$ around the HRT surfaces in the Euclidean (or complex) saddles need not coincide.

We can now use this observation to construct a saddle $g$ for the bulk path integral for \eqref{eq:AppD} by gluing $n-2$ copies of $g_1$ in sequence with two copies of $g_2$; see again \figref{fig:cond2}. Assuming that this $g$ is the dominant saddle, the left-hand side of \eqref{eq:AppD} is calculated semiclassically as
\be
\label{eq:g123}
\Tr_R\( \td\s^\a_R (\td\r^\a_R)^{n-2} \td\s^\a_R \)
\simeq
\e_{window}^{1-n}\fr{e^{-I[g]}}{e^{-(n-2)I[g_1]-2I[g_2]}}
\tr_r\( \s^\a_r (\r^\a_r)^{n-2} \s^\a_r\).
\ee
Here $I[g]$ denotes the classical action, while the factor $\e_{window}^{1-n}$ accounts for the normalization of the area-window constraints. Away from the HRT surface, $I[g]$ precisely cancels against $(n-2)I[g_1]+2I[g_2]$, so it suffices to compare their contributions at the HRT surface. This is similar to a calculation in \cite{Dong:2018seb}. The result is
\ba
& I[g]-(n-2)I[g_1]-2I[g_2] \\
={}& \fr{A}{8\pi G} \[((n-2)\p_1 +2\p_2 -2\pi) -(n-2)(\p_1-2\pi) -2(\p_2-2\pi) \] \\
={}& \fr{A}{4G}(n-1).
\ea
Replacing the area-window midpoint $A$ by $A^\a=A+4G\log\e_{window}$ then gives \er{eq:AppD}, as desired.

It is worth pointing out that we are assuming that analytic continuation gives a reliable result, at least in the situation we use it in (where the right-hand side of \er{treq2} is bounded by $1/\etail$ and is not exponentially large); in other words, we assume that the error made in the semiclassical approximation \er{eq:g123} does not become large in the $n\to 1$ limit. This type of assumption has been made in similar contexts, including the Lewkowycz-Maldacena derivation of the RT formula, although some readers may be concerned that here we are analytically continuing to a new regime involving inverses of density operators. It would be useful in the future to analyze error bounds in such situations.

The above argument assumed a simple dominant saddle $g$, which is the natural saddle in the presence of a single candidate extremal surface. In more general situations, where we have multiple candidate HRT surfaces, e.g., two surfaces with areas $A_1$ and $A_2$ such that $A_1<A_2$, there can be other subleading saddles (see \cite{Dong:2020iod,Marolf:2020vsi,Akers:2020pmf,Kudler-Flam:2022jwd} for discussion about these saddles). In this case, the analytic continuation is harder to do, but we expect such effects to be suppressed by factors of $e^{(A_1-A_2)/4G}$.

\section{Proof of Lemma~\re{lemmacondre}}
\label{app:lemmaproof}
Here we prove Lemma~\re{lemmacondre} (which we first restate below).

{
\renewcommand{\thenlemma}{\re{lemmacondre}}
\begin{nlemma}[Restated]
Let $\r$, $\s$ be states in some Hilbert space, not necessarily of finite dimension. Suppose
\be\la{trrssbd3}
\<\r^{-1}\>_{\s^2}\leq1/\etail.
\ee
Then for every $\e \in [0,1)$ and every real $s$, we have
\be\la{condreboundapp}
\lt|\fr{1}{1+is} \[\Tr \(\r_\e^{1+is}\) - \Tr \(\r^{1+is}\)\] - \e \Tr \[\r^{is} (\s - \r)\] \rt| \leq \fr{|s| \e^2}{2(1-\e)\etail},
\ee
where $\r_{\e}:=\r+\e(\s-\r)$.
\end{nlemma}
\addtocounter{nlemma}{-1}
}

\begin{proof}
First suppose that the Hilbert space is finite-dimensional and that $\r$ is invertible. Define $\d\r := \e(\s-\r)$. Our goal is to bound
\be
\D := \fr{1}{1+is} \[\Tr \(\r_\e^{1+is}\) - \Tr \(\r^{1+is}\)\] - \Tr \(\r^{is} \d\r\).
\ee
To do so, we first recall a formula for the derivative of the exponential of an operator $A(x)$,
\be
\fr{d}{dx} e^{A(x)} = \int_0^1 d\q\, e^{(1-\q)A} \(\fr{d A}{dx}\) e^{\q A},
\ee
which can be obtained from (10.15) of \cite{Higham:2008}. We also recall a similar formula for the logarithm of a positive-definite operator $A(x)$,
\be
\fr{d}{dx} \log A(x) = \int_0^\infty dt \(A+t\)^{-1} \(\fr{d A}{dx}\) \(A+t\)^{-1},
\ee
which can be obtained from (11.10) of \cite{Higham:2008} by a change of variables. Combining these two formulas and applying them to $\r_{x\e}^{1+is}$ for $x\in[0,1]$ gives
\be\la{dr1is}
\fr{d}{dx} \r_{x\e}^{1+is} = (1+is) \int_0^1 d\q \int_0^\infty dt\, \r_{x\e}^{(1+is)(1-\q)} \(\r_{x\e}+t\)^{-1} \d\r \(\r_{x\e}+t\)^{-1} \r_{x\e}^{(1+is)\q}.
\ee
Its trace is given by
\be
\fr{d}{dx} \Tr \r_{x\e}^{1+is} = (1+is) \int_0^1 d\q \int_0^\infty dt\, \Tr\[ \r_{x\e}^{1+is} \(\r_{x\e}+t\)^{-2} \d\r \] = (1+is) \Tr\[ \r_{x\e}^{is} \d\r \],
\ee
where we have used cyclicity of the trace and commutativity of any two functions of $\rho_{x\e}$.
This useful formula allows us to rewrite $\D$ as
\ba
\D &= \int_0^1 dx \Tr\[ \(\r_{x\e}^{is} - \r^{is}\) \d\r \] = \int_0^1 dx \int_0^x dy \fr{d}{dy} \Tr\( \r_{y\e}^{is} \d\r \)\\
&= \int_0^1 dx (1-x) \fr{d}{dx} \Tr\( \r_{x\e}^{is} \d\r \). \la{drewrite}
\ea

To bound this, we replace $1+is$ by $is$ in \er{dr1is} and multiply it with $\d\r$ to obtain
\be\la{drisdr}
\fr{d \r_{x\e}^{is}}{dx} \d\r = is \int_0^1 d\q \int_0^\infty dt\, \r_{x\e}^{is(1-\q)} \(\r_{x\e}+t\)^{-1} \d\r\, \r_{x\e}^{is\q} \(\r_{x\e}+t\)^{-1} \d\r.
\ee
For any $t>0$, H\"older's inequality \er{holder} gives the bound
\ba
\lt\| \r_{x\e}^{is(1-\q)} \(\r_{x\e}+t\)^{-1} \d\r\, \r_{x\e}^{is\q} \(\r_{x\e}+t\)^{-1} \d\r \rt\|_1 &\leq \lt\| \r_{x\e}^{is(1-\q)} \rt\|_\infty \lt\| \r_{x\e}^{is\q} \rt\|_\infty \lt\| \(\r_{x\e}+t\)^{-1} \d\r \rt\|_2^2 \\
& = \Tr \[ \d\r \(\r_{x\e}+t\)^{-2} \d\r\].
\ea
Taking the trace of \er{drisdr} and using this bound gives
\ba
\lt| \fr{d}{dx} \Tr\( \r_{x\e}^{is} \d\r \) \rt| &\leq |s| \int_0^1 d\q \int_0^\infty dt\, \Tr \[ \d\r \(\r_{x\e}+t\)^{-2} \d\r\] = |s| \Tr \[ \d\r\, \r_{x\e}^{-1} \d\r\]\\
&\leq \fr{|s|}{1-x\e} \Tr \[ \d\r\, \r^{-1} \d\r\] = \fr{|s| \e^2}{1-x\e} \(\<\r^{-1}\>_{\s^2}-1\) \leq \fr{|s| \e^2}{(1-\e)\etail}, \la{trdrisdr}
\ea
where in passing to the second line we used $\r_{x\e} =(1-x\e) \r+x\e\s \geq (1-x\e)\r$ and thus $\r_{x\e}^{-1}\leq(1-x\e)^{-1}\r^{-1}$. In the last step we used $x\in[0,1]$ and the assumed condition \er{trrssbd3}.
Integrating \er{trdrisdr} and using \er{drewrite} then gives the desired bound:
\be\la{deltabound}
|\D| \leq \int_0^1 dx (1-x) \fr{|s| \e^2}{(1-\e)\etail} = \fr{|s| \e^2}{2(1-\e)\etail}.
\ee

For the general case, Lemma~\re{lemmainversehs} shows that the support of $\s$ is contained in the support of $\r$ and that $\r^{-1/2}\s$ is Hilbert--Schmidt. We work on the support $\cH$ of $\r$. For $\L\geq1$, let $P_\L:=\mathbb 1_{[1/\L,1]}(\r)$ and $\bar P_\L:=\mathbb 1-P_\L$. Since $\r$ is trace-class, $P_\L$ has finite rank. Choose a normalized vector $|0\>$ orthogonal to $\cH$. On the finite-dimensional space $P_\L\cH\oplus\mathbb C|0\>$, define the trace-preserving compressions
\be
\r_\L:=P_\L\r P_\L+r_\L|0\>\<0|,\qqu
\s_\L:=P_\L\s P_\L+s_\L|0\>\<0|,
\ee
where $r_\L:=\Tr(\bar P_\L\r)$ and $s_\L:=\Tr(\bar P_\L\s)$. If $r_\L=0$, support inclusion gives $s_\L=0$, and we omit the $|0\>$ block and its contributions below. In either case, we have
\be\la{eq:Q2compression}
\lt\<\r_\L^{-1}\rt\>_{\s_\L^2}=\Tr\!\left[P_\L\s P_\L(P_\L\r P_\L)^{-1}P_\L\s P_\L\right]+\fr{s_\L^2}{r_\L}\leq\<\r^{-1}\>_{\s^2},
\ee
because the first term is the squared norm $\|P_\L\r^{-1/2}\s P_\L\|_2^2$, and for the second term, Cauchy--Schwarz gives
\be
s_\L^2=\left|\Tr\!\left[(\bar P_\L\r^{1/2})(\bar P_\L\r^{-1/2}\s\bar P_\L)\right]\right|^2
\leq\bigl\|\bar P_\L\r^{1/2}\bigr\|_2^2\bigl\|\bar P_\L\r^{-1/2}\s\bar P_\L\bigr\|_2^2=r_\L\bigl\|\bar P_\L\r^{-1/2}\s\bar P_\L\bigr\|_2^2.
\ee
Thus the finite-dimensional bound \er{deltabound} applies to the states $\r_\L$, $\s_\L$ with the same right-hand side.

As $\L\to\infty$, the states $\r_\L$, $\s_\L$, and their mixture $\r_{\e,\L}:=(1-\e)\r_\L+\e\s_\L$ converge in trace norm to $\r$, $\s$, and $\r_\e$, respectively, after extending all states by zero to $\cH\oplus\mathbb C|0\>$.
It remains to justify passing to the limit on the left-hand side of \er{condreboundapp}.
Recall that, for density operators $\t_\L, \t$, convergence $\t_\L\to\t$ in trace norm implies that their decreasing eigenvalue lists $\lambda_j^{(\L)},\lambda_j$ converge in the $\ell^1$ norm. Since $x^{1+is}$ (extended by zero at $x=0$) satisfies $|x_1^{1+is}-x_2^{1+is}|\le\sqrt{1+s^2}|x_1-x_2|$ for $x_1,x_2\geq0$, we find
\be
\left|\Tr\(\t_\L^{1+is}\)-\Tr\(\t^{1+is}\)\right|\leq\sqrt{1+s^2}\sum_j\bigl|\lambda_j^{(\L)}-\lambda_j\bigr|\longrightarrow0.
\ee
We apply this to $\r_\L$ and $\r_{\e,\L}$. Convergence of the remaining linear term follows from
\be
\Tr\!\left[\r_\L^{is}(\s_\L-\r_\L)\right]=\Tr\!\left[\r^{is}P_\L(\s-\r)P_\L\right]+r_\L^{is}(s_\L-r_\L) \longrightarrow \Tr\!\left[\r^{is}(\s-\r)\right].
\ee
Taking $\L\to\infty$ in the finite-dimensional bound proves the lemma.
\end{proof}

\addcontentsline{toc}{section}{References}
\bibliographystyle{JHEP}
\bibliography{references}

\end{document}